# Symmetry breaking of three self-organization rules:

# A general theory for the origin of complexity

Wen-Hao Wu[1], Ze-Zheng Li[1], Wen-Xu Wang[1,2,3,4,*]

1. School of Systems Science, Beijing Normal University, Beijing, 100875, China;
2. State Key Laboratory of Cognitive Neuroscience and Learning, Beijing Normal University, Beijing 100875, China;
3. DG/McGovern Institute for Brain Research, Beijing Normal University, Beijing 100875, China;
4. Chinese Institute for Brain Research, Beijing, 102206, China.

*Email: wenxuwang@bnu.edu.cn

## Abstract

Complex spatiotemporal patterns in nature significantly challenge reductionism-based modern science. The lack of a paradigm beyond reductionism hinders our understanding of the emergence of complexity. The diversity of countless patterns undermines any notion of universal mechanisms. Here, however, we show that breaking the symmetry of three simple and self-organization rules give rise to nearly all patterns in nature, such as a wide variety of Turing patterns, fractals, spiral, target and plane waves, as well as chaotic patterns. The symmetry breaking is rooted in basic physical quantities, such as positive and negative forces, space, time and bounds. Besides reproducing the hallmarks of complexity, we discover some novel phenomena, such as abrupt percolation of Turing patterns, phase transition between fractals and chaos, chaotic edge in travelling waves, etc. Our asymmetric self-organization theory established a simple and unified framework for the origin of complexity in all fields, and unveiled a deep relationship between the first principles of physics and the complex world.

## I Introduction

Complex spatiotemporal patterns found in nature challenges reductionism in the sense that even though we have full knowledge of ingredients in a complex system, its emergent behaviors remain highly unpredictable [Anderson, 1972; Barabási, 2012]. The mystery was firstly unveiled by Turing in 1952 via his reaction-diffusion model [Turing, 1952]. Interestingly, complex patterns can emerge from diffusive processes and reactions. Since then, much effort has been dedicated to exploring diverse patterns that arise from complex dynamics, including the Belousov–Zhabotinsky chemical-oscillation reaction[Zhabotinsky, 1964] , spiral and target waves in cardiac dynamics and species coexistence [Davidenko et al., 1992; Reichenbach et al., 2007; Takagi et al., 2003; Wang et al., 2011], various biological (Turing) patterns accompanied by embryonic development [Meinhardt, 2009; Negrete & Oates, 2021; Pourquié, 2003], hexagonal patterns from Rayleigh–Bénard convection [Ahlers et al., 2009; Grossmann et al., 2016], spatial patterns in bacterial colonies [Kerr et al., 2002; Palacci et al., 2013], hypercolumn organization in the primary visual cortex [Gilbert & Wiesel, 1989; Hubel & Wiesel, 1977; Livingstone & Hubel, 1984], and many fractal patterns on the earth [Mandelbrot, 1984].

Despite a variety of mathematical models, we still lack a deep understanding of mechanisms underlying complex patterns and dynamics. For example, the relevance of reaction-diffusion models to embryo development is still debated in the biological field [Kondo & Miura, 2010]. Furthermore, merely providing mathematical approximations and phenomenological fittings is not the ultimate goal of studying complex systems [Wolfram, 2002]. Instead, the key to understanding complexity is to uncover self-organization mechanisms from first principles. A fundamental question is if there exist any general mechanisms underlying diverse complex systems in many fields. There has been a conventional belief that it is unlikely to propose a general theory for interpreting and predicting all complex phenomena, because every complex system has its own emergent mechanisms and has to be studied individually. This belief has led to Philip W. Anderson's well-known article "More is different" [Anderson, 1972; Barabási, 2012]. In fact, it has been a consensus that the goal of studying complex systems is to seek out their unique mechanisms.

In this paper, we challenge the common view of complexity and call for a paradigm shift in exploring complex phenomena across disciplines. In particular, we propose a general theory by integrating the first principles of physics and three self-organization rules, i.e., individuals' positive and negative forces, linear combination of received forces and updating bounded states driven by joint force. The key to shaping complex patterns lies in the symmetry breaking in certain aspects of the three simple rules, including breaking the symmetry of positive-negative (excitation-inhibition) forces, breaking spatial symmetry of media (substrates), breaking time symmetry of forces, breaking the symmetry of initial configurations and breaking the symmetry of interrelationships. Empirical evidence of the symmetry breaking is ubiquitous in biological, chemical and physical systems, such as in molecular regulation during biological development [Tsiairis & Aulehla, 2016], differences between excitatory and inhibitory neurons [Luo, 2020], nonlinear chemical reactions [Kuramoto, 1984], and the fundamental forces in physics. The symmetry breaking can occur spontaneously in physical and chemical systems in certain circumstances, or encoded in DNA in simple but unknown ways. Our theory based on the repetition of microscopic rules cannot be captured by differential equations, which in essence differs from the models resting on reaction-diffusion dynamics [Gierer & Meinhardt, 1972; Kondo & Miura, 2010; Kondo et al., 2021; Krause et al., 2021]. We substantiate the theory by simulations of both static patterns, such as Turing patterns and fractals, and dynamic patterns, such as travelling waves and chaotic patterns. Crucial features of the patterns and waves, such as phase transitions, abrupt pattern percolation, chaotic dynamics and wave properties, are accurately predicted by our analytical results. Our theory shares the same perspective of emergence with Cellular Automata in that evolution governed by simple rules can generate unpredictable and incredible patterns. However, our asymmetric self-organization theory consists of a single set of rules rather than uncounted rules in Cellular Automata. In other words, we discover universal and fundamental principles for the origin of complex patterns and phenomena in the world.

## II The asymmetric self-organization theory

### The three self-organization rules

**Rule I:** Spatially local positive and negative forces. $N$ individuals fully occupy a two-dimension discrete space with scale $L \times L = N$. Every individual exerts both positive and negative forces to its local neighbors within a circle range (see Fig. 1(a)). There are four parameters that determine the range and strength of interactions, denoted by $r_+$ and $w_+$ for the radius and strength of positive forces, respectively, and by $r_-$ and $w_-$ for negative forces, respectively. In general, certain functions can be assigned to the strengths, such as a decreasing function from the center. However, we found that detailed function forms do not affect qualitative features of patterns. Thus, for simplicity, we assume that $w_+$ and $w_-$ are constant within the interaction range (see Fig. 1(a)). Either the positive or negative force from an individual is the product of the individual' s state $s$ and the strength $w_+$ or $w_-$. The premise of both local positive and negative forces is self-evident in nature and human societies.

**Rule II:** Linear combination of received forces on individuals. Each individual at an arbitrary location (site) receives both positive and negative forces from its local neighbors. Apparently, the range of receiving positive and negative forces is as well $r_+$ and $r_-$ with constant strength $w_+$ and $w_-$. There could be many ways to define a joint force. Without loss of generality, we still employ a linear combination as the simplest and most general representative. The joint force $F_i$ received by an arbitrary individual, say, $i$ at step $t$ can be formulated as

$$F_i(t) = \sum_{|r_j - r_i| < r_+} s_j(t) \cdot w_+ \; - \sum_{|r_k - r_i| < r_-} s_k(t) \cdot w_-,$$

where $|\boldsymbol{r}_j - \boldsymbol{r}_i|$ is the spatial distance between individual $j$ and $i$, $r_+$ and $r_-$ are the range (radius) of positive and negative forces, respectively, $w_+$ and $w_-$ are the strength of positive and negative forces, respectively, $s_j(t)$ and $s_k(t)$ are the state value of individual $j$ and $k$ at step $t$, respectively, $s_j(t)w_+$ and $s_k(t)w_-$ are the positive and negative force from individual $j$ and $k$, respectively. The first and second sum on the right-hand side are the combined positive and negative forces from individuals within the range $r_+$ and $r_-$ centered at individual $i$ (see Fig. 1(a)).

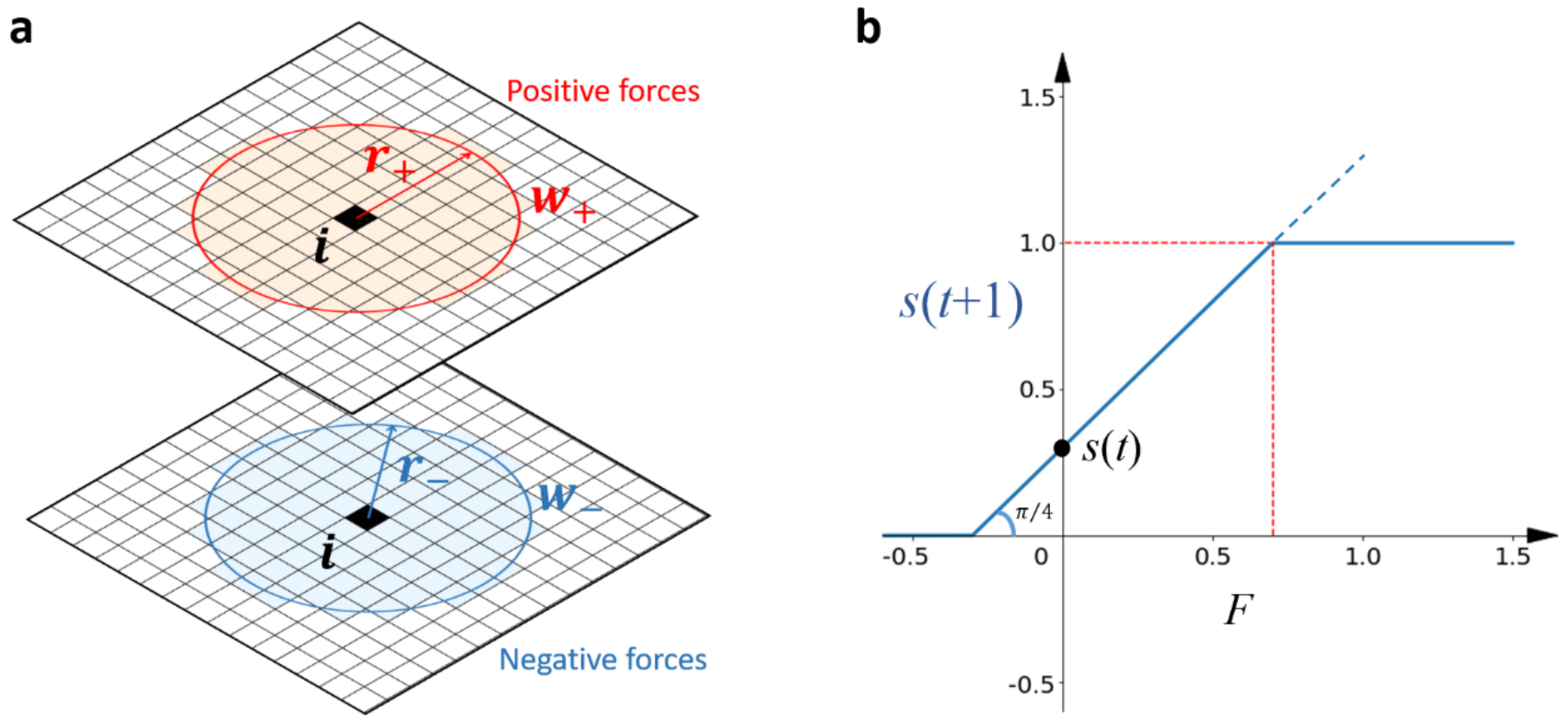

**Fig. 1: Schematic illustrations of forces and their joint effect on state updating.** (**a**) Strength $w_+$, $w_-$ and range $r_+$, $r_-$ of both positive and negative forces from an arbitrary individual, say, $i$. Colorized individuals (sites) within a circle receive both positive and negative force from $i$. In the meantime, all individuals exert forces to their local individuals with the same $w_+$, $w_-$, $r_+$ and $r_-$. (**b**) State $s(t+1)$ at step $t+1$ as a function of the joint force $F$ added to state $s(t)$ at step $t$. All states are bounded between zero and one, or, the upper bound is released.

**Rule III:** Updating bounded states driven by joint forces. State $s_i(t+1)$ of individual $i$ at time $t+1$ is determined by the joint force $F_i(t)$ it received at time step $t$ and its state $s_i(t)$ at time $t$. Analogous to the joint force, we still adopt a linear relationship between individual states and received joint forces, but with a lower and an upper bound (or only a lower bound) of state. $s_i(t+1)$ is calculated via

$$\begin{aligned} s_i(t+1) &= s_i(t) + f[F_i(t)] \\ &= s_i(t) + F_i(t), \qquad (0 \le s_i \le 1 \;\text{ or }\; 0 \le s_i < \infty) \end{aligned}$$

where $f[F_i(t)] = F_i(t)$ is the contribution of joint force $F_i(t)$ received by individual $i$ at time $t$ to the change of its state at $t+1$ (see Fig. 1(b) for the linear function with a lower and a possible upper bound). The formula means that the state of an individual at an arbitrary step is determined by the individual's state at a previous step in linear combination with the joint force received at a previous step. In the presence of both bounds, i.e., the bound symmetry is hold, individual states are confined to the region $0 \leq s_i \leq 1$; otherwise $0 \leq s_i < \infty$ if the upper bound is released. States of all individuals are updated simultaneously.

The upper bound of state corresponds to the saturation effect in biological and chemical systems, and energy dispersion and dissipation in physical systems. The lower bound is set to zero for representing full silence, and negative values are meaningless. The state bounds introduce nonlinear effects on self-organization, and are crucial to complexity. Sometimes, the bound symmetry might be broken, and the upper bound is released. The bounded linear function is not the only option, and many nonlinear monotonic functions can play a similar role, such as a sigmoid function. Here we employ a simplest form to facilitate our theoretical analyses.

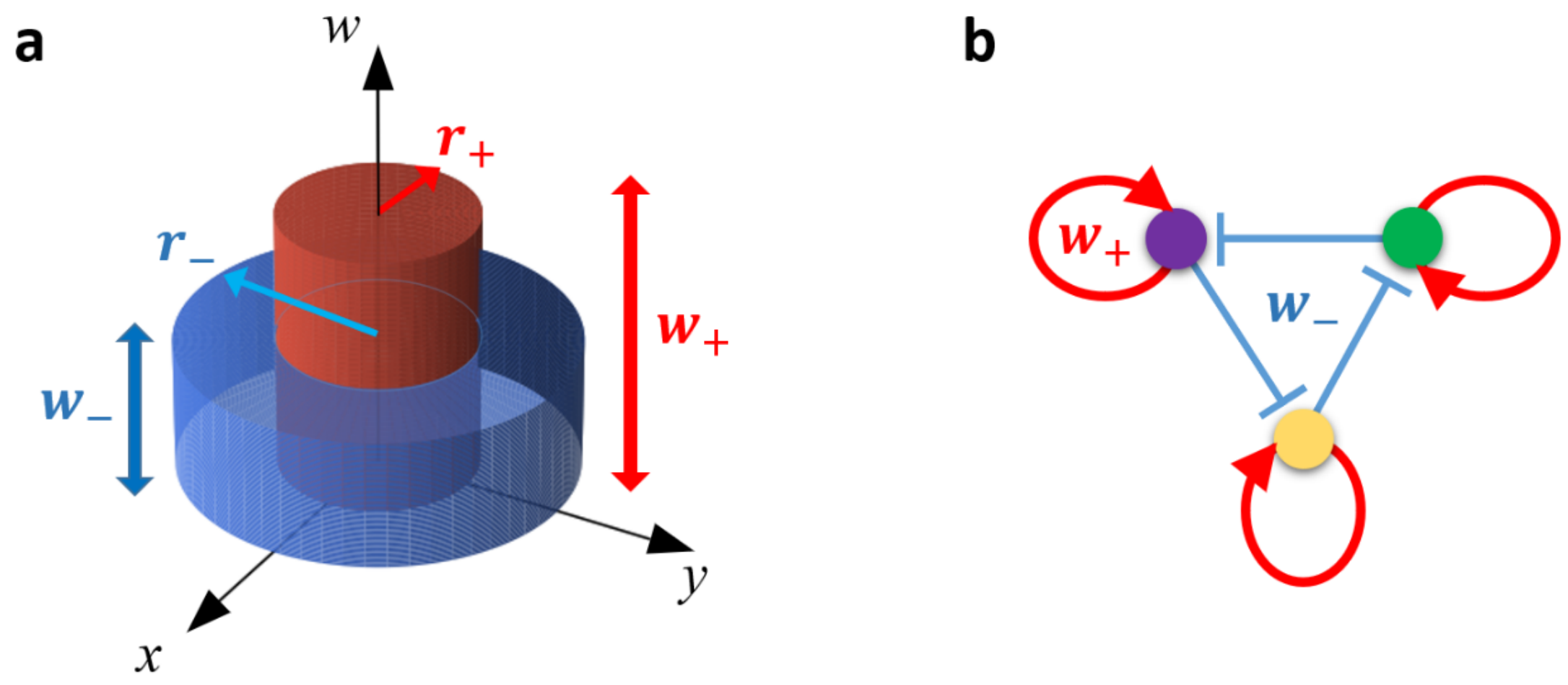

**Fig. 2: Schematic illustrations of two categories of symmetry breaking in self-organization.** (**a**) Symmetry breaking between positive (excitatory) and negative (inhibitory) forces. The range $r_-$ of negative forces (blue) is larger than $r_+$ of positive forces (red). The strength $w_-$ of negative forces is smaller than $w_+$ of positive forces. $x$ and $y$ are spatial coordinates. (**b**) Symmetry breaking of mutual inhibitions (bidirectional regulation) among three types of individuals. Within each type, individuals mutually excite each other within a local range, displayed as the red self-loops; in between two types of individuals, mutual inhibitions are broken with only unilateral inhibition, represented by the blue inhibitory links. The strength $w_+$, $w_-$ and range $r_+$, $r_-$ of forces in (**b**) are identical without symmetry breaking.

**Symmetry breaking in the three self-organization rules**

We classify all kinds of symmetry breaking in the self-organization into four categories. Within each category, there might be a few subtypes (see Table 1.). Countless complex patterns are the result of a single or a combination of some types of symmetry breaking.

**Category I:** Positive-negative (excitation-inhibition) asymmetry. There are three subtypes in Category I: Subtype (A) excitation-inhibition strength asymmetry, Subtype (B) excitation-inhibition range asymmetry, and Subtype (C) action-time asymmetry between excitation and inhibition. Fig. 2(a) illustrates a combination of Subtype A and B. In order to generate and stabilize Turing patterns, Subtype A and B symmetry breaking cannot be arbitrarily combined. In particular, Turing patterns stem from a single combination, i.e., excitation strength is larger than inhibition strength but the range of excitation is smaller than that of inhibition (see Fig. 2(a)). Empirical evidence of the specific combination of Subtype A and B has been observed in the genetic regulation of monkey flowers and morphogenesis of animals [Ding et al., 2020]. In Subtype C, the simultaneity of applying excitatory and inhibitory forces is broken. We will elaborate on how the time asymmetry leads to fractal and chaotic patterns.

**Category II:** State asymmetry. There are two subtypes in Category II: Subtype (A) initial-state asymmetry and Subtype (B) state-bound asymmetry. In Subtype A, the symmetry breaking of initials-state configurations is able to initiate engaging patterns with certain global orders, such as multi-armed spiral waves and target waves. Such initial heterogeneity of state distribution may result from noise-induced fluctuations in physical and chemical systems, or from a certain distribution of progenitor cells encoded in DNA. In Subtype B, the symmetry between the upper and lower bound of individual state is broken and the upper bound is released, i.e., $0 \leq s_i \leq \infty$ (see Fig. 1(b)). This symmetry breaking might occurs in systems with low energy dispersion and dissipation.

**Category III:** Spatial-medium (substrate) asymmetry. Category III symmetry breaking includes four subtypes: Subtype (A) symmetry breaking in orthogonal coordinates, Subtype (B) symmetry breaking in polar coordinates, Subtype (C) symmetry breaking in the gradient of media and Subtype (D) symmetry breaking in mutation. In Subtype A, the strength and range of forces between horizontal and vertical directions become asymmetric, which can be simply implemented by a linear transformation on orthogonal coordinates. In Subtype B, the strength and range of forces between radial and tangential direction are different; in Subtype C, there is a gradient along a certain direction, in which both the strength and range of positive and negative forces gradually decrease. Category III Subtype A to C can be affected by external fields, such as Rayleigh–Bénard convection induced by the gravitational field, and the asymmetric diffusion and transportation of growth factors affected by the gravitational field in plants. In Subtype D, there is a relatively high mutation rate at some locations. In general, mutations during cell replication are rare and eliminated by immune systems, but with two counter-examples, e.g., immune cells and progenitor cells. The diversity of immunoglobulins stems from gene rearrangement of immune cells. The differentiation of progenitor cells by means of epigenetics leads to diverse mature cells associated with specific functions. These mutations are categorized into Subtype D.

**Category IV:** Symmetry breaking of interrelationships (bilateral regulation) among multi-types individuals. In the presence of multiple kinds of individuals, such as pigment cells with different colors, there might exist bidirectional regulation between two types of individuals. The symmetry breaking refers to unilateral regulation. For instance, for three kinds of individuals, there are unilateral inhibitions among different types of individuals, whereas within each kind, individuals mutually excite each other, as exemplified in Fig. 2(b). For organisms, the symmetry breaking can be coded in DNA at the molecular scale. Category IV symmetry breaking is key to generating a wide variety of traveling waves.

**Table I:** Four categories of symmetry breaking and their subtypes in self-organization.

| Category | Subtype | Brief description |
|---|---|---|
| I: positive-negative (Excitation-inhibition) asymmetry | A: Strength asymmetry | Excitatory strength differs from inhibitory strength. |
| | B: Range asymmetry | The range of excitation differs from that of inhibition. |
| | C: Time asymmetry | Excitation and inhibition are not simultaneously exerted, and their duration might be different as well. |
| II: State asymmetry | A: Initial-state asymmetry | Inhomogeneous spatial distribution of individuals' initial states. |
| | B: State-bound asymmetry | The upper bound of individual state is released. |
| III: Spatial-medium (substrate) asymmetry | A: Asymmetry in orthogonal coordinates | The strength and range of both excitation and inhibition in the horizontal direction differ from that in the vertical direction. |
| | B: Asymmetry in polar coordinates | The strength and range of both excitation and inhibition along radial and tangential directions are different. |
| | C: Gradient asymmetry | There is a gradient of media along a certain direction, regulated by external fields. |
| | D: Mutation asymmetry | Mutation rate at some locations (sites) is relatively higher. |
| IV: Interrelationship asymmetry | | For multi-types of individuals, the symmetry of interrelationships (mutual regulation) among distinct types is broken with only unilateral actions left. |

## III Results

We systematically show a variety of static and dynamic patterns resulting from a single or some combinations of the symmetry breaking. The patterns are associated with organism development, chemical reactions and physical processes. Interestingly, phase transitions, pattern percolation, fractals and chaotic phenomena, biodiversity and the origin of order occur in the course of pattern evolution. We provide theoretical results for the primary properties of both static and dynamic patterns. All of the results demonstrate the generality of our asymmetric self-organization underlying complex systems in nature.

**Random Turing patterns from positive-negative force (strength and range) asymmetry [Category-I(AB)].**

The combination of Category I(AB) symmetry breaking accounts for a large variety of random Turing patterns, as shown in Fig. 3. A necessary requirement is that the range of negative force is larger and the strength of positive force is higher (see Fig. 2(a)). This constraint is in agreement with genetic regulation of pigment cells in both animals and plants [Watanabe & Kondo, 2015].

In general, local Turing patterns can be classified into spots and irregular stripes. The size and location of spots are jointly determined by initial configurations and the synergy of local forces. Some initially excited individuals become seeds of spots. Their influences diffuse outward and activate individuals in their vicinities, leading to the growth of excitatory patches. The growth might halt when spots reach a certain size. Within a spot, excitation dominates and individual states approach the upper bound, while outside spots inhibition takes charge and individuals are silent (see Fig. 3). For sufficiently low ratios ($w_+/w_-$ and $r_+/r_-$) of excitation to inhibition, there exist spots only (Fig. 3 (a) (e)).

Irregular stripes stem from the merging of local spots. As we increase the ratio $w_+/w_-$ or $r_+/r_-$, excitatory spots expand and merge into irregular stripes (Fig. 3(b) (f)). When we continuously augment the ratios, small stripes will merge into large stripes. At a critical ratio, a huge cluster containing all stripes arises and percolation transition occurs (Fig. 3 (c) (g), see theoretical analyses for more details). For sufficiently large ratios, excitation dominates, leaving only silent spots (Fig. 3(d) (h)). Although these types of Turing patterns can be generated by reaction-diffusion models as well, our self-organization rules offer a much simpler and intuitive fashion based exclusively on the symmetry breaking between excitation and inhibition. The simplicity allows us to discover the new pattern percolation that essentially differs from traditional site- and bond-percolation. In addition, diffusion is not imperative for the formation of Turing patterns but local interactions are.

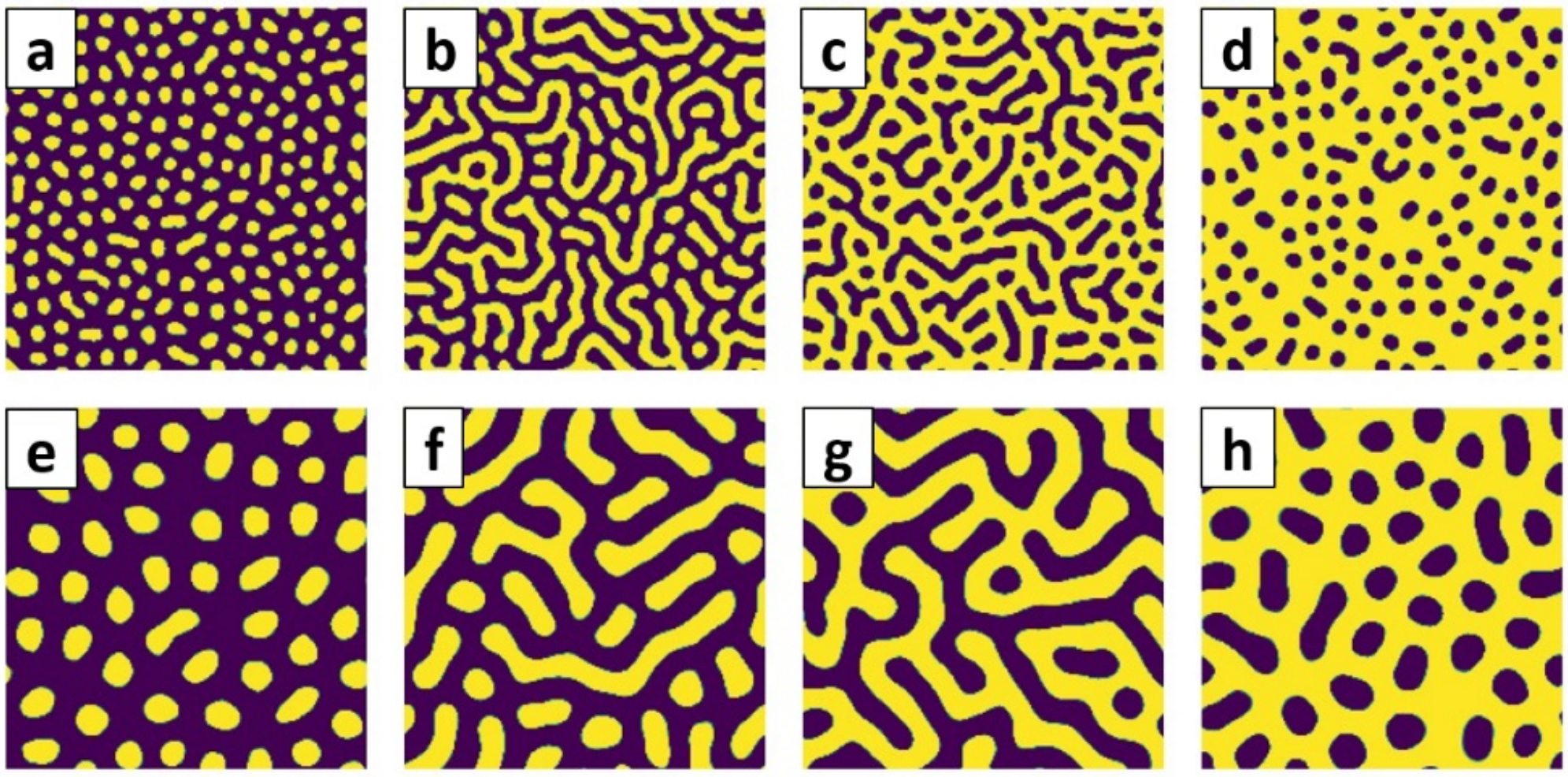

**Fig. 3: Random Turing patterns from breaking the symmetry of excitation-inhibition strength and range (Category I(AB)).** From (**a**) to (**d**), $r_+ = 5$, $r_- = 8$; from (**e**) to (**h**), $r_+ = 10$, $r_- = 16$. $w_+$ in all panels is $0.1$, and $w_-$ in (**a**) and (**e**) is $0.05$, in (**b**) and (**f**) is $0.042$, in (**c**) and (**g**) is $0.039$, and in (**d**) and (**h**) is $0.037$. The patterns are mainly determined by the ratio $r_+/r_-$ and $w_+/w_-$.The color represents state values of individuals. Except at some rare locations, individual states are polarized and approach the upper and lower bounds with yellow ($1.0$) and purple ($0.0$) color. The space scale is $N = 201 \times 201$.

**Turing patterns with certain order from the combination of force asymmetry, initial-state asymmetry and medium asymmetry [Category I(AB) + II(A) + III(ABC)].**

If initial-state asymmetry [Category II(A)] is incorporated into Category-I(AB) symmetry breaking, Turing patterns with certain orders arise (Fig. 4). As shown in Fig. 4(a) and (f), if initially excitatory individuals are aligned to the vertices of a hexagonal grid, we obtain local hexagonal shapes arranged in a spatial order, resembling the patterns of tortoiseshells, compound eyes, corals, basalt columns and bubbles. If we take the damping of growth factors with time into account, we can have patterns on the skin of leopards, as shown in Fig. 4(b) and (g).

The symmetry breaking of media (Category III) plays a significant role in global order in, such as the stripes of zebras, and the concentric and radial patterns around the eyes of pufferfish, as shown in Fig. 4(c) to (e) and (h) to (j). During embryonic development, asymmetric media (substrates), rooted in the secretion and transportation of growth factors, shape vertical stripes of zebra and horizontal stripes of fishes. If growth factors are produced around a small region, such as around the fish eyes, a central symmetry of patterns emerges, such as the concentric and radial patterns. In addition, external fields might cause symmetry breaking of media as well, such as Rayleigh–Bénard convection elicited by the gravitational field. Basalt columns with the hexagonal symmetry are the outcome of Rayleigh–Bénard convection in magma.

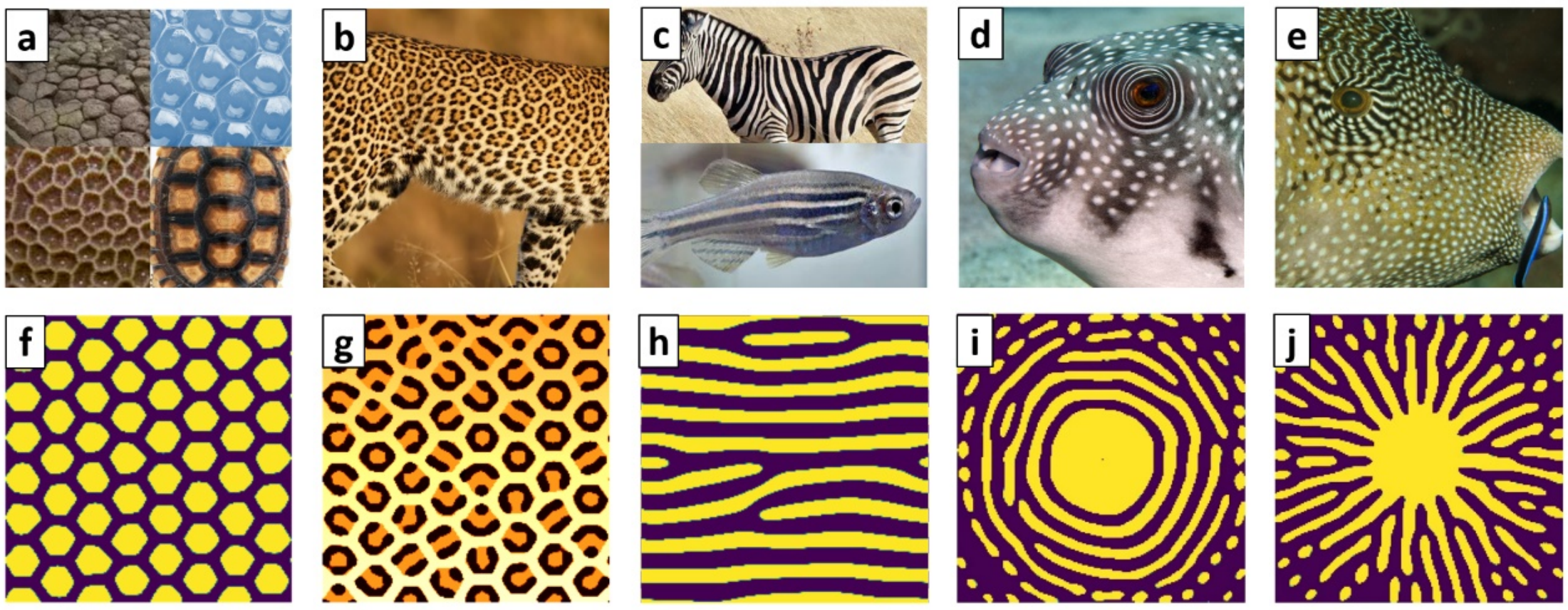

**Fig. 4: Turing patterns from the combination of force asymmetry, initial-state asymmetry and medium asymmetry.** (**a**)-(**e**) Representative patterns in nature, where (**a**) shows hexagonal patterns of basalt columns, bubbles, coral bones and a tortoise shell. (**f**)-(**j**) Relevant patterns generated by our asymmetric self-organization rules. (**f**)-(**g**) are from the combination of strength and range asymmetry and initial-state asymmetry; (**h**)-(**j**) are from the combination of strength and range asymmetry, initial-state asymmetry and medium asymmetry, where (**h**) is from asymmetries in orthogonal coordinates, and (**i**) and (**j**) are from asymmetries in polar coordinates. In (**f**), $w_- = 0.044$, $r_+ = 6$ and $r_- = 9$, and in (**g**) $w_-$ is changed from $0.044$ to $0.055$, and $r_+ = 6$ and $r_- = 9$. In (**h**), $w_-$, $r_+$ and $r_-$ along vertical direction is $0.04$, $3$ and $7$, respectively, and along horizontal direction are $0.04$, $6$ and $7$, respectively. The space scale is $N = 201 \times 201$. The detailed settings in (**f**) to (**j**) are elucidated in Appendix. $w_+ = 0.1$ in all the panels. The pictures in (**a**) to (**c**) are from open-source galleries, and in (**d**) and (**e**) are from https://www.quanjing.com/imgbuy/QJ5101109725.html.

From a mathematical point of view, the two types of medium asymmetries resemble transformations of orthogonal and polar coordinates, respectively. Globally horizontal stripes can be shaped by proportionally enlarging the four force parameters along the horizontal direction. Likewise, we can have vertical stripes by reinforcing the vertical direction. In polar coordinates, the concentric and radial patterns stem from the dominance of tangential and radial directions, respectively. It is worth noting that the two kinds of coordinate transformations are not mutually exclusive, due to the coexistence of many growth factors (see Appendix for more details).

**Branching fractals from the combination of force asymmetry and gradient asymmetry [Category-I(AB) + III(C)].**

Branching fractals are ubiquitous in nature, such as trees, lungs, vascular networks, river networks, snowflakes etc. The branching fractals arise from the combination of Category-I(AB) and III(C) symmetry breaking. Note that Category-I(AB) can jointly produce excitatory spots. On the one hand, the spots can appear on the skin of animals; on the other hand, the spots can act as the growth points of branches on plants, lungs and blood vessels. To be specific, a plane full of spots can be transformed into a cylindrical surface by connecting two parallel boundaries with a periodic condition. This transformation gives rise to a branch with a number of growth points on its surface. At each spot, a new branch sprouts and extends until growth factors fade away. Finally, a branching fractal emerges. During the course, the gradient of growth factors determines both the length of branches and the size of spots on branches. Moreover, the gradient of growth factors might be regulated by external fields. For instance, regulated by the gravitational field, the upper and lower sides of each branch of plants undergo different growth rates, allowing branches to grow upwards. Different responses to the gravitational field yield a great diversity of plants (see Methods and Appendix for details).

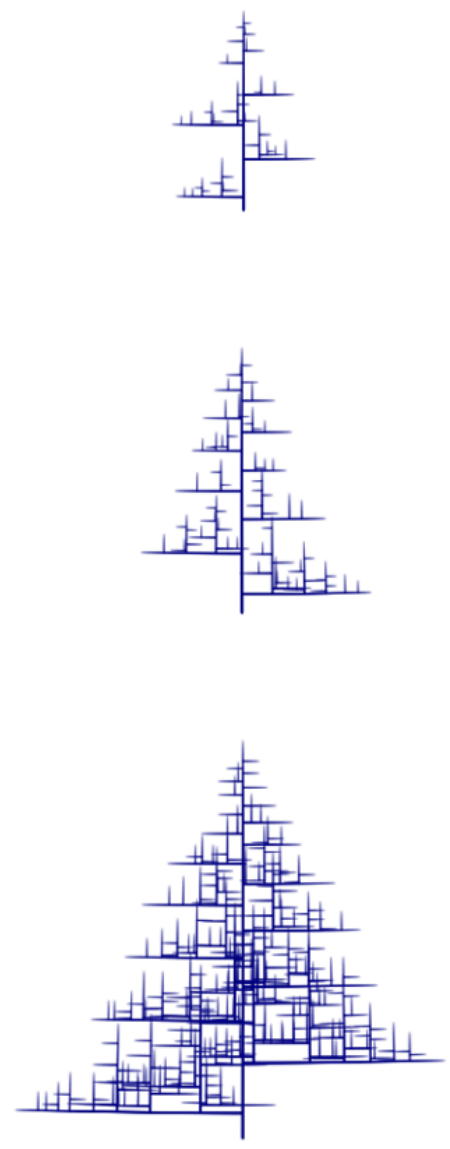

**Fig. 5: Branching fractals from the combination of force asymmetry and gradient asymmetry.** See Appendix for parameter values of interaction strength and interaction range, the gradient of growth factors and the sensitivity of growth factors to the gravitational field.

As shown in Fig. 5, if the concentration of growth factors declines faster from the root, the plant will be shorter with a smaller number of branches. If growth factors are sensitive to the gravitational field, the plant will have a slimmer canopy; otherwise, the canopy will be closer to that of a Christmas tree, in which each branch is vertical to its parent branch (see Fig. 5). The diversity is a natural outcome of our self-organization rules with explicit physical and biological relevance. Likewise, the structure of lungs, vascular networks and dendrite networks of neurons etc. emerges in the same vein. It is noteworthy that all involved symmetry breaking is based on physically and biologically feasible mechanisms rather than artificially iterating a branching operator at different scales for generating fractals.

**Fractals, chaos and kaleidoscopes from interaction time and strength asymmetry (Category-I(AC))**

We show another kind of fractal pattern on seashells based on the symmetry breaking of interaction time and strength (Category-I(AC)). As shown in Fig. 6(a) to (d), the two-dimension fractals are composed of a space dimension and a time dimension, in accordance with the growth process of seashells. The simultaneity of applying excitatory and inhibitory forces is broken, but the interaction ranges $r_+$ and $r_-$ are the same and confined to nearest neighbors. Combining Category-II(A) with Category-I(AC) symmetry breaking, we acquire two types of fractal patterns on seashells: a regular Sierpinski triangle (gasket) initiated from a single excited individual, and a random fractal pattern consisting of stochastically located triangles from random initial states. Although some Cellular Automata can yield similar patterns, our self-organization are more relevant to the biological and physical mechanisms. As we increase the strength $w_+$ of excitation, chaotic patterns emerge with stochastic states of individuals (Fig. 6(e) and (f)). We will elucidate the transition to chaos in more detail by the aid of theoretical analyses.

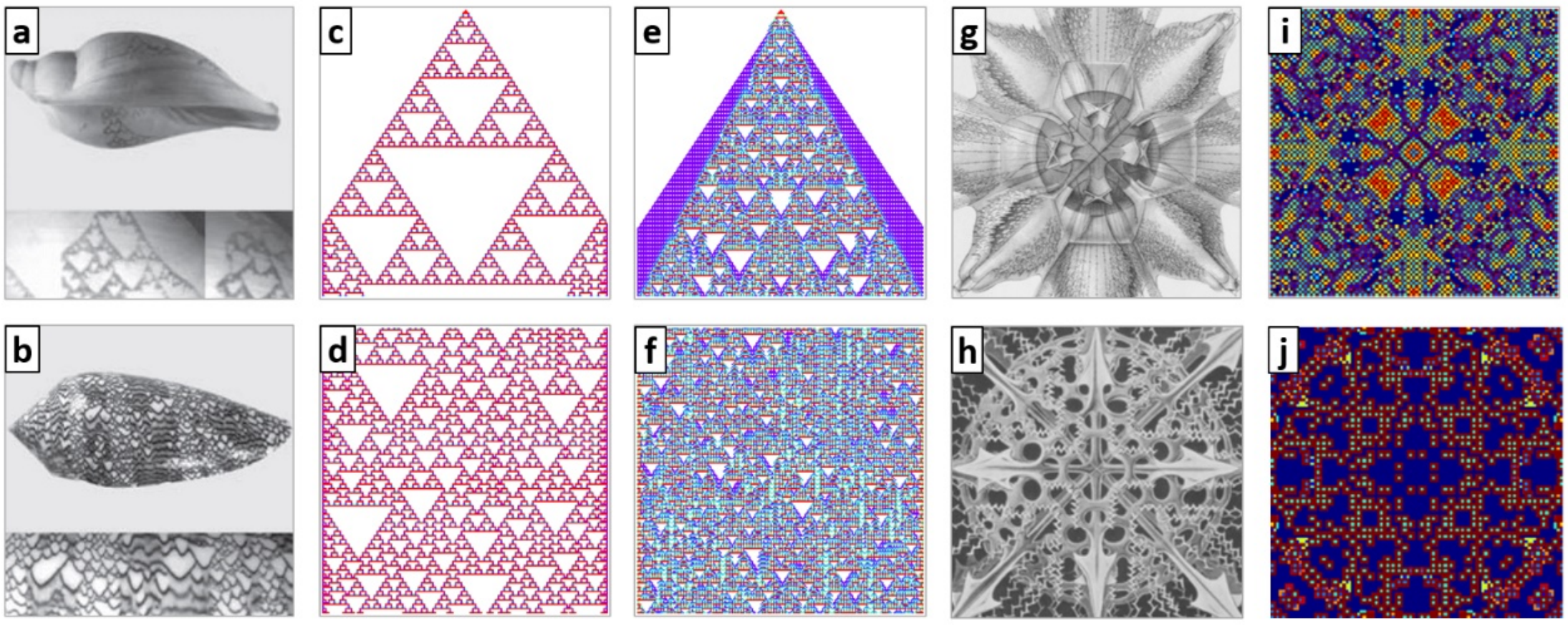

**Fig. 6: Fractals, chaotic patterns and kaleidoscopes from breaking time and strength symmetry.** (**a**)(**b**) Fractal patterns on two kinds of seashells from [Wolfram, 2002]. (**c**) A fractal pattern based on time asymmetry between excitation and inhibition initiated from a single excited individual for mimicking the fractal pattern in (**a**). (**d**) A fractal pattern based on time asymmetry from a random initial condition for emulating the pattern in (**b**). (**e**) A chaotic pattern from a single excited individual when the ratio $w_+/w_-$ of excitatory strength to inhibitory strength exceeds a critical value, as compared with (**c**). (**f**) A chaotic pattern from a random initial condition when $w_+/w_-$ exceeds a critical value, as compared to (**d**). In (**c**) to (**f**), the patterns consist of a space dimension and a time dimension. (**g**)(**h**) Two types of marine Protozoa from [Haeckel, 1974]. (**i**)(**j**) A snapshot of dynamic kaleidoscopes for (**i**) four and (**j**) eight neighbors in two-dimension physical space, respectively. For the simulated patterns, two rounds of excitatory interactions are followed by one round of inhibitory interactions. In (**c**) and (**d**), $w_+ = \sqrt{2}$, and in (**e**) and (**f**) $w_+ = \sqrt{2} - 0.1$. In (**c**) to (**f**), $w_- = 0.5$. In (**i**), $w_+ = 3$ and $w_- = 1/3$, and in (**j**) $w_+ = 4$ and $w_- = 1/4$. In all the panels, $r_+ = r_- = 1$. The scale of one-dimension physical space is $N = 201$ and of two dimensions is $N = 101 \times 101$. See Methods and Appendix for details of the critical values and theoretical analyses.

In two-dimension physical space, sophisticated dynamic patterns arise from the time- and strength-asymmetry between excitation and inhibition. A single excited individual initiates nonrepetitive, chaotic kaleidoscopes for both four- and eight-neighbor situations. Some snapshots are similar to diverse marine Protozoa, as exemplified in Fig. 6(g) to (j). The intriguing patterns are reminiscent of Conway' s game of life as a canonical example of the emergence of complexity from simplicity [Conway, 2000]. Our asymmetric self-organization with more general and fundamental rules than Conway' s game of life, offers strong evidence for the emergence of complexity from fundamental physical forces, beyond biological and ecological interactions.

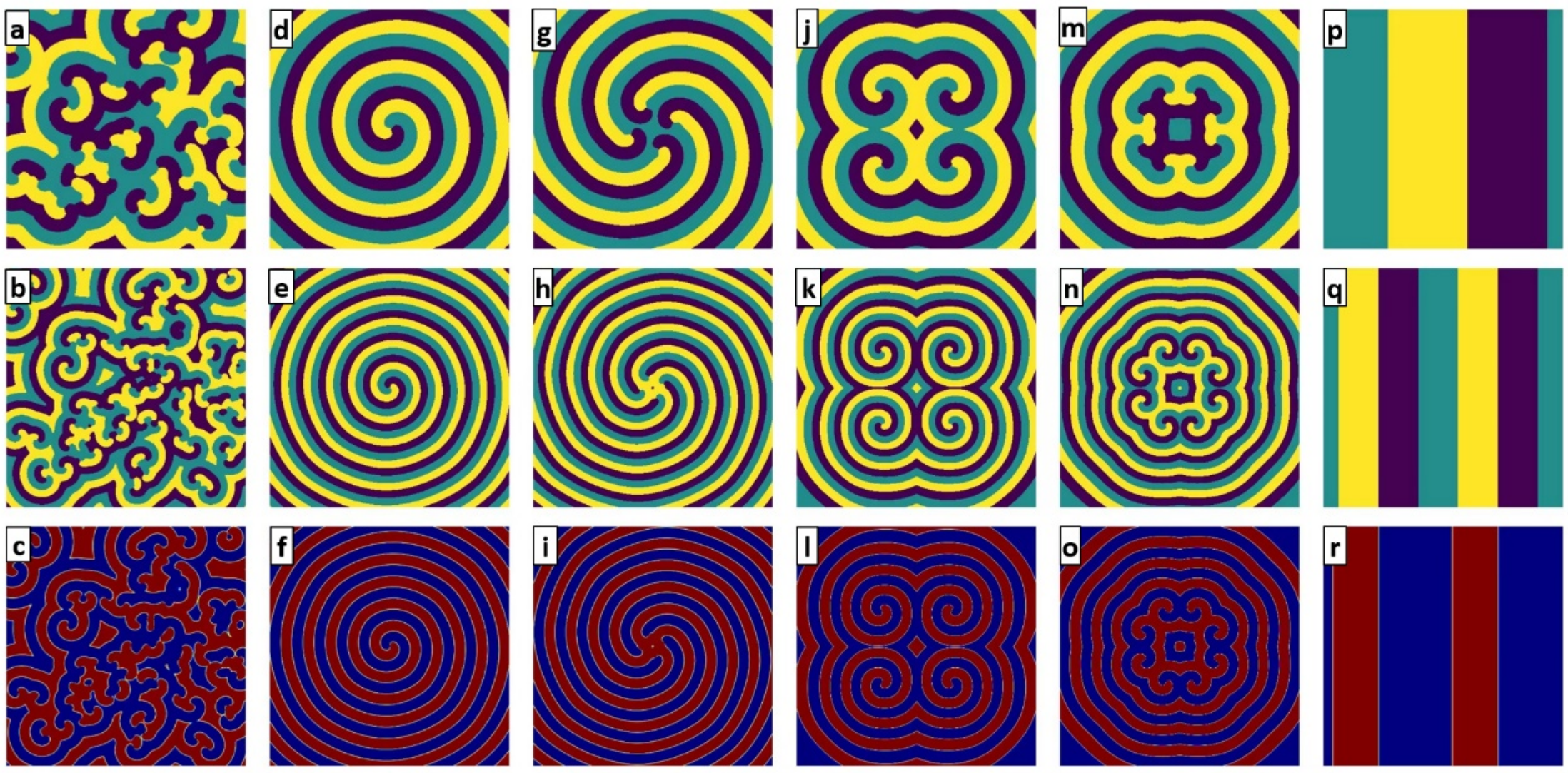

**Fig. 7: Random and regular isometric spiral waves and plane waves based on interrelationship asymmetry.** (**a**)(**b**)(**c**) Random spiral waves from random initial conditions for two interaction ranges ($r_+ = r_-$). (**d**)(**e**)(**f**) A single spiral for two interaction ranges ($r_+ = r_-$). (**g**)(**h**)(**i**) A four-armed spiral for two values of $r_+$ ($r_+ = r_-$). (**j**)(**k**)(**l**) Two pairs of spirals with inverse rotational directions for two values of $r_+$ ($r_+ = r_-$). (**m**)(**n**)(**o**) Four pairs of spirals with inverse rotational directions for two values of $r_+$ ($r_+ = r_-$). (**p**)(**q**)(**r**) Plane waves from two initial conditions. Random spirals are from the symmetry breaking of interrelationships; multi-armed spirals, spiral pairs and plane waves are from the combination of interrelationship asymmetry and initial-state asymmetry. In the upper panels $r_+ = r_- = 3$ and in the middle panels $r_+ = r_- = 6$. The panels in the bottom row are the same as that in the middle row, but only show the states of a single type of individuals (individual states close to edges are chaotic). In all the panels, $w_+ = w_- = 0.1$. The space scale is $N = 201 \times 201$.

**Random isometric spiral waves from interrelationship asymmetry (Category IV)**

If there are multi-type individuals, symmetry breaking might occur on bidirectional actions among different types of individuals, while the range, strength and time of excitatory and inhibitory forces are identical (see Fig. 2(b)). This symmetry breaking of interrelationships is classified into Category IV. In the presence of three types of individuals, Category IV symmetry breaking gives rise to isometric spiral waves, resembling nonlinear chemical reactions, biological traveling waves and turbulence. As shown in Fig. 7(a) to (c), several spiral waves randomly distribute in the space. Each spiral is composed of three types of individuals, and they rotate around a meeting point. The spirals are robust against external perturbations and their features primarily depends on the range ($r_+ = r_-$) of interactions.

Category IV symmetry breaking underpins biodiversity, the first principle of biology. In general, organisms adapt to their ecological niches via competition and cooperation, akin to the excitatory and inhibitory forces. Accompanied by the broken symmetry of interrelationships, species stably coexist in the form of traveling waves. The effect of symmetry breaking on facilitating biodiversity has been echoed in culture dishes, in which three bacterial colonies with cyclic competitions are mixed. The experiment showed that bacterial colonies coexist and exhibit spatial patterns [Kerr,Riley,Feldman & Bohannan, 2002].

**Regular isometric spiral waves and plane waves from the combination of initial-state asymmetry and interrelationship asymmetry (Category-II(A) + IV)**

We incorporate Category-II(A) symmetry breaking into Category IV to create traveling patterns with global orders, where Category IV symmetry breaking generates spiral waves, and Category-II(A) symmetry breaking elicits global orders. The regular spiral waves can be categorized into two types, i.e., multi-armed spirals and spiral pairs with opposite rotation-directions. Note that a spiral emerges insofar as three types of individuals meet at a site. This fact allows us to devise spatial configurations for initiating a single spiral. To be concrete, we assign each kind of individuals a small patch, and three patches spatially constitute a regular triangle (see Appendix). In the beginning, mutual excitation drives expansion of the patches until they encounter each other. Subsequently, the cyclic inhibitions among them motivate their rotations around the meeting location, producing a single spiral, as shown in Fig. 7(d) to (f).

Based on a single spiral, it is convenient to figure out the initial configurations of multi-armed spirals. We dub three different patches a triple-component. The number of triple-components determine the number of arms (see Appendix). Fig. 7(g) to (i) exemplify spirals with four arms. The global order stems from the initial ordered arrangement of triple-components around a circle with equal intervals between adjacent patches. A typical example of multi-armed spirals is the Milky Way galaxy with approximately six major spiral arms, and our earth is located at the Orion Arm.

Sometimes a pair of spirals with opposite rotational directions appear in nonlinear chemical reactions and biological waves, such as in the Belousov–Zhabotinsky reaction [Zhabotinsky, 1964] and the social amoeba *Dictyostelium discoideum* [Kondo & Miura, 2010]. The spiral pairs can evolve from specific initial configurations, slightly different from that of multi-armed spirals. We dub two different initial patches a double-component. Initially, a number of double-components are placed around a circle with a third patch located at the center (see Appendix). As shown in Fig. 7(j) to (o), the number of spiral pairs equals the number of double-components, and the global order of the spiral pairs relies on initial arrangements.

Plane waves can be obtained by preventing the meeting of three types of individuals. For example, initially placing three different patches along a straight-line yields plane waves, as shown in Fig. 7(p) to (r). In the course of traveling, the curvature borders of plane waves progressively become straighter, and entirely straight borders appear eventually (see Appendix Fig. S46). Spiral waves are more stable than plane waves, and the presence of spirals disrupts plane waves.

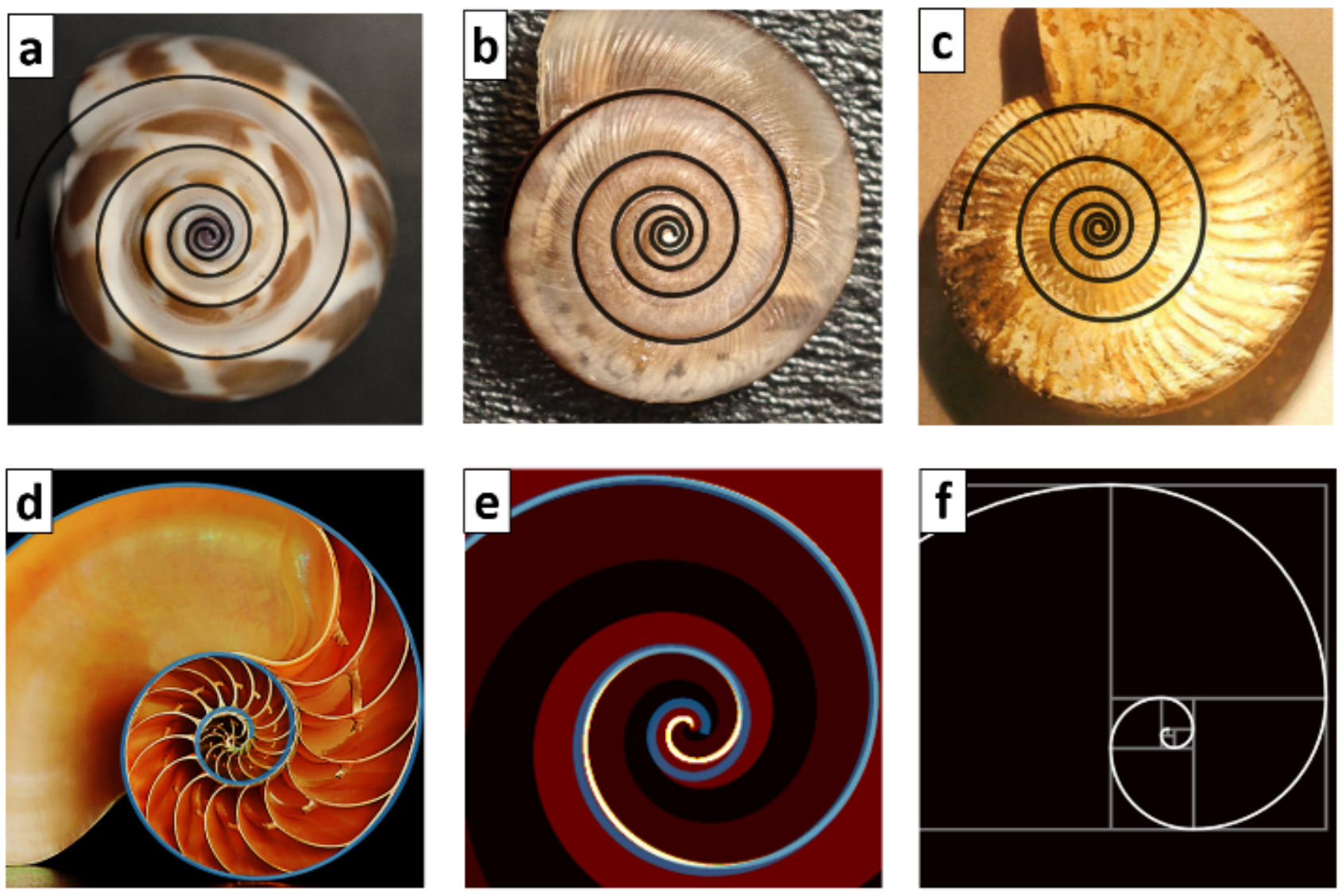

**Fig. 8: Spacing-expanding spirals from breaking state-bound symmetry and compared to the golden spiral.** The shell of (**a**) a sea snail, (**b**) a terrestrial snail and (**c**) an extinct ammonite. (**d**) A spiral inside a seashell. (**e**) A simulation result of spacing-expanding spirals from releasing the upper bound of individual state and the asymmetry of interrelationships. (**f**) The golden spiral based on the standard Fibonacci sequence. The black spirals in (**a**) to (**c**) and the blue coil in (**d**) and (**e**) are from a modified Fibonacci sequence with considering death. The white spiral in (**e**) is the highlight of a simulated spiral. See Appendix for more details of the modified Fibonacci number. The shell images are from open-source galleries.

### Spacing-expanding spirals from the combination of state-bound asymmetry and interrelationship asymmetry (Category-II(B) + IV)

Other than isometric spirals, spirals with incremental spacing are common as well, such as the shape of snail shells and hurricanes. The expanding spirals can be generated by integrating state-bound asymmetry (Category-II(B)) and interrelationship asymmetry. As shown in Fig. 8, the spacing between adjacent spiral coils gradually becomes larger from the center to outer coils, akin to the shell spirals for accommodating a growing body. The release of upper bound allows for a continuous increment of individual states, which captures the proliferation of cells and body growth, as well as energy dispersion and dissipation from the center to the outer of hurricanes. As a result, spiral spacing expands. We make a comparison between the expanding spirals and the golden spiral that is rooted in the Fibonacci sequence with biological relevance. However, as shown in Fig. 8, we find the golden spiral shows a faster expansion than shell spirals. Instead, a modified Fibonacci sequence with taking the death of individuals into account can better mimic shell spirals than the standard golden spiral. The effect of death is embodied by the inhibitory force in our asymmetric self-organization. As displayed in Fig. 8, our simulated spacing-expanding spirals are in good agreement with the modified golden spiral and shell spirals (see Appendix for more details of the modified Fibonacci sequence and modified golden spiral).

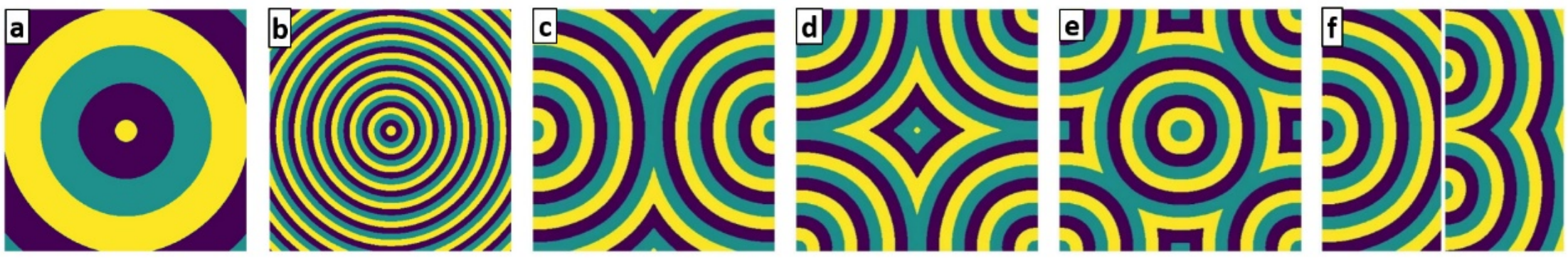

**Fig.9: Target waves based on the combination of mutation and interrelationship asymmetry.** (**a**)(**b**) A single target wave with a single mutation seed centered at the space for two interaction ranges $(r_+ = r_-)$. (**c**) Two mutation seeds at the middle of the left and right boundary. (**d**) Four mutation seeds at four corners. (**e**) Five mutation seeds with an additional seed at the center. (**f**) A single mutation seed at the middle of the left boundary, and a vertical obstacle in the middle with two slits that connect the left and right space. The two slits become the center of two target waves in the right space. In (**a**) and (**b**), $r_+ = r_- = 3$, and in (**c**) to (**f**) $r_+ = r_- = 5$. In all the panels, $w_+ = w_- = 0.1$. The mutation interval $\Delta t$ in (**a**) is $30$, in (**b**) to (**f**) is $5$. The space scale is $N = 401 \times 401$. The size of each mutation patch equals to $r_+$ and $r_-$.

**Target waves from the combination of mutation asymmetry and interrelationship asymmetry (Category-III(D) + IV)**

Category-III(D) symmetry breaking (individual mutations) is necessary for producing target waves. Mutational locations are the source of target waves, and the spacing (width) of rings is determined by both the mutation rate and interaction range, as shown in Fig.9 (see theoretical analyses and Appendix for more details). Initiated from a single central mutation source composed of a patch of individuals, a classic target wave arises (Fig.9(a) (b)); while with multiple mutation sources at specific locations, intriguing patterns with global orders emerge (Fig.9(c) to (e)). In the presence of a spatial obstacle with two slits, target waves can pass through the two slits, giving rise to two coexistent target waves. Contrary to Yang' s double-slit experiment, the two target waves cannot cross each other and interference is absent (Fig.9(f)).

Target waves are closely related with morphogenesis and cardiac dynamics. The mutation process resembles differentiations of pigment cells from progenitor cells in morphogenesis, where mutations are relevant to epigenetic changes of progenitor cell's DNA expressions. The target waves can mimic target patterns of gene-expression waves in the experiment of tail-bud explants [Negrete & Oates, 2021; Tsiairis & Aulehla, 2016]. Furthermore, if we extend the concept of biological mutation to state changes in dynamics, the mutation source of target waves is analogous to the sinoatrial node of heart. The mutations are akin to pulses generated from the sinoatrial node for exciting cardiac muscle cells. In cardiac dynamics, regular traveling waves from the sinoatrial node is key to the normal functioning of hearts. Heart diseases are often associated with broken waves and disorders, which can be reproduced by our simulations. Note that similar to plane waves, target waves are less stable than spirals. To maintain a global target wave, the condition of initiating spirals should be forbidden, i.e., the meeting of three types of individuals at a site. Suppose that an external shock imposes on the global target wave, such that rings distort and the global target wave is broken into scattered spirals (see Fig. 7(a) (b)). The only way to regain a target wave is resetting the whole space via external intervention. This operation is similar to the use of drugs and defibrillators to restore normal heart functioning from an arrhythmia or ventricular fibrillation.

## Theoretical analyses

### Pattern percolation and phase transitions in Turing patterns

The asymmetric self-organization gives birth to a new type of percolation, characterized by the emergence of a large excitatory cluster across the space. As shown in Fig. 10(a), as the strength ratio $w_+/w_-$ increases, the normalized size $A/N$ of the largest excitatory cluster of different system scales $N$ meets at a single intersection point, a hallmark of phase transition with finite sizes. A standard finite-size scaling analysis further confirms the phase transition [Rittenberg, 1983], where all sizes of the largest cluster for different system sizes collapse and overlap with each other (see the inset of Fig. 10(a)). In the thermodynamic limit ($N \to \infty$), the largest excitatory cluster will undergo an abrupt transition from a negligible size to the whole space at the critical strength ratio $(w_+/w_-)_{C_2}$. To our knowledge, such abrupt percolation transition of Turing patterns has not been reported in the literature, and the percolation essentially differs from the conventional site- and bond-percolation in lattices and complex networks [Grimmett, 1999].

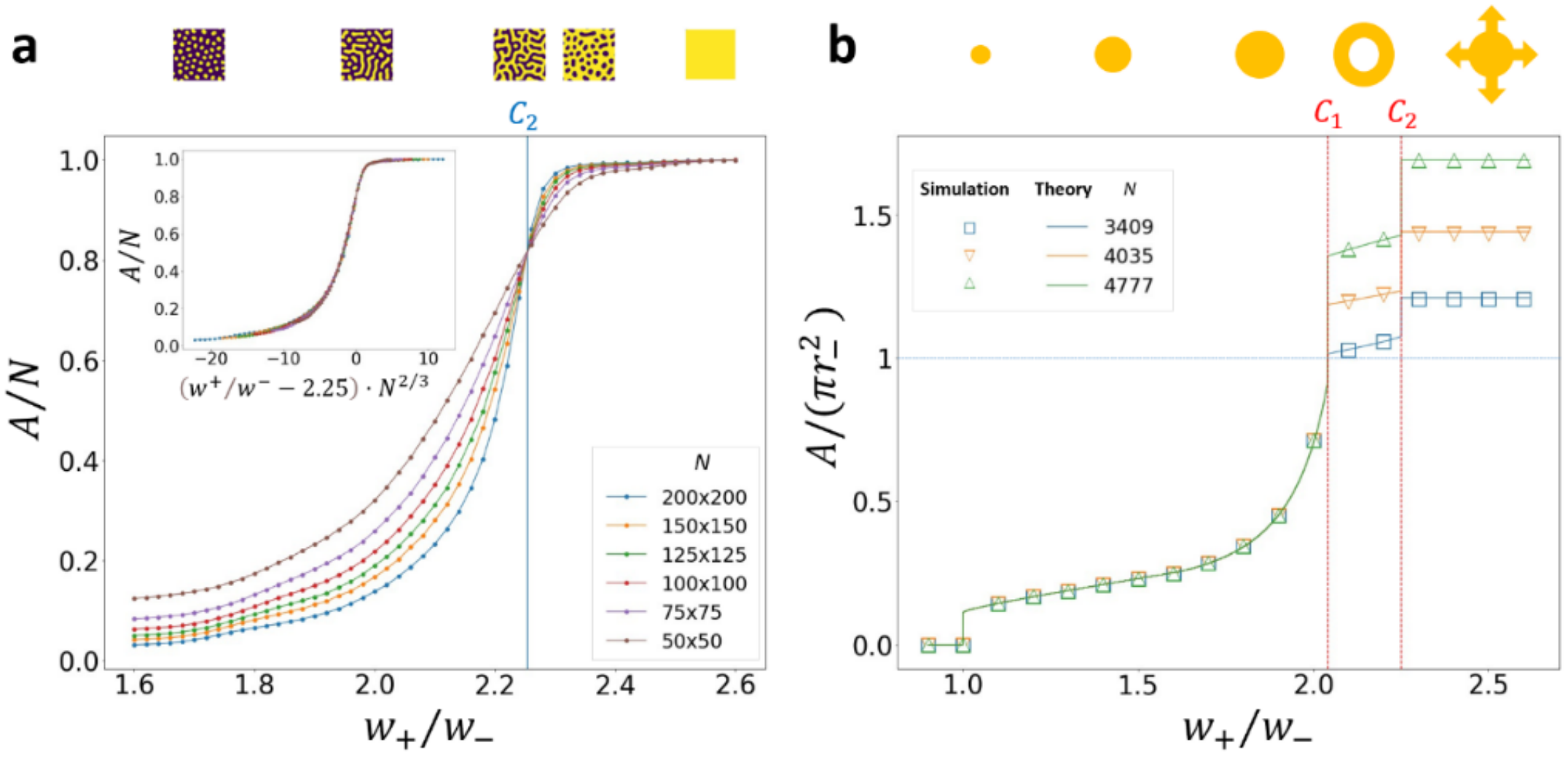

**Fig. 10: Pattern percolation and phase transitions of Turing patterns.** (**a**) Area $A$ of the largest excitatory cluster normalized by population size $N$ as a function of the strength ratio $w_+/w_-$. The inset is the finite-size scaling analysis of different population sizes (see Methods). The vertical line at $w_+/w_- = (r_-/r_+)^2 = 2.25$ is the analytical result of the percolation transition point $C_2$. (**b**) The area $A$ of a single spot divided by the area $\pi r_-^2$ of inhibitory range as a function of the strength ratio $w_+/w_-$ initiated from a single excited individual. $C_1$ and $C_2$ are two phase-transition points derived from theoretical analyses and the curves in (**b**) are theoretical results for different population sizes. Typical patterns associated with different values of $w_+/w_-$ are illustrated on top of each panel. Simulation results are obtained by 1000 independent realizations. See Methods and Appendix for details of the theoretical results.

We put forward a spatial-stability analysis to calculate the percolation threshold. Instead of evolving from random initial conditions, we consider a single excited individual to facilitate our analysis. Since the strength of excitation is higher than that of inhibition, the initial seed will trigger an expanding spot. Due to the relatively larger inhibitory range, a balance might be achieved when the spot expands to a certain size. The equilibrium at the spot' s border allows us to derive spot size based on a geometrical analysis (see Methods and Appendix for details). However, the border balance will be broken if excitatory strength exceeds a critical value $C_1$. In this case, excitation dominates the border and the spot expands ceaselessly. Meanwhile, at $C_1$, the combined force at spot center is reversed from positive to negative, rendering a silent hole inside. Taken together, a phase transition occurs at $C_1$ from a stable spot to an outstretched ring (see Fig. 10(b)). If we continuously augment excitatory strength from $C_1$, there arises a second phase transition point $C_2$, at which the centered joint force reverses from negative to positive again, and the expanding ring turns into an expanding spot. Our spatial-stability analysis precisely predicts the two phase-transition points and the size of spots and rings in a finite-size space, as shown in Fig. 10(b).

The percolation transition of Turing patterns occurs exactly at the second phase transition point $C_2$, where expanding spots emerge. In the presence of multiple expanding spots at $C_2$, a new equilibrium with two possible scenarios occurs. The first is the formation of stripes from the merging of small adjacent spots; the second is the resistance to merging when spots and stripes become relatively large. In the first scenario, excitatory forces dominate the gaps among small spots. In the second, negative forces become overwhelming at the gaps among large spots and strips, such that further merging ceases. As a result of the two scenarios, a giant cluster composed of irregular strips emerges at $C_2$, the hallmark of pattern percolation. Beyond $C_2$, excitatory forces prevail, and for sufficiently large values of $w_+/w_-$, the whole space is full of excited sites, in the absence of silent sites.

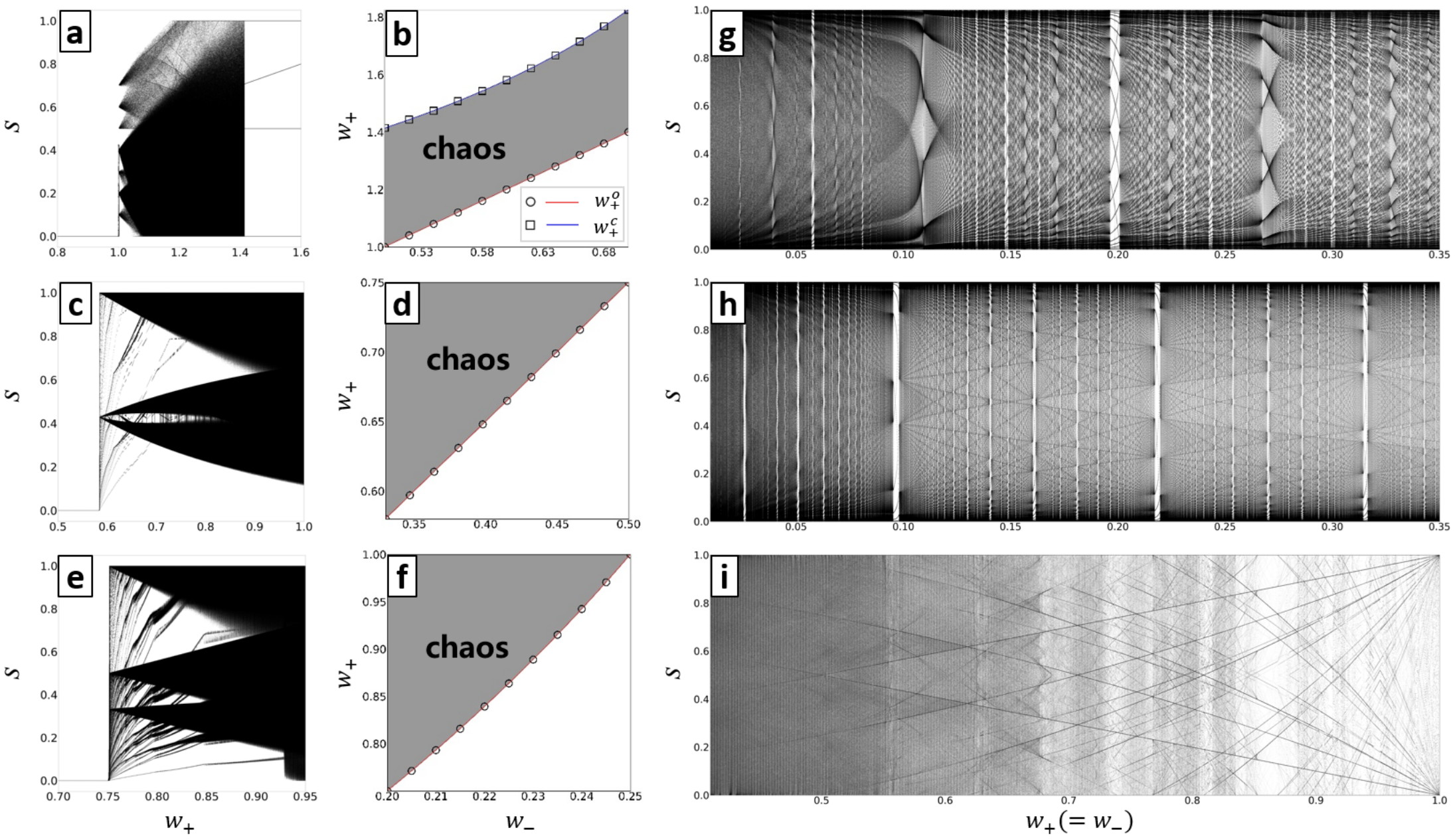

**Fig. 11: Chaos, fractals and their phase transitions in dynamic patterns and traveling waves.** (**a**) Individual state $s$ as a function of $w_+$ and (**b**) relevant phase diagram ($w_+$ versus $w_-$) of time-asymmetry patterns in one-dimension physical space. (**c**) $s$ as a function of $w_+$ and (**d**) relevant phase diagram ($w_+$ versus $w_-$) of time-asymmetry kaleidoscopes in two-dimension physical space with four neighbors. (**e**) $s$ as a function of $w_+$ and (**f**) relevant phase diagram ($w_+$ versus $w_-$) of time-asymmetry kaleidoscopes in two-dimension physical space with eight neighbors. (**g**)-(**i**) $s$ as a function of $w_+$ for (**g**) a plane wave with $r_+ = r_- = 2$, (**h**) a plane wave with $r_+ = r_- = 3$, and (**i**) a spiral wave with $r_+ = r_- = 2$. In (**a**), (**c**) and (**e**), $w_- = 0.5$, and $r_+ = r_- = 1$. In the phase diagrams (**b**) (**d**) and (**f**), the gray region is the chaotic phase, the red and blue curves (lines) are the analytical onset $w_+^o$ of chaos and the analytical phase transition $w_+^c$ between chaos and fractals, respectively (see Methods for the analytical results and Appendix for more details), and symbols are relevant simulation results. $s$ is recorded during 500 steps after 1000 steps.

## Chaos in dynamic patterns and traveling waves

we employ a traditional approach to exploring routes to chaos and phase transitions [Ott, 2002]. As shown in Fig. **11**(a), there are three phases in the diagram of individual states versus excitatory strength $w_+$ for time-asymmetry patterns in one-dimension physical space. When $w^+$ is less than an onset $w_+^o$, the space is silent; as $w^+$ exceeds the onset $w_+^o$, the system enters a chaotic phase with bifurcation-like behaviors; after $w^+$ exceeds a second critical value $w_+^c$, the system enters a fractal phase with self-similarity features. There are only four individual states in the fractal phase. Typical patterns in the fractal and chaotic phase are exemplified in Fig. **6**(c) (d) and Fig. **6**(e) (f), respectively. Moreover, the critical values $w_+^o$ and $w_+^c$ depend on $w_-$, as shown in the phase diagram in Fig. **11** (b). We analytically derived the boundaries among silent, chaotic and fractal phases, and the analytical results (see Methods and Appendix) are in exact agreement with simulation results ( Fig. **11** (b)).

For time-asymmetry kaleidoscopes in two-dimension physical space, there exist a single transition from silence to chaotic phase ( Fig. **11**(c) to (f)). At the onset $w_+^o$ of chaos, a bifurcation-like behavior arises as well for either four- or eight-neighbor situations (their relevant chaotic patterns are exemplified in Fig. 6(i) (j)). In the chaotic phase, dynamic patterns are nonrepetitive and unpredictable, albeit based on simple self-organization rules. The phase-transition points at $w_+^o$ are as well dependent on $w_-$, as shown in the phase diagrams in Fig. **11**(d) and (f). The phase boundaries are precisely predicted by our analytical results (see Methods and Appendix).

In traveling waves, including spiral, target and plane waves, chaos arises in the vicinity of moving edges between different types of individuals, as exemplified in the bottom panels in Fig. **7**. As shown in Fig. **11**(g) (h) (i), in a wide range of interaction strength ($w_+ = w_-$), individual states are nearly ergodic with periodic windows embedded. In each window, individual states switch among a finite number of values. Moreover, the subtle and intricate structures in the state-parameter diagram are a hallmark of chaos as well. A paradox is that despite the chaos at moving edges, all kinds of traveling waves are statistically stable and their primary features are predictable (see theoretical predictions of dynamic patterns). The paradox is of great significance for adapting to environmental changes and maintaining biodiversity. Specifically, the chaos induces mutations to a small part of individuals at frontiers confronted with more uncertainty, while the characteristics of the species remain unchanged. This is the most effective strategy for a species to explore new evolutionary possibilities. For an individual, the strategy is implemented via sexual reproduction for introducing compatible mutations.

## Analytical features of traveling waves

Exact theoretical results of traveling waves were obtained by numerically solving transcendental equations. However, with some reasonable approximations, analytical results of the primary features of traveling waves can be formulated (see Methods) and compared to simulation results. As shown in Fig. **12**, the analytical predictions, including the traveling velocity of plane waves, target waves and isometric spiral waves, the rotational velocity of isometric spiral waves, and the spacing of isometric spiral waves, target waves, and expanding spirals, are all in good agreement with simulation results. The theory of expanding spirals differs from that of the standard golden spiral based on the Fibonacci sequence. The process of deriving a modified Fibonacci sequence and its general-term formula by considering lifespan is detailed in Appendix.

The analytical results reveal a crucial condition of the coexistence of different individuals through traveling waves. Note that the spacing of spiral and target waves is positively related with the interaction range $r_+$ and $r_-$. For sufficiently large interaction ranges, the spacing will exceed the space scale, such that traveling waves can no longer exist and only a single type of individuals remains in the end. This analysis demonstrates that local interactions are necessary for biodiversity and the emergence of complexity. On the contrary, global interactions progressively amplify minor differences, thereby disrupting any local equilibriums and patterns in the long run.

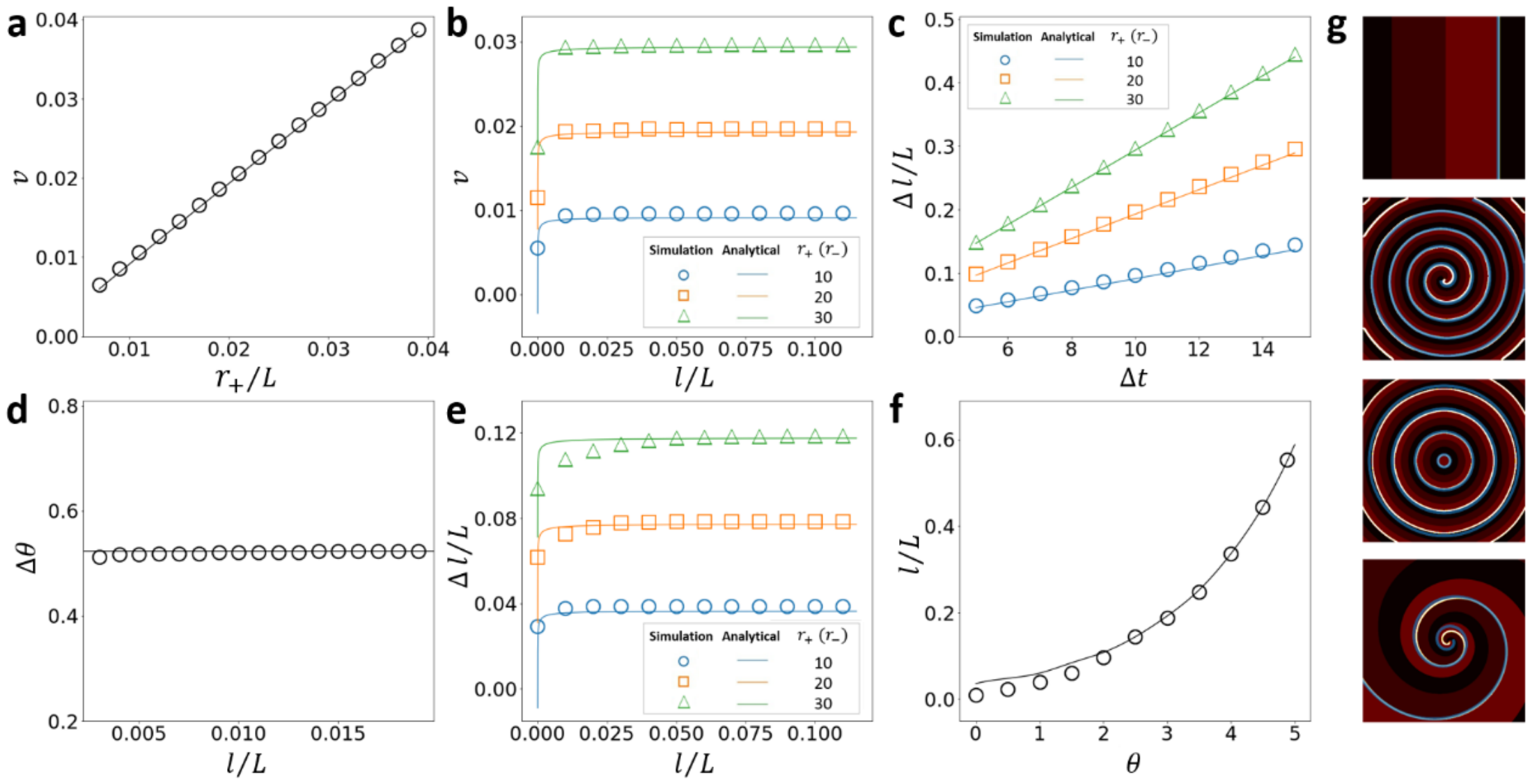

**Fig. 12: Simulation and analytical results of traveling waves.** (**a**) The velocity $v$ of plane waves as a function of normalized interaction range $r_+/L$ ($r_+ = r_-$). (**b**) The radial velocity $v$ of target waves as a function of normalized distance $l/L$ from the center for different interaction ranges ($r_+ = r_-$) (**c**) The spacing $\Delta l$ of target waves as a function of mutation interval $\Delta t$ for different interaction ranges ($r_+ = r_-$). (**d**) The rotational velocity $\Delta\theta$ of isometric spiral waves as a function of normalized distance $l/L$ from the center. (**e**) The spacing $\Delta l$ of isometric spirals as a function of normalized distance $l/L$ from the center for different interaction ranges ($r_+ = r_-$). (**f**) The spacing $\Delta l$ of spacing-expanding spirals as a function of rotational angle $\theta$. (**g**) Illustrations of the comparison between simulated waves and analytical waves. The lines and curves in (**a**) to (**f**) are analytical results (see Methods) and symbols are simulation results averaged over 2000 independent realizations. The variance is smaller than symbol size. In (**g**), the white curves are simulated boundaries between two types of individuals, and the blue curves are analytical wave boundaries. In all the panels, $w_+ = w_- = 0.2$.

## IV Conclusion and Discussion

Whether innumerable complex systems share any common mechanisms remains elusive in spite of advances in contemporary science. Inspired by the success of physics, researchers are tempted to pursue universal mechanisms or principles underlying complex systems. However, ranging from bacterial colonies to galaxies, complex systems are rather different from ideal objects tackled by physics. Diverse and complex phenomena preclude us from believing in any universal rules and principles. As an alternative, one has to build specific models for studying complex systems from different views. As a result, a large number of models and hypotheses presented in the literature. However, as more models are being accumulated, we become more and more perplexed, and wonder how such a variety of possible mechanisms in models are relevant to basic physical and chemical laws? How are the models designed by human brains related to the real situations without central intelligence? Are self-organization and emergence merely obscure concepts or, can they be embodied as explicit as physical laws? Due to the lack of fundamental principles about exploring complexity, we believe these questions puzzle not only us but also many others confronted with complex systems in different fields.

Our work addresses the fundamental issues in complexity science and suggests a paradigm shift. Surprisingly, we found that the real complex world shares underlying principles with ideal physical systems, and the first principles of physics, i.e., symmetry and symmetry breaking, immediately dominate the complex world as well. There exist three general and self-organization rules that are rooted in basic physical quantities. Symmetry breaking in certain aspects of the three self-organization rules accounts for nearly all complex spatiotemporal patterns in nature, including numerous Turing patterns, fractals, chaotic patterns and traveling waves. The physical, chemical and biological relevance to the self-organization and its symmetry breaking is explicit with lots of empirical evidence. The asymmetric self-organization is so simple and general that it is in essence feasible and insightful in any disciplines. Self-organization is no longer an ambiguous concept, since we endowed self-organization with a general and simple set of rules. Furthermore, the asymmetric self-organization builds an inherent connection between animate and inanimate matter, because the three self-organization rules are ubiquitous and underpinned by fundamental physical forces, chemical reactions, gene-protein regulation, morphogenesis, ecological adaptation, geological processes, social interactions and so on. The presence of more diversiform patterns in biology is attributed to the diversity of molecular structures based on the four covalent bonds of carbon. The biodiversity allows for richer combinations of symmetry breaking, inducing dazzling biological patterns.

Symmetry became inseparable from various conservation laws after Noether proved her groundbreaking theorem [Noether, 1927]. Since then, symmetry became one of the first principles in physics. Symmetry plays a significant role in particle physics, crystal physics, quantum theory, relativity and so on, and has implications in many areas using group theory [Schwichtenberg, 2018]. Symmetry breaking is also a fundamental concept in physics [Landau

& Lifshitz, 1980]. It is deemed that the violation of mirror symmetry or parity in weak interactions might account for diversity in the universe [Wu, 2008]. Symmetry breaking is also responsible for phase transitions in many physical phenomena [Landau, 2008]. Our theory by integrating self-organization and its symmetry breaking challenges a traditional perspective on emergence. Prior to our theory, it is believed that with the rise of discipline hierarchy from physics at the bottom to social science at the top, new laws and principles emerge in each layer, such that fundamental laws in the bottom layer is no longer available in high layers. For example, it is hard to use physical laws to explain and predict biological and social phenomena. However, our work explicitly points out limitations of the conventional perspective on emergence. In fact, symmetry and symmetry breaking as the first principles of physics penetrate a number of discipline layers and dominate the origin of complex patterns in many fields. Although central intelligence is absent in nature, the symmetry breaking of self-organization resembles a few switch buttons on a central control panel. Nearly all complex spatiotemporal patterns stem from on-off combinations of these buttons. The outcomes are often unpredictable without computers, but can be selected by Mother Nature. Only can those patterns that are resistant to external disturbance and competitions present, such as Turing patterns for camouflage and bifurcation fractals for the exchange of substance and energy. Taken together, symmetry and its breaking are reinterpreted in terms of self-organization, and can be regarded as the first principle in science, not merely in physics.

Accompanied by the new wave of complexity science, whether reductionism is expired becomes a heated debate. It has been a common belief that reductionism restricts our understanding of complex systems, in spite of its feats in modern science. An aphorism in complexity science is "the whole is greater than the sum of its parts." This implies that reduction to elementary ingredients is useless to understand complex system as a whole. Instead, systems science is recommended as a replacement of reductionism. Unfortunately, the framework of systems science is yet to be constructed and its fundamental principles are ambiguous. We contend that reductionism is not only indispensable for exploring complex systems, but also the foundation of complexity science, akin to the other disciplines. Indeed, emergent behaviors are usually unpredictable from a single individual. The exclusive way to comprehend emergence is to figure out underlying self-organization mechanisms. However, a prerequisite for reverse engineering self-organization is reduction to elements and their immediate interactions. In this perspective, reductionism underpins complexity science. The misunderstanding of reductionism lies in a missing link between reduction and self-organization. It is indeed not possible to fully understand emergence based exclusively on reductionism. The missing link is mainly ascribed to a lack of knowledge about self-organization rather than ascribed to reductionism. Despite often mentioned in the literature, the connotations and implications of emergence and self-organization are vague for a long time. In this circumstance, it is difficult to believe that reductionism is imperative to complex systems. Our theory suggests a novel framework for exploring complexity in all disciplines and areas by integrating reduction and self-organization, where the reduction takes charge of individual behaviors, and the self-organization based on the reduction is responsible for collective and emergent behaviors.

## Methods

### Colorization of patterns

For Turing patterns composed of a single type of individuals, the color of a site (location) denotes state value of the individual at the site. The binary-color patterns indicate that the states of individuals in Turing patterns are polarized and approach either the lower or the upper bounds, i.e., $s = 1$ or $s = 0$.

For time-asymmetry patterns on one- and two-dimension physical space, the color of sites denotes state values of individuals, ranging from $s = 0$ to $s = 1$. For the fractal patterns (Sierpinski gasket), there are only a small number of different states; while for the chaotic kaleidoscopes, an infinite number of states present.

For traveling waves with three types of individuals, each site is simultaneously occupied by three different individuals, and their states range from zero to one. For three-color waves, the color of a site is determined by the color of an individual with the highest state value. If two individuals have an identical state, the site follows the color of the individual who inhibits the other. If occasionally three individuals are of the same state value, the site is set to be white. If we show only one type of individuals, the color denotes state values of the individuals (see the bottom panels of Fig. 7 and Fig. 8(e)).

### Generating branching fractals

Besides the strength and range of excitation and inhibition, there are two additional parameters for regulating branching structures, i.e., the damping of a growth factor and its response to the gravitational field. The angle of a new branch pertaining to its parent branch is determined by the gradient of the growth factor regulated by the gravitational field. For the bud of each branch, the gradient differs between proximal and distal sides to the root, due to the restriction of the gravitational field to the transportation of growth factors. As a result, the growth rate of proximal side is higher than that of distal side, leading to an upward-growing trend. Associated with the growth of branches, the growth factor' s gradient affected by the gravitational field causes a nonlinear effect on branch bending (see Appendix for more details).

Our approach can naturally create three-dimension branching fractals based on the self-organization rules. However, for better visualization, we only show two-dimension cases. For real plants, there might be other fine-tuning mechanisms in addition to the two kinds of symmetry breaking, e.g., leaf growth at the end of small branches. Nonetheless, the gradient of growth factors regulated by the gravitational field is the primary mechanism underlying plant growth, shaping a great diversity of plants.

Other than organisms, our self-organization can reproduce non-living branching fractals, such as river networks and snowflakes. In analogy with plants, the potential energy at the sources of rivers propels their growth, akin to the effect of growth factors. Most rocks along the riverbanks are resistant to the impact of water strikes, resembling the long-range dominance of inhibitory forces. At sporadic locations, rocks are fragile, similar to the local dominance of excitation. At these sites, new river branches develop and eventually a tree-like river network arises. Likewise, the fractals of snowflakes are products of the symmetry breaking between attraction and repulsion during the collision of ice crystals in the air.

**Time asymmetry between excitation and inhibition**

Time asymmetry can occur on duration time and working sequence of forces. For time-asymmetry patterns and chaotic Kaleidoscopes in one- and two-dimension physical space, we allow excitatory forces to occur two-rounds earlier than inhibition. After excitation continues for two rounds, inhibition begins to work for one round. Repetitions of this process generate fractal and chaotic patterns. It is noteworthy that there are many possible settings of time asymmetries and interaction ranges, and we have not enumerated all the scenarios. In general, the time asymmetry is capable of generating much more patterns in addition to the Sierpinski gasket and chaotic kaleidoscopes.

**Finite-size scaling analysis**

In the thermodynamic limit, there is a scaling behavior near the percolation transition point, i.e., $\frac{A}{N} \sim \left[\frac{w_+}{w_-} - \left(\frac{w_+}{w_-}\right)_{c_2}\right]^{\beta}$, where $\left(\frac{w_+}{w_-}\right)_{c_2}$ is the phase transition point and $\beta$ is a critical exponent. For a finite size $N$, the scaling behavior becomes $\frac{A}{N^{1+\frac{\beta}{\nu}}} = \mathcal{F}\left\{\left[\frac{w_+}{w_-} - \left(\frac{w_+}{w_-}\right)_{c_2}\right] N^{\frac{1}{\nu}}\right\}$, where the critical exponent $\nu$ characterizes the divergence of correlation length. At the phase transition point $\left(\frac{w_+}{w_-}\right)_{c_2}$, the function $\mathcal{F}$ is independent of $N$, indicating that the curves of order parameter $\frac{A}{N}$ versus $\frac{w^+}{w^-}$ for different sizes $N$ meet at a single intersection point $\left(\frac{w_+}{w_-}\right)_{c_2}$. With respect to the invariance of $\mathcal{F}$, if we plot $\frac{A}{N^{1+\frac{\beta}{\nu}}}$ versus $\left[\frac{w_+}{w_-} - \left(\frac{w_+}{w_-}\right)_{c_2}\right] N^{\frac{1}{\nu}}$, all curves ought to collapse and overlap with each other. Indeed, the finite-size scale analysis gives a single collapsed curve with the exponents $\beta = 0$ and $\nu = 1.5$. The phase transition point $\left(\frac{w_+}{w_-}\right)_{c_2} = \left(\frac{r_-}{r_+}\right)^2$ is an analytical result base on our spatial-stability analysis (see Methods and Appendix).

**Spatial-stability analyses for phase transitions and pattern percolation**

We formulate phase transitions from stable spots to unstable outstretched rings and spots. The stability of a spot can be examined by state changes of the individuals at the border. The evolution of an arbitrary individual' s state, say $s_i$, can be estimated as

$$\frac{\mathrm{d}s_i}{\mathrm{d}t} \sim F_i,$$

where $F_i$ is the joint force received by $i$. Note that the formula might be violated if $s_i$ is close to either the upper or lower bound or $F_i$ is sufficiently large. The nonlinear effect from bounds precludes us from precisely predicting states of individuals. However, the formula can help us to analyze the stability of individuals at the border of spots or rings (see Appendix for detailed mathematical treatment.) To be specific, a spot is stable if $\mathrm{d}s_i/\mathrm{d}t = 0$ at its border; a spot continuously expands if $\mathrm{d}s_i/\mathrm{d}t > 0$ at its current border; an unstable ring expands if $\mathrm{d}s_i/\mathrm{d}t > 0$ at its outer border and $\mathrm{d}s_i/\mathrm{d}t < 0$ at its inner border. The stability conditions yield a few transcendental equations without analytical solutions, but we can numerically solve the equations. The first phase transition occurs at the transition from $\mathrm{d}s_i/\mathrm{d}t = 0$ to $\mathrm{d}s_i/\mathrm{d}t > 0$ at the border, and $\mathrm{d}s_i/\mathrm{d}t < 0$ at the center. The second phase transition occurs from $\mathrm{d}s_i/\mathrm{d}t < 0$ to $\mathrm{d}s_i/\mathrm{d}t > 0$ at the center of an unstable ring. At the second phase transition point $C_2$, unstable rings turn into expanding spots, and pattern percolation occurs. The forces can be captured by geometric features of interaction ranges due to the simple setting of forces, allowing for theoretical predictions of the phase transitions (see Appendix for theoretical analyses).

**Analytical results of pattern features**

we provide analytical results about the percolation transition of Turing patterns, phase transitions between chaos and fractal patterns in one-dimension physical space, onset of chaos in one- and two-dimension spatial chaos, and primary features of traveling waves, as shown in Table II. We obtain the analytical results based on some approximations applied to transcendent equations (see Appendix for more details).

**Table II: Analytical results of the crucial features of static and dynamic patterns.**

| | |
|---|---|
| Pattern percolation point $c_2$ | $\left(\frac{w_+}{w_-}\right)_{c_2} = \left(\frac{r_-}{r_+}\right)^2,\ \left(\frac{r_-}{r_+}\right)_{c_2} = \sqrt{\frac{w_+}{w_-}},$ |
| Phase transition $w_+^c$ between chaos and fractals in one-dimension physical space | $w_+^c = \sqrt{\frac{1}{1-w_-}},\ w_- \geq 0.5,$ |
| Onset $w_+^o$ of chaos in one-dimension physical space | $w_+^o = 2w_-, \qquad w_- \geq 0.5,$ |
| Onset $w_+^o$ of chaos in two-dimension physical space with 4 neighbors | $w_+^o = w_- + \frac{1}{4}$ |
| Onset $w_+^o$ of chaos in two-dimension physical space with 8 neighbors | $w_+^o = \frac{1+4w_-}{4-8w_-}$ |
| Velocity $v$ of plane waves | $v = r_+ - \sqrt[3]{\frac{9}{32r_+ \cdot w_+^2}},$ $(r_+ = r_-,\ w_+ = w_-)$ |
| Rotational velocity $\Delta\theta$ of isometric spirals | $\Delta\theta = \frac{\pi}{6}$ |
| Radial velocity $v$ of target waves and isometric spirals at distance $l$ from the center | $v = r_+ - \sqrt[3]{\frac{9(l+r_+)}{32r_+ \cdot l \cdot w_+^2}},$ $(r_+ = r_-, w_+ = w_-)$ |
| Spacing $\Delta l$ of isometric spirals at distance $l$ from the center | $$\Delta l = \frac{2\pi}{3 \cdot \Delta\theta}\left(r_+ - \sqrt[3]{\frac{9(l + r_+)}{32 r_+ \cdot l \cdot w_+^2}}\right),$$ $(r_+ = r_-,\ w_+ = w_-)$ |
| Spacing $\Delta l$ of target waves with mutation interval $\Delta t$ | $$\Delta l = \Delta t\left(r_+ - \sqrt[3]{\frac{9}{32 r_+ \cdot w_+^2}}\right),$$ $(r_+ = r_-,\ w_+ = w_-)$ |

## References

Ahlers, G., Grossmann, S. & Lohse, D. [2009] "Heat Transfer and Large Scale Dynamics In Turbulent Rayleigh-Bénard Convection,"*Rev. Mod. Phys.* **81**, 503-537.

Anderson, P. W. [1972] "More Is Different,"*Science* **177**, 393-396.

Barabási, A.-L. [2012] "The Network Takeover,"*Nat. Phys.* **8**, 14-16.

Conway, J. H. [2000] *On Numbers and Games* (A K Peters/CRC Press, New York).

Davidenko, J. M., Pertsov, A. V., Salomonsz, R., Baxter, W. & Jalife, J. [1992] "Stationary and Drifting Spiral Waves of Excitation in Isolated Cardiac Muscle,"*Nature* **355**, 349-351.

Ding, B. Q., Patterson, E. L., Holalu, S. V., Li, J. J., Johnson, G. A., Stanley, L. E., Greenlee, A. B., Peng, F., Bradshaw, H. D., Blinov, M. L., Blackman, B. K. & Yuan, Y. W. [2020] "Two MYB Proteins in a Self-Organizing Activator-Inhibitor System Produce Spotted Pigmentation Patterns,"*Curr. Biol.* **30**, 802-814.e8.

Gierer, A. & Meinhardt, H. [1972] "A Theory of Biological Pattern Formation,"*Kybernetik* **12**, 30-39.

Gilbert, C. D. & Wiesel, T. N. [1989] "Columnar Specificity of Intrinsic Horizontal and Corticocortical Connections in Cat Visual-Cortex,"*J. Neurosci.* **9**, 2432-2442.

Grimmett, G. [1999] *Percolation* (Springer Berlin, Heidelberg).

Grossmann, S., Lohse, D. & Sun, C. [2016] "High-Reynolds Number Taylor-Couette Turbulence,"*Annu. Rev. Fluid Mech.* **48**, 53-80.

Haeckel, E. [1974] *Art-Forms in Nature* (Dover Publications).

Hubel, D. H. & Wiesel, T. N. [1977] "Functional Architecture of Macaque Monkey Visual-Cortex,"*Proc. R. Soc. B* **198**, 1-59.

Kerr, B., Riley, M. A., Feldman, M. W. & Bohannan, B. J. M. [2002] "Local Dispersal Promotes Biodiversity in A Real-Life Game of Rock–Paper–Scissors,"*Nature* **418**, 171-174.

Kondo, S. & Miura, T. [2010] "Reaction-Diffusion Model as a Framework for Understanding Biological Pattern Formation,"*Science* **329**, 1616-1620.

Kondo, S., Watanabe, M. & Miyazawa, S. [2021] "Studies of Turing Pattern Formation in Zebrafish Skin,"*Philos. Trans. R. Soc., A* **379**, 20200274.

Krause, A. L., Gaffney, E. A., Maini, P. K. & Klika, V. [2021] "Modern Perspectives on Near-Equilibrium Analysis of Turing Systems,"*Philos. Trans. R. Soc., A* **379**, 202-268.

Kuramoto, Y. [1984] *Chemical Oscillations, Waves, and Turbulence* (Springer Berlin, Heidelberg).

Landau, L. D. [2008] "On the Theory of Phase Transitions,"*Ukr. J. Phys.* **53**, 11.

Landau, L. D. & Lifshitz, E. M. [1980] *Statistical Physics* (Butterworth-Heinemann).

Livingstone, M. S. & Hubel, D. H. [1984] "Anatomy and Physiology of a Color System in The Primate Visual-Cortex,"*J. Neurosci.* **4**, 309-356.

Luo, L. Q. [2020] *Principles of Neurobiology (2nd ed.)* ( Garland Science).
Mandelbrot, B. B. [1984] *The Fractal Geometry of Nature* (W. H. Freeman and Company, San Francisco).
Meinhardt, H. [2009] *The Algorithmic Beauty of Sea Shells* (Springer Berlin, Heidelberg).
Negrete, J. & Oates, A. C. [2021] "Towards a Physical Understanding of Developmental Patterning,"*Nat. Rev. Genet.* **22**, 518-531.
Noether, E. [1927] "Abstrakter Aufbau der Idealtheorie in Algebraischen Zahl- und Funktionenkörpern,"*Math. Ann.* **96**, 26-61.
Ott, E. [2002] *Chaos in Dynamical Systems* (Cambridge University Press, Cambridge).
Palacci, J., Sacanna, S., Steinberg, A. P., Pine, D. J. & Chaikin, P. M. [2013] "Living Crystals of Light-Activated Colloidal Surfers,"*Science* **339**, 936-940.
Pourquié, O. [2003] "The Segmentation Clock: Converting Embryonic Time into Spatial Pattern,"*Science* **301**, 328-330.
Reichenbach, T., Mobilia, M. & Frey, E. [2007] "Mobility Promotes and Jeopardizes Biodiversity in Rock–Paper–Scissors Games,"*Nature* **448**, 1046-1049.
Rittenberg, V. [1983] *Finite-Size Scaling Theory* (Springer US, Boston, MA).
Schwichtenberg, J. [2018] *Physics from Symmetry* (Springer Cham).
Takagi, S., Pumir, A., Kramer, L. & Krinsky, V. [2003] "Mechanism of Standing Wave Patterns in Cardiac Muscle,"*Phys. Rev. Lett.* **90**, 124101.
Tsiairis, C. D. & Aulehla, A. [2016] "Self-Organization of Embryonic Genetic Oscillators into Spatiotemporal Wave Patterns,"*Cell* **164**, 656-667.
Turing, A. M. [1952] "The Chemical Basis of Morphogenesis,"*Philos. Trans. R. Soc., B* **237**, 37-72.
Wang, W. X., Ni, X., Lai, Y. C. & Grebogi, C. [2011] "Pattern Formation, Synchronization, and Outbreak of Biodiversity in Cyclically Competing Games,"*Phys. Rev. E* **83**, 011917.
Watanabe, M. & Kondo, S. [2015] "Is Pigment Patterning in Fish Skin Determined by the Turing Mechanism?,"*Trends. Genet.* **31**, 88-96.
Wolfram, S. [2002] *A New Kind of Science* (Wolfram Media).
Wu, C. S. [2008] *The Discovery of the Parity Violation in Weak Interactions and Its Recent Developments* (Springer Japan, Tokyo).
Zhabotinsky, A. M. [1964] "Periodic liquid phase reactions,"*Proc. Ac. Sci. USSR* **157**, 392-95.

## Acknowledgements

We thank Xue-Er Cui for preparing schematic figures in Appendix and Prof. Wu Li for valuable discussion. This study was supported by STI2030-Major Projects (2022ZD0204600), the National Natural Science Foundation of China (No. 82021004), NSFC under Grant No. 11975049 and Collaborative Research Fund of Chinese Institute for Brain Research, Beijing.

## Contributions

W.H.W. and W.X.W. designed the study. W.H.W. carried out simulations. W.X.W. wrote the manuscript. W.H.W., Z.Z.L. and W.X.W. prepared Appendix. W.H.W., Z.Z.L. and W.X.W. calculated theoretical results. All the authors reviewed and edited the manuscript.

# Appendix

## Table of Contents

# Explanation of pattern generation

## Leopard pattern

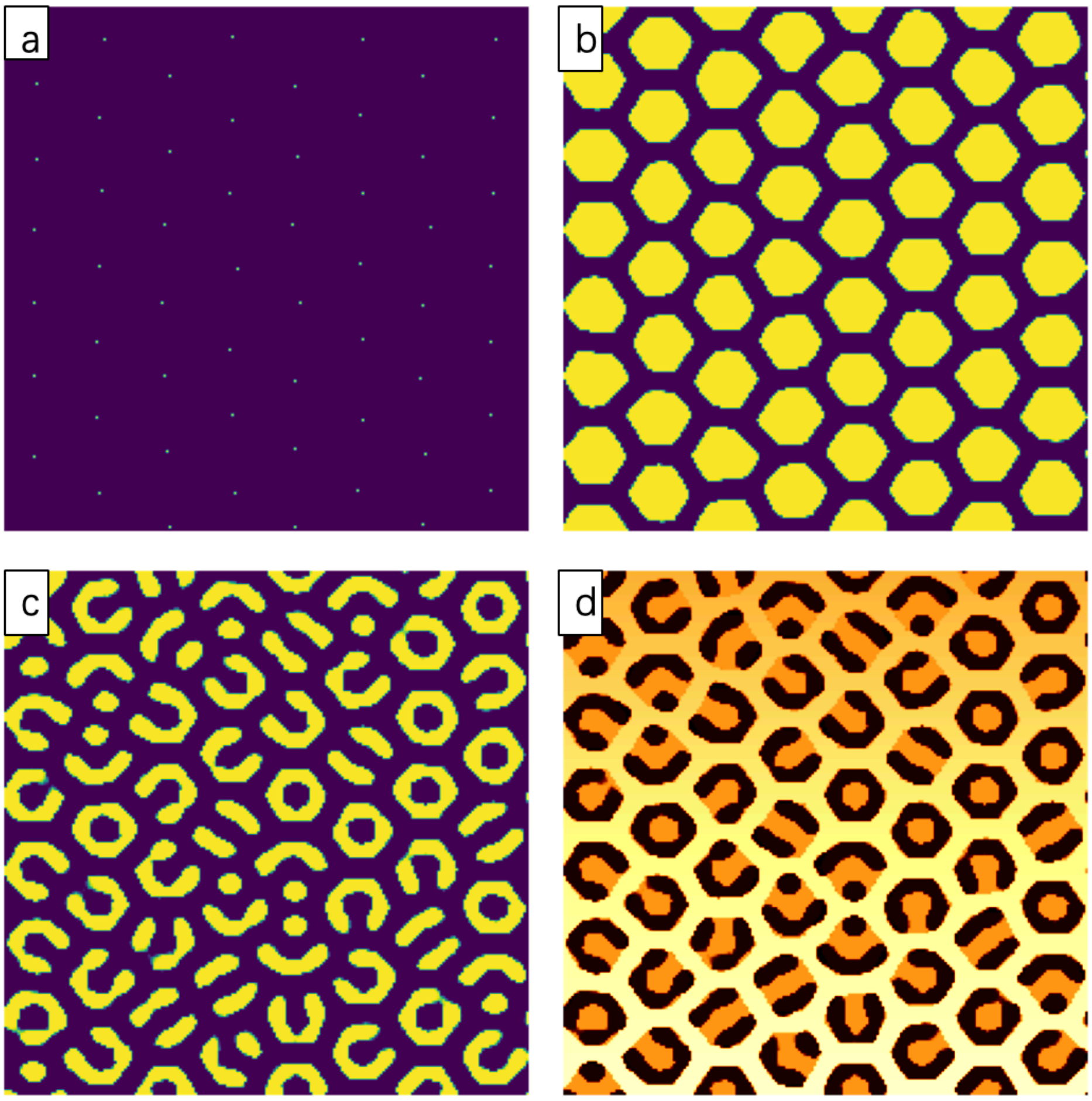

**Fig. S1: The process of generating a leopard pattern.** (**a**) Random perturbations taken from a standard normal distribution, (**b**) $w_+/w_- = 2.25$, (**c**) $w_+/w_- = 1.83$. In all the panels, the scale of physical space is $N = 201 \times 201$, $r_+ = 6$, $r_- = 9$.

The key to generating leopard patterns lies in the decay of transporting growth factors with time. In particular, we suppose that the decay mainly affects the ratio $w_+/w_-$ of excitatory to inhibitory strength over time. Initially, excited individuals are aligned to the vertices of a hexagonal lattice, as shown in Fig. S1(a). After a number of steps, patterns in Fig. S1(b) arise and become stable. Subsequently, by keeping the range of excitation and inhibition fixed, and decreasing the ratio of excitatory to inhibitory strength from 2.25 to 1.83, the pattern in Fig. S1(c) arises. Next, the patterns in Fig. S1(b) are combined with Fig. S1(c) and they are represented by different colors. Finally, a gradually varied background along the vertical axis is embedded to yield the leopard pattern in Fig. S1(d).

# Fish-eye patterns

Fish-eye patterns, including the concentric and radial patterns, as shown in Fig. 4(i) in the main text, arise from the symmetry breaking in polar coordinates. We illustrate a unit range in polar coordinates shown in Fig. S2, where $r^{\perp}$ represents a small distance along the radial direction and $r^{\parallel}$ represents a small distance along the tangential direction. We assume that $w_{+}$ linearly decreases from the origin (center) because of the damping of growth factors from the center, which is formulated as

$$w_{+}(l) = w_{+}(0) - kl,$$

where $k$ is the rate of decay, and $w_{+}(l)$ is the interaction strength along both the tangential and radial direction at distance $l$ to the origin (center).

The symmetry breaking between excitatory range $r_{+}^{\perp}$ and $r_{+}^{\parallel}$ along tangential and radial directions account for the concentric and radial patterns (see Fig. S2). In particular, for the concentric pattern, $r_{+}^{\perp} = 1.6\pi$ and $r_{+}^{\parallel} = 3$, and for the radial pattern, $r_{+}^{\perp} = \pi$ and $r_{+}^{\parallel} = 5$. The inhibitory range of both concentric and radial patterns are identical with $r_{-}^{\perp} = 3\pi$ and $r_{-}^{\parallel} = 9$. The other parameters shared by both the concentric and radial patterns are $k = 5.5 \times 10^{-4}$, $w_{+}(0) = 0.085$, $w_{-} = 0.01$ and $N = 201 \times 201$.

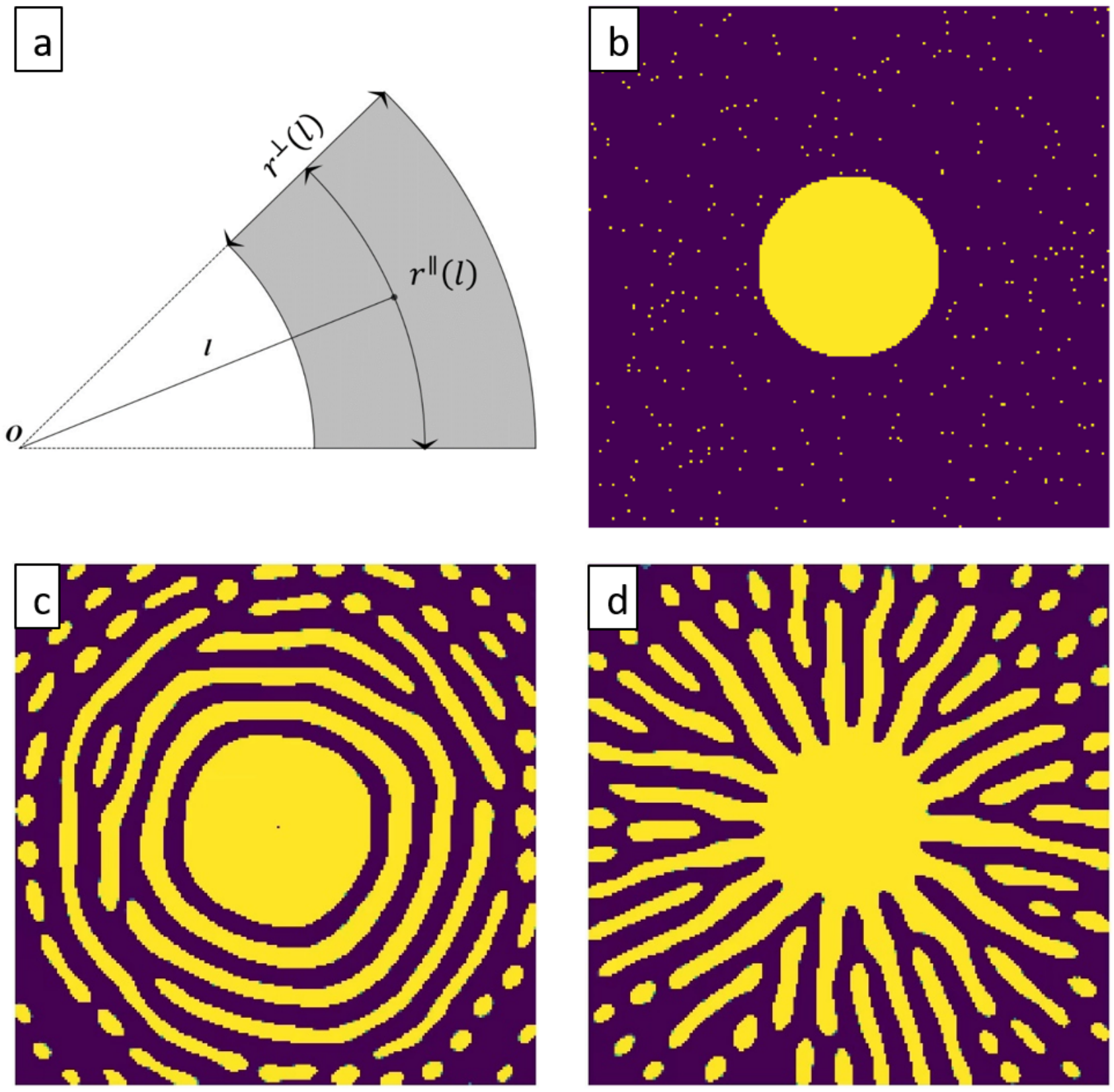

**Fig. S2: Illustration of (a) a unit range in polar coordinates and (b) the initial condition of (c) concentric and (d) radial pattern around fish eyes.**

# Two-dimension branching fractals

Branching fractals emerge from the combination of excitation-inhibition asymmetry and medium asymmetry. To be concrete, the surface of each branch resembles a plane full of excitatory spots with a periodic boundary condition. In other words, a plane is converted into a cone with many spots on its surface. Within each spot cells are excited and outside spots cells are silent. The spots serve as the growth points, at which new branches sprout. Due to the decay of growth factors from the root to the treetop, spots gradually become smaller, as well as the width of branches from the spots. In general, the gravitational field regulates the transportation of growth factors, shaping the diversity of plants. The diversity allows plants to seek their own ecological niches for competing solar energy in a forest. In other words, plants exploit the gravitational field to regulate the distribution of their growth factors for the formation of their specific shapes. In this regard, in addition to the symmetry breaking between excitation and inhibition, there are two parameters adjusting the pattern of branching fractals, i.e., parameter $\sigma$ for capturing the transportation and decay of growth factors and parameter $\varepsilon$ for the sensitivity of growth factors in response to the gravitational field. In short, the two parameters determine the length of branches and their growth directions, respectively. The breaking of growth symmetries as characterized by the two parameters is classified into spatial-medium asymmetry (Category II-C), where the medium refers to spatial distribution of the growth factors. It is worth noting that the two parameters for producing branching fractals are of explicitly physical and biological meanings rather than artificially repeating a fractal operator at different spatial scales in traditional approaches.

For simplicity and without loss of generality, we simulate two-dimension branching fractals, but our approach is applicable to any dimensions as well. The shape of branches is characterized by two physical quantities, thickness and direction at different locations. Branch thickness is immediately determined by the concentration of growth factors, and branch direction is determined by growth differences between the upper and lower side of a branch. The growth differences are determined by the concentration of growth factors as well and are regulated by the gravitational field. For simplicity, we assume a single growth factor that stands for the synergy of a group of actual growth factors affecting branch growth. Moreover, the growth factor gradually decays from the root to the treetop; and along the reverse direction of the gravitational field, the growth factor is transported to allow plants to grow upward. The gravitational field plays a role in guiding the transportation of the growth factor within a certain angel on both sides to the vertical direction (the angle is a tunable parameter for different plants).

In the real situation, branches grow through cell division and proliferation toward distal end layer by layer. However, a realistic simulation of the real developmental process is time-consuming and not necessary for understanding branching fractals. As an alternative, we simplify the cell-division process and use a single big cell to represent a group of original cells. The single cell is of a round shape and adjacent cells can overlap along a branch (see Fig. S3). The size of a single cell is determined by the concentration of the growth factor. Moreover, the location of any cell is determined by its nearest upstream cell and the growth-factor concentrations of both sides along its branch. Specifically, the growth direction of a cell is regulated by the effect of the gravitational field on the growth factor. A few biological quantities are involved in the development process associated with the distribution of the growth factor, such as the width of trunk and the height of a tree, the width and length of each branch, the density of branches, the growth direction of each cell, and so on.

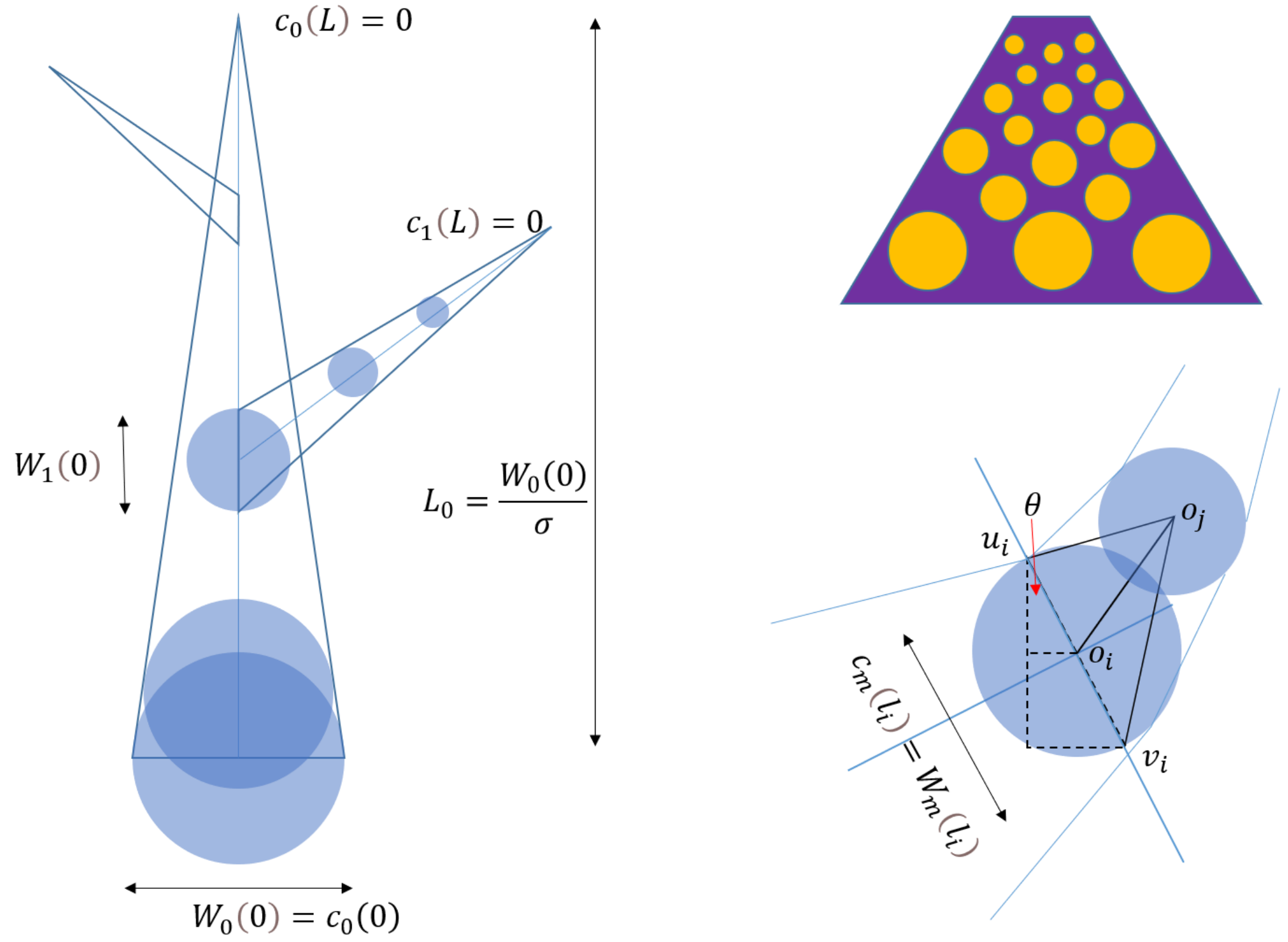

**Fig. S3: Settings of generating two-dimension branching fractals.**

Let us firstly derive the damping of the growth factor from the root without considering the gravitational field. Along with the decay of the growth factor from the root to the treetop and to the end of each branch, cells will become smaller and smaller, until their sizes become negligible. Without loss of generality, we assume that the growth factor decreases linearly from the root and from the growth point of each branch. As shown in Fig. S3, we formulate the concentration of the growth factor as

$$c_i(l) = c_i(0) - \sigma l,$$

where subscript $i$ denotes the number of a specific branch and the trunk is called number zero, parameter $\sigma$ is the damping rate of the growth factor along branches, and $l$ is the distance from the root or the growth point of each branch. For simplicity, we assume that concentration $c_0(0)$ at the root equals the width $W_0(0)$ of root. Thus, the height $L_0$ of a tree can be calculated with respect to the fact that at the top the concentration and width both equal zero, i.e.,

$$c_0(L_0) = c_0(0) - \sigma L_0 = W_0(0) - \sigma L_0 = 0,$$

which yields

$$L_0 = \frac{W_0(0)}{\sigma}.$$

The width $W_i(0)$ at the root of each branch is determined by the size of its growth spot, where the growth spot is determined by the excitatory-inhibitory forces and the concentration of the growth factor at the spot together. Let us take the number one branch as a typical example. For two-dimension trees, the surface space is one dimension; and in the one-dimension space, for a given set of interaction strength $w_+$, $w_-$ and range $r_+$, $r_-$, we obtain a number of one-dimension spot on the surface of the trunk as growth points of branches, according to the asymmetric self-organization. Subsequently, at the growth points of the number one branch, the width $W_1(0)$ of its root is

$$W_1(0) = d_1 \cdot \frac{c_0(l_1)}{W_0(0)},$$

where $d_1$ is the diameter of the growth spot of the number one branch, and $c_0(l_1)/w_0(0)$ is the normalized concentration of the growth factor at location $l_1$ of the trunk. The initial concentration of the number one branch equals to its initial width $W_1(0)$, and its length $L_1$ is determined by $W_1(0)$ as well, i.e.,

$$L_1 = \frac{W_1(0)}{\sigma} = \frac{d_1 \cdot c_0(l_1)}{\sigma \cdot W_0(0)}.$$

In the same vein, the location, initial width and length of all branches are determined insofar as we know the width of tree root (see Fig. S3).

Secondly, we include the effect of the gravitational field and elucidate how the concentration of the growth factor decays along the vertical direction, as shown in Fig. S3. For an arbitrary cell, say $i$ centered at $o_i$ with diameter $W_m(o_i)$ in branch number $m$, the concentration $c_m(o_i)$ at its center equals its diameter, i.e., $c_m(o_i) = W_m(o_i)$, and $c_m(o_i)$ is determined by the distance of $o_i$ from the growth point along the branch. For the next downstream cell $j$ centered at $o_j$, the growth rate of its two sides differs from each other and regulated by the gravitational field. We denote the two side points of $i$ by $u_i$ and $v_i$. Note that the concentration at a location equals the width of a branch at the location, e.g., the length $W_m(o_i)$ of diameter $(u_i v_i)$ equals $c_m(o_i)$. Thus, we employ a geometric analysis to derive concentrations of the growth factor so as to determine the location of cell $j$. The concentration difference between $u_i$ and $v_i$ accounts for the deviation of direction between cell $j$ and $i$. In addition, the differences stem from the gravitational field. The concentration at $u_i$ and $v_i$ can be estimated by referring to the concentration at location $o_i$. Because that the transportation process of growth factors is suppressed by the gravitational potential, the concentration at $u_i$ is lower than that at $v_i$. For simplicity, we assume that the decrease of concentration is proportional to the difference of the gravitational potential along the vertical direction. As shown in Fig. S3, the concentration at $u_i$ and $v_i$ can be expressed as

$$c_m(u_i) = c_m(o_i) - \frac{1}{2}\varepsilon \cdot c_m(o_i) \cdot \cos\theta,$$

$$c_m(v_i) = c_m(o_i) + \frac{1}{2}\varepsilon \cdot c_m(o_i) \cdot \cos\theta,$$

where $\theta$ is the angle between the direction of diameter $(u_i v_i)$ and the vertical direction, $\varepsilon$ is the damping of the growth factor regulated by the gravitational field along the vertical direction, and the second term on the right-hand side is the projection of the length of radius onto the vertical direction. The central location $o_j$ of cell $j$ is determined by $c_m(u_i)$ and $c_m(v_i)$ together, as shown in Fig. S3, where the length of $(u_i o_j)$ and $(v_i o_j)$ is proportional to $c_m(u_i)$ and $c_m(v_i)$, respectively. To ensure that $(u_i o_j)$ and $(v_i o_j)$ have an intersection, we multiply $c_m(u_i)$ and $c_m(v_i)$ with a scale factor $2r_+$ to increase the length of $(u_i o_j)$ and $(v_i o_j)$, i.e.,

$$L_{(u_i o_j)} = 2r_+ \cdot c_m(u_i),$$
$$L_{(v_i o_j)} = 2r_+ \cdot c_m(v_i),$$

where the scale factor $2r_+$ is the diameter of the excitatory range, and we believe this factor reflects the growth process of branches. The diameter of cell $j$ is determined by the distance from $o_j$ to the root of the branch according to formula

$$c_m(o_j) = c_m(0) - \sigma l_{o_j},$$

where $l_{o_j}$ is the distance at $o_j$ along the branch, and $c_m(o_j)$ equals the diameter $W_m(o_j)$.

Since we have obtained both the center and diameter of cell $j$, the spatial location of cell $j$ is fully determined. Subsequently, we iterate these processes to calculate the downstream cell next to $j$, until the diameter of the last cell along the branch becomes zero. A number of trees by tuning parameter $\sigma$ and $\varepsilon$ are shown in Fig. S4 and in Fig. 5 in the main text.

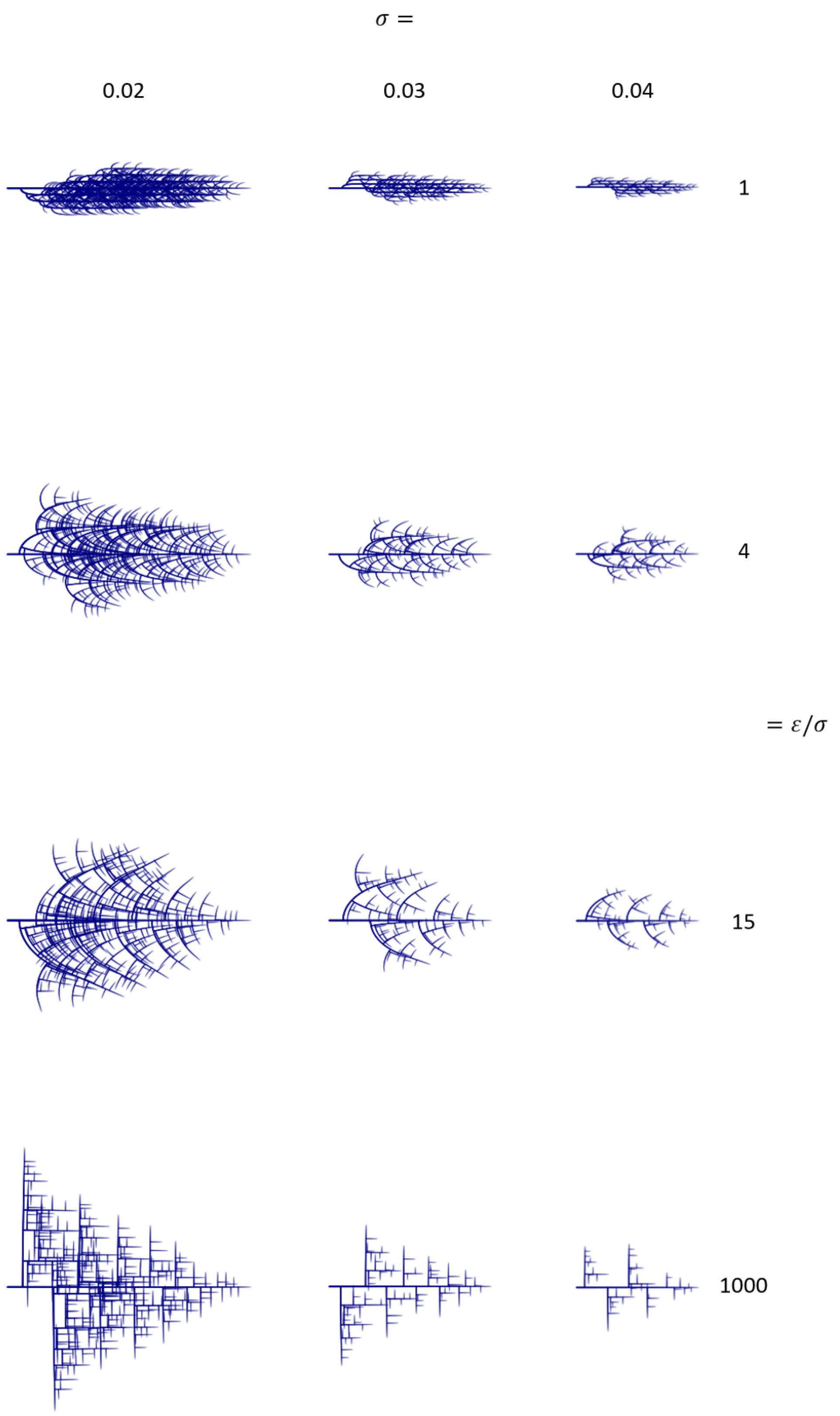

**Fig. S4: Branching fractals by adjusting parameter $\sigma$ and $\varepsilon$.** Other parameter values are $W_0(0) = 40$ pixels, $r_+ = 20$, $r_- = 40$, $w_+ = 0.01$, and $w_- = 0.0067$.

# Spatial-stability analyses of Turing patterns

## Spatial-stability analyses of spots

Let us derive spot sizes based on the geometrical properties of interactions. The size of a spot depends on both the excitatory and inhibitory forces. At the border of a spot, the forces reach an equilibrium. Thus, if we are able to identify locations where the forces are balanced, the locations determine spot sizes. In particular, forces received by every individual are jointly determined by the states of neighbors within interaction ranges. We examine individual states to seek ways of estimating joint forces. Fortunately, we find that the states of individuals in any static patterns are polarized and approach either the lower or the upper bound, i.e., $s = 1$ or $s = 0$. And the number of individuals, according to the discrete space arrangement, is just the size of the area of interactions. The joint force defined as the sum of the product of individual state and interaction strength, is thus proportional to the area of interactions. In principle, given a spot, we can ascertain both the positive and negative interaction areas of an arbitrary individual at the border. However, regarding to the relationship between the interaction ranges and the size of spot, we have to discuss two scenarios separately. We classify stable spots into Case I, and Case I contains Subcase A and B, as displayed in Table S1.

**Case I-A**: $r < 2R_0$. As shown in Fig. S5, suppose that a spot initiated from a single excited individual is of radius $R_0$. The blue circle represents the border of the spot, and the black circle represents the range of individual interactions (either positive or negative) with the radius of interact range $r$. The gray area where the two circles overlap represents the area of interaction, denoted as $A^{\mathrm{IA}}$. We can calculate the interaction area with radius $r$ via

$$A^{\mathrm{IA}}(r) = 2\theta R_0^2 + \left(\frac{\pi}{2} - \theta\right) r^2 - R_0 r \cos\theta,$$

where

$$\theta = \arcsin\left(\frac{r}{2R_0}\right).$$

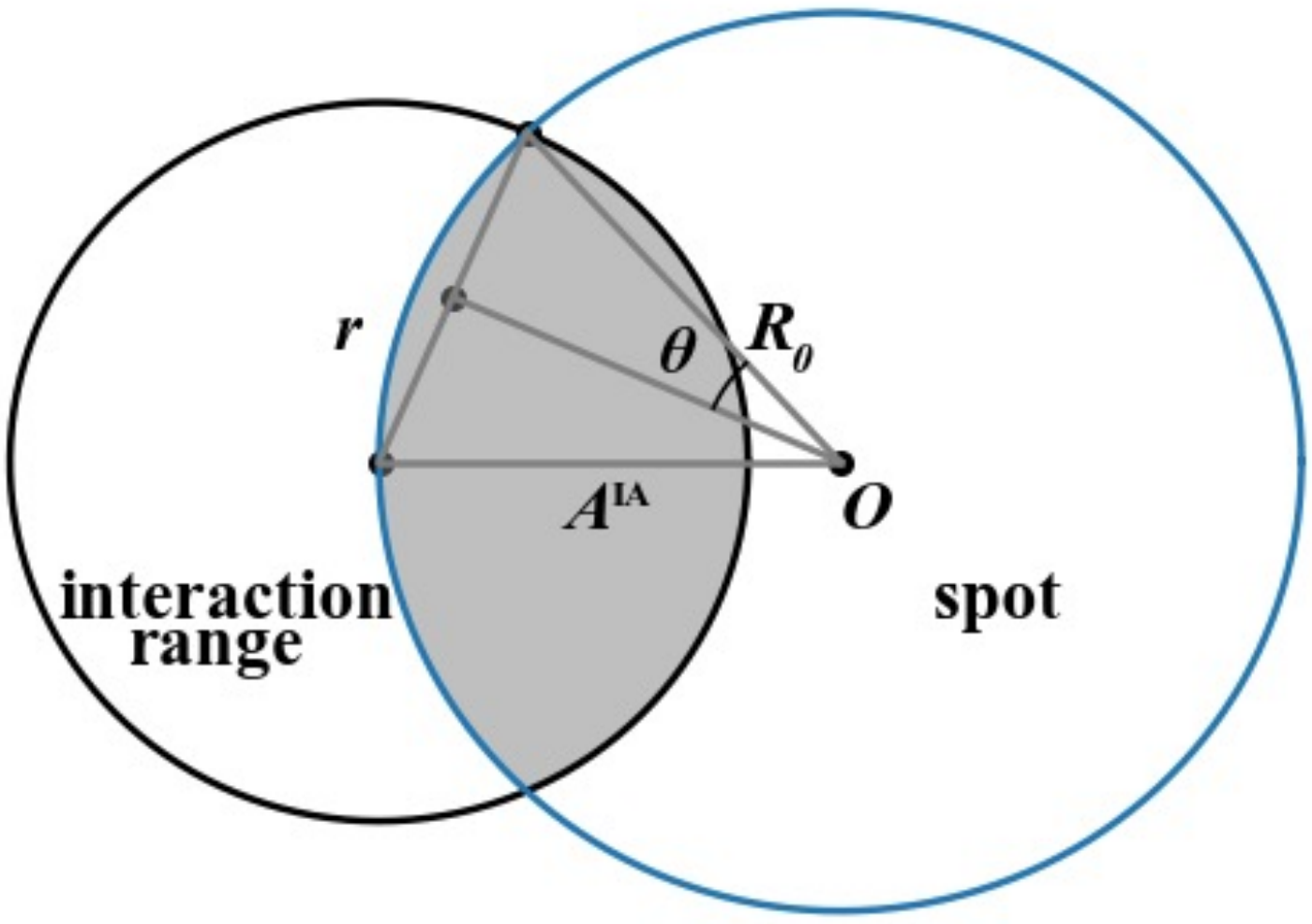

**Fig. S5. Illustration of interaction area and relevant variables.** The black circle represents an interaction range centered at a border individual, the blue circle is the border of a spot and the gray area denotes the interaction area.

**Case I-B:** $r \geq 2R_0$. As shown in Fig. S6, the gray area $A^{\mathrm{IB}}$ represents the area of interaction. We have

$$A^{\mathrm{IB}}(r) = \pi R_0^2.$$

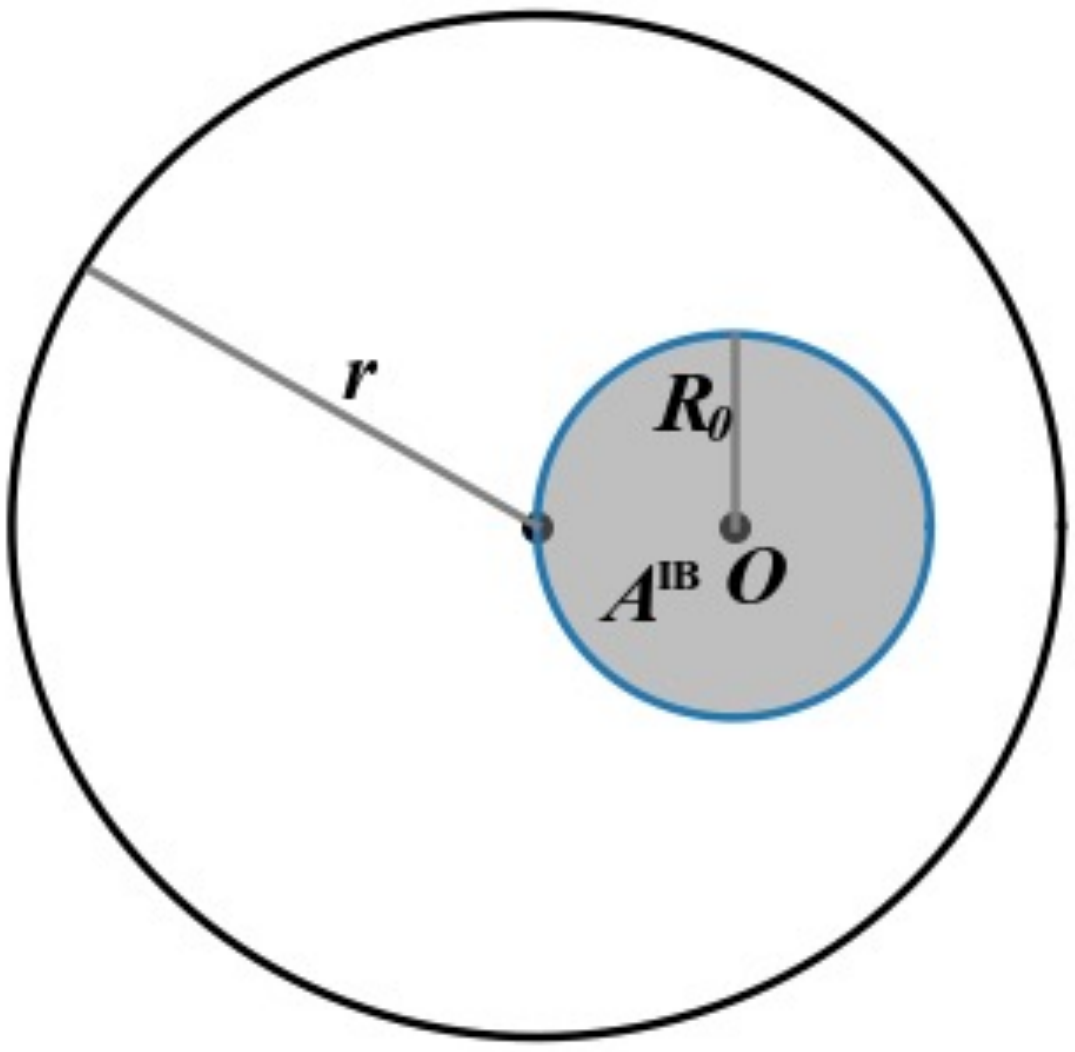

**Fig. S6. Illustration of interaction area and relevant variables.**

Let us examine the locations where the excitatory and inhibitory forces are equal, in order to derive the size of spots. According to the definition, both excitatory and inhibitory forces received by individuals located at the border of a spot are $A(r_+)w_+$ and $A(r_-)w_-$, respectively. At the border, an equilibrium is achieved, i.e.,

$$A(r_+)w_+ = A(r_-)w_-.$$

Taking into account the fact that the range of both positive and negative forces belongs to either Case I-A and I-B, there are three possible combinations of forces at the border associated with equilibrium, as follows:

$$\begin{cases} A^{\mathrm{IA}}(r_+)w_+ = A^{\mathrm{IA}}(r_-)w_-, & 2R_0 > r_- > r_+, \\ A^{\mathrm{IA}}(r_+)w_+ = A^{\mathrm{IB}}(r_-)w_-, & r_- > 2R_0 > r_+, \\ A^{\mathrm{IB}}(r_+)w_+ = A^{\mathrm{IB}}(r_-)w_-, & r_- > r_+ > 2R_0. \end{cases}$$

Because these are transcendent equations without analytical results, we have to solve the equations numerically.

On the one hand, if the range ratio $r_-/r_+$ is fixed, the numerical solution of spot size $R_0/r_-$ depends exclusively on the strength ratio $w_+/w_-$, as shown in Fig. S7. In particular, the red-dash-line ratio $w_+/w_-$ is an asymptotic line corresponding to an infinite-size spot.

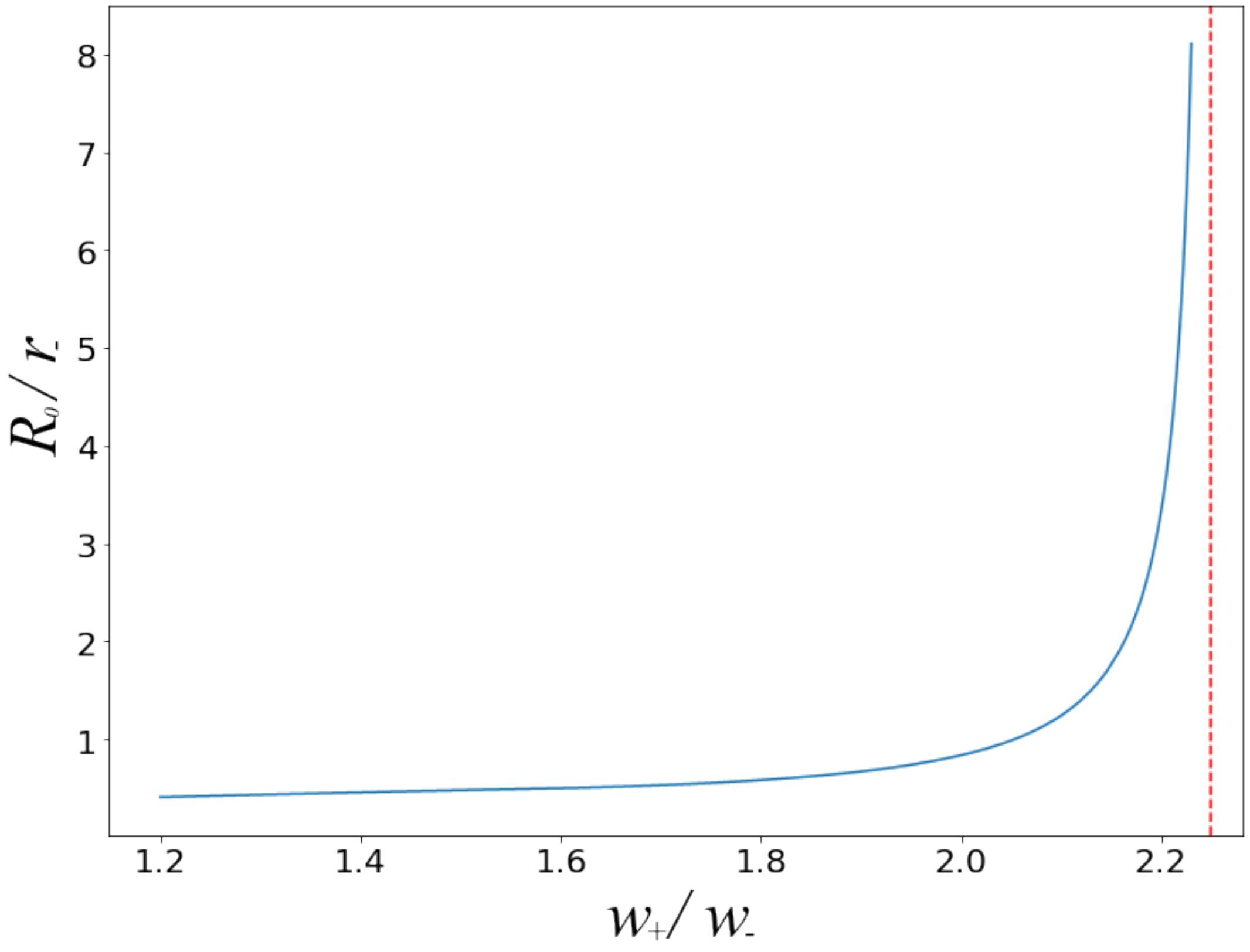

**Fig. S7: Theoretical radius of spots.** The radius $R_0$ of a single spot normalized by the radius $r_-$ of inhibition as a function of the strength ratio $w_+/w_-$. Spots initiate from a single excited individual. The red-dash-line ratio $w_+/w_-$ is an asymptotic line corresponding to an infinite-size spot.

On the other hand, if the strength ratio $w_+/w_-$ is fixed, the spot size $R_0/r_-$ depends on the range ratio $r_-/r_+$, as shown in Fig. S7. In particular, the red-dash-line ratio $r_-/r_+$ corresponds to an infinite-size spot.

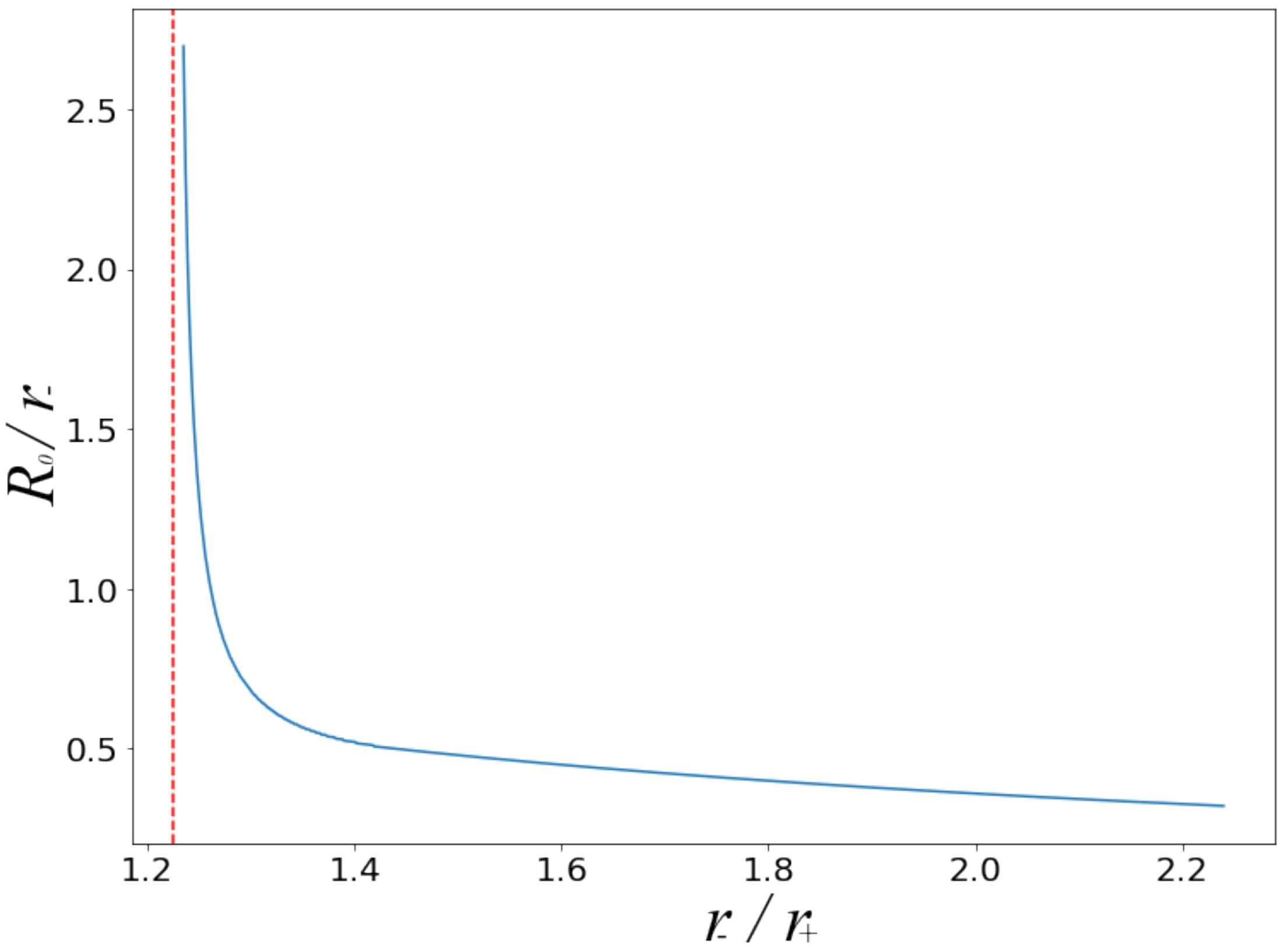

**Fig. S8: Theoretical radius of spots.** The radius $R_0$ of a single spot normalized by the radius $r_-$ of inhibition as a function of the radius ratio $r_-/r_+$. Spots initiate from a single excited individual. The red-dash-line ratio $r_-/r_+$ corresponds to an infinite-size spot.

# Phase-transition points and pattern percolation

Let us first derive the critical point $C_2$, at which pattern percolation occurs. In fact, the percolation transition arises when a single spot tends be infinite in an unlimited space. In this case, the border of an infinite-size spot approaches a straight line, and positive and negative forces balance at the straight border.

Let us denote the critical point as $(w_+/w_-)_\infty$ and $(r_-/r_+)_\infty$. As shown in Fig. S9, when $R_0$ tends to infinity, the interaction area of both positive and negative forces centered at a border individual becomes a half-circle. Based on the immediate relationship between areas and interaction forces, we can have the equilibrium condition, as follows

$$w_+ \cdot \frac{1}{2}\pi r_+^2 = w_- \cdot \frac{1}{2}\pi r_-^2, \qquad (R_0 \to \infty)$$

which gives

$$\left(\frac{w_+}{w_-}\right)_\infty = \left(\frac{r_-}{r_+}\right)^2,$$

or

$$\left(\frac{r_-}{r_+}\right)_\infty = \sqrt{\frac{w_+}{w_-}}.$$

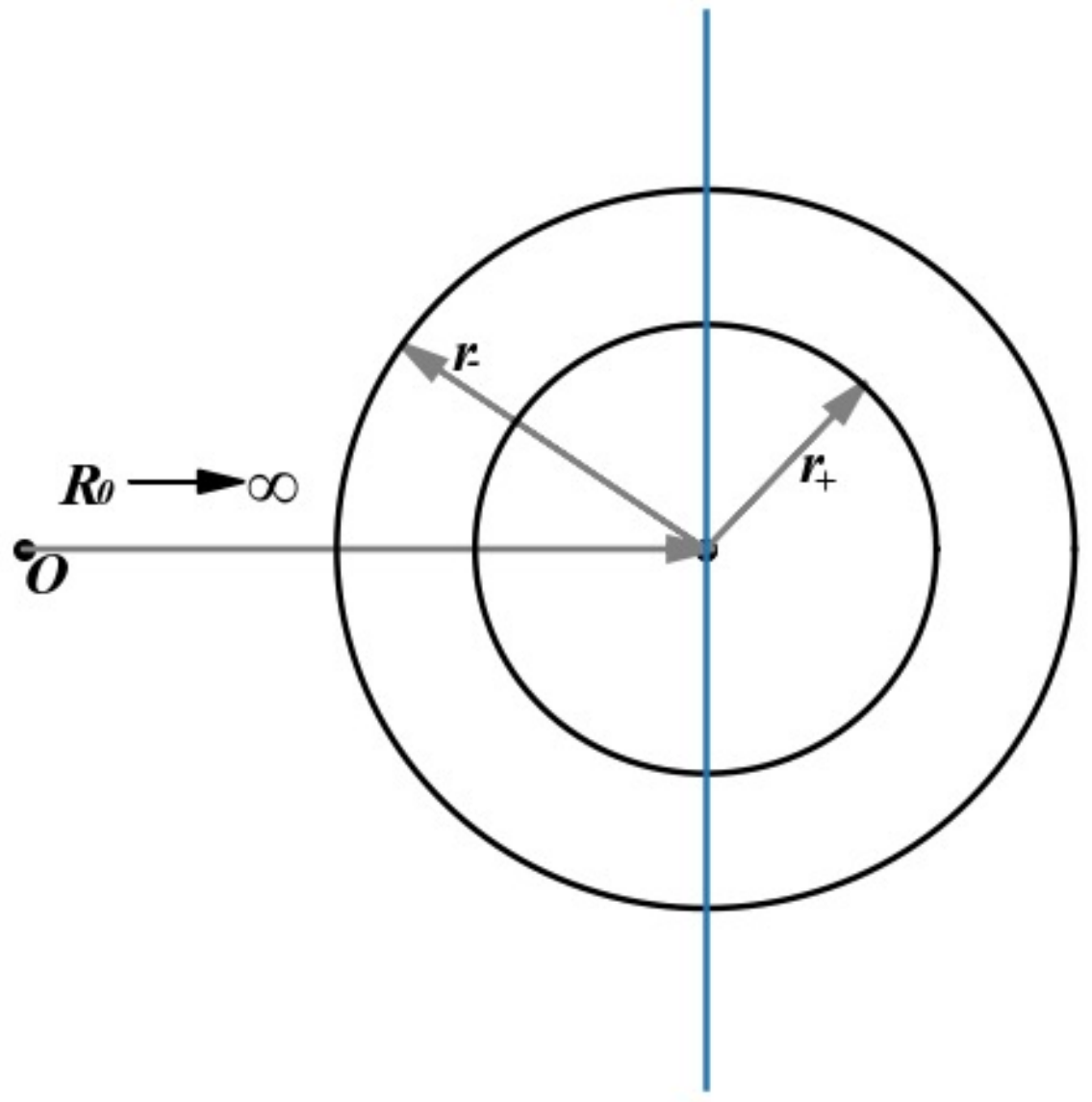

**Fig. S9: Illustration of interaction ranges at the border of an infinite-size spot.**

Let us continuously derive the critical point where a finite-size spot turns to a ring. At the phase transition point, the positive-force dominance at the center of a spot reverse to negative-force dominance, giving rise to a silent hole inside. We thus formulate forces at the center to identify the transition point. Specifically, the forces as well depend on the interaction areas centered at the center of a spot.

We denote the spot radius at the point of transition as $R_{\mathrm{rev}}$. We can infer that $r_+ < R_{\mathrm{rev}} < r_-$. To be concrete, if $r_+ > R_{\mathrm{rev}}$, the positive force is always dominant at the spot center, rendering a ring impossible. If $R_{\mathrm{rev}} > r_-$, the joint force at the center is independent of the spot size $R_{\mathrm{rev}}$, indicating that the central force cannot be reversed. Taken together, at the transition point from a spot to a ring, the size of spot ought to satisfy the condition $r_+ < R_{\mathrm{rev}} < r_-$.

As shown in Fig. S10, the equilibrium condition at the center is

$$w_+ \cdot \pi r_+^2 = w_- \cdot \pi R_{\mathrm{Rev}}^2, \qquad (r_+ < R_{\mathrm{rev}} < r_-)$$

which yields the critical transition size from a spot to a ring

$$R_{\mathrm{rev}} = r_+ \sqrt{\frac{w_+}{w_-}}.$$

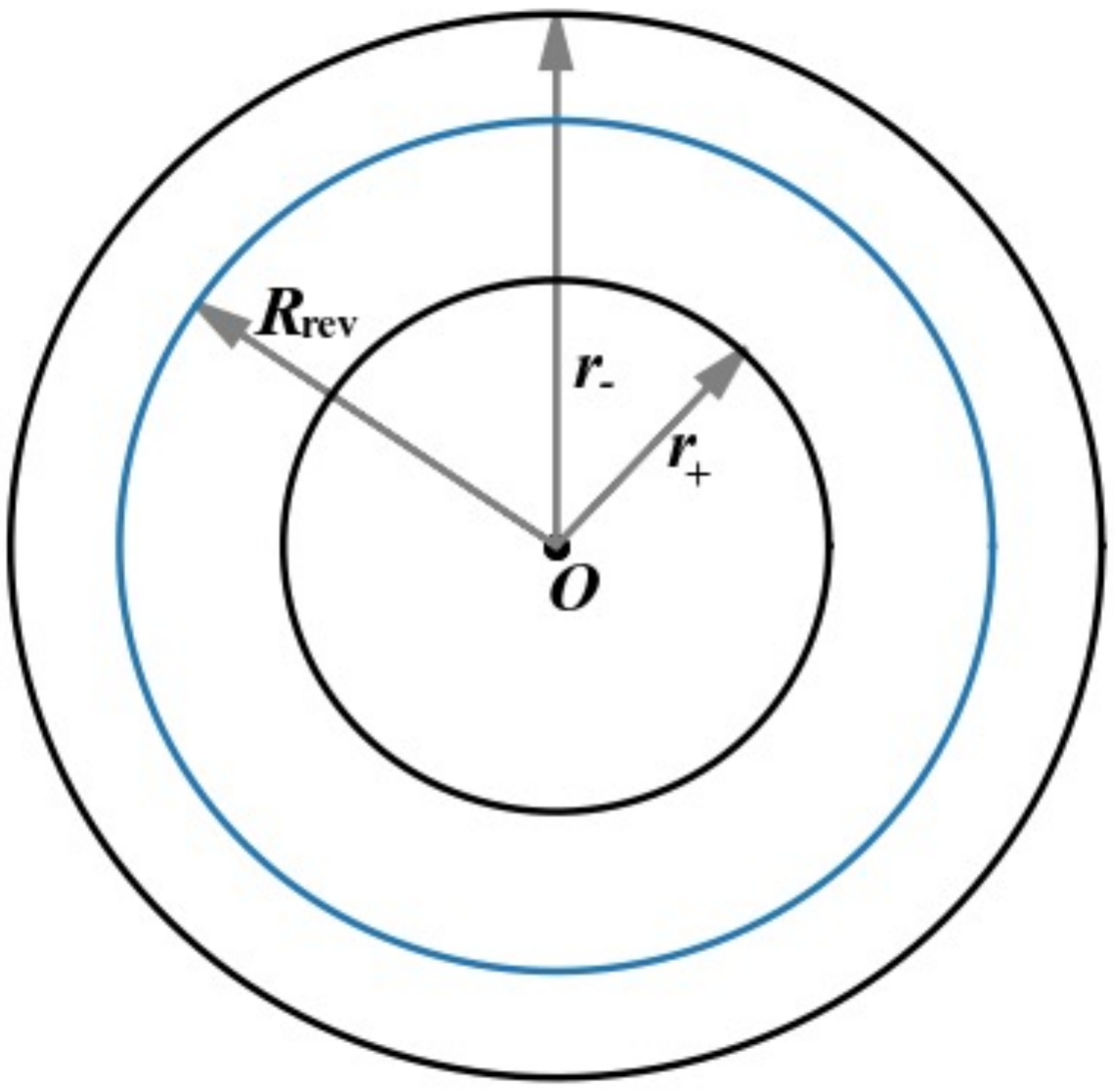

**Fig. S10: Illustration of interaction ranges at the transition from a spot to a ring.**

However, the ring-formation condition might not always be satisfied. As shown in Fig. S11, when $w_+/w_-$ is relatively small, the actual spot size is less than the ring-formation condition $R_{\mathrm{rev}}$, such that no transition occurs. Only if $w_+/w_-$ is larger than the critical ratio $(w_+/w_-)_{\mathrm{ring}}$, $R_{\mathrm{rev}}$ can be reached, resulting in the phase transition from a spot to a ring. The numerical solution of the phase transition point is in good agreement with simulation results, as shown in Fig. 10(b) in the main text.

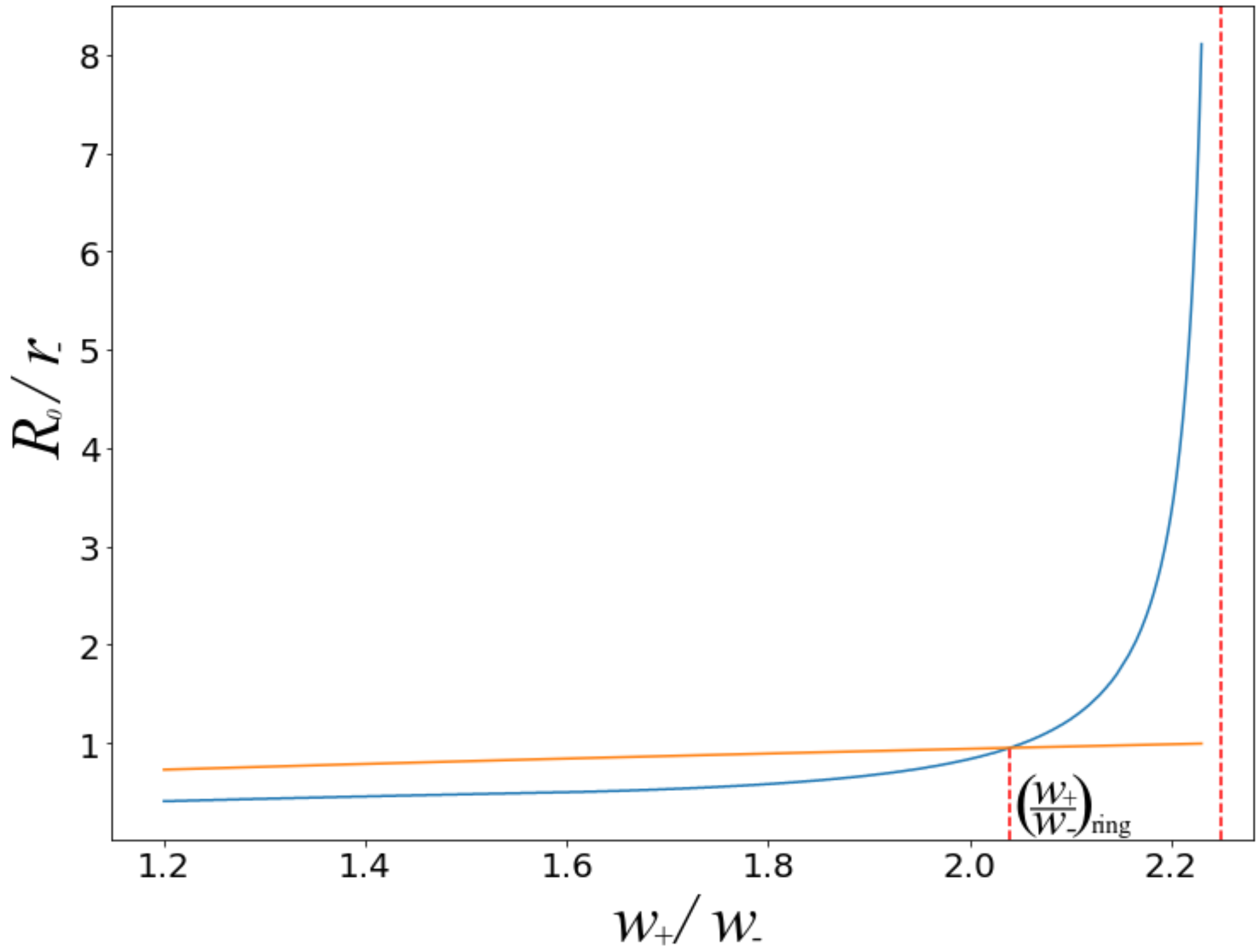

**Fig. S11: Graphical solution of the phase-transition from a spot to a ring.** The blue curve corresponds to the radius $R_0$ of a single spot divided by inhibitory range $r_-$ as a function of the strength ratio $w_+/w_-$ initiated from a single excited individual. The yellow curve corresponds to the critical radius $R_{\mathrm{rev}}$ of the ring-formation condition divided by $r_-$ as a function of $w_+/w_-$. The intersection of the yellow with the blue curve is the phase transition point from a spot to a ring.

## Spatial-stability analyses of rings

A ring is stable if both its inner and outer borders are stable. In this regard, we examine the stability of inner and outer border separately. Their stability is determined by the joint forces as well, and can be converted into a geometrical problem. Fig. S12 shows variables involved in our analysis, where $R_0$ and $R_1$ represent the radius of inner and outer borders of the ring, respectively, and $r$ represents the range of individual interactions (either positive or negative). The gray area $A_1$ is the overlap between the interaction range (either positive or negative) centered at an arbitrary inner-border individual and the outer circle of the ring. The blue area $A_2$ is the overlap between the interaction range $r$ (either positive or negative) centered at an arbitrary inner-border individual and the inner circle of the ring. The difference $A_1 - A_2$ between $A_1$ and $A_2$ is the actual area where interactions occur.

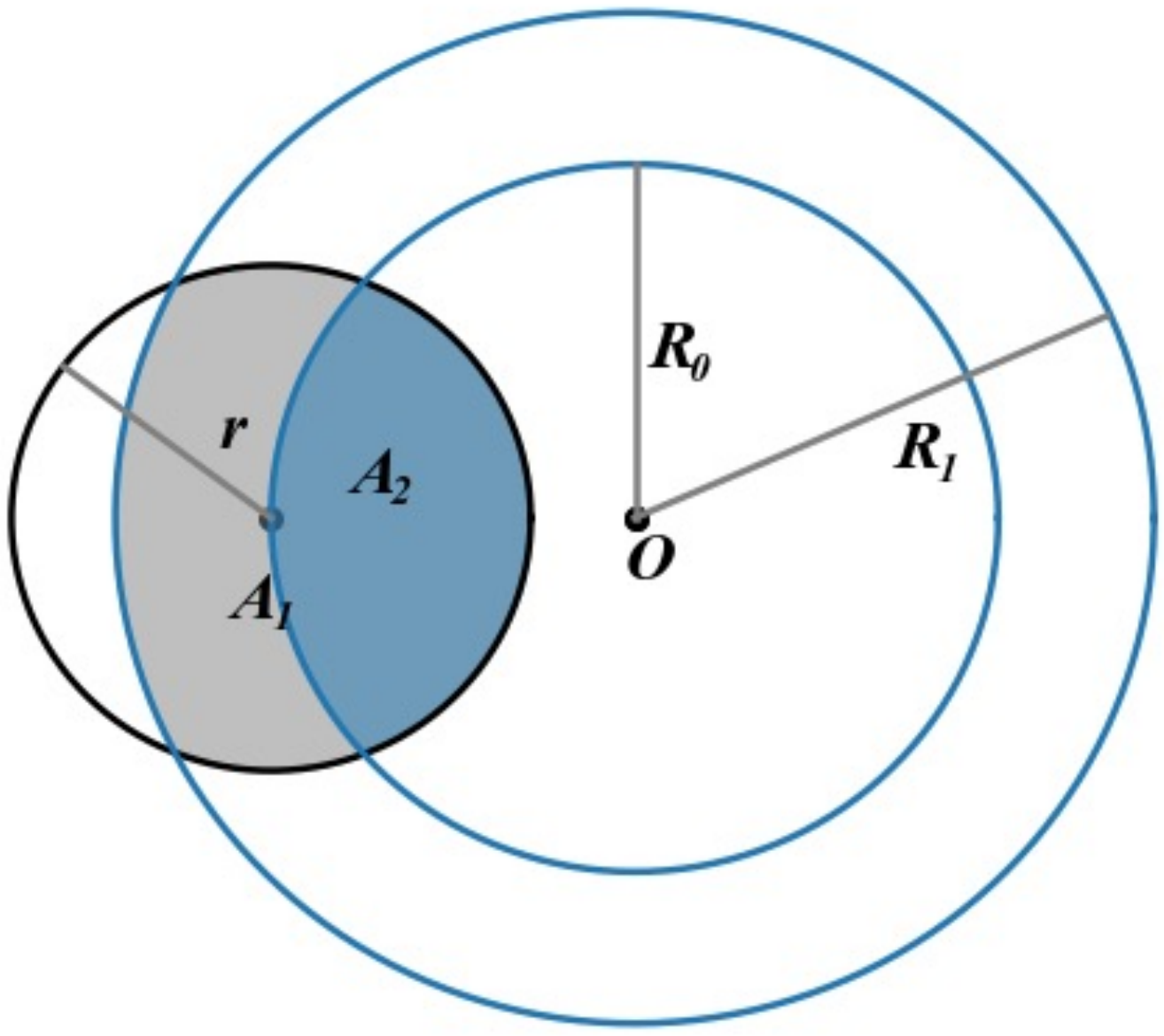

**Fig. S12: Variables involved in the analysis of inner-ring stability.**

The geometric features associated with the inner and outer borders and their interaction ranges are classified into two cases (Case II for the inner border and Case III for the outer border) with a few subcases, as displayed in Table S1. We firstly examine the stability of inner border, as follows.

**Case II(A)**: $r \leq R_1 - R_0$. As shown in Fig. S13, we have

$$A_1^{\mathrm{IIA}}(r) = \pi r^2.$$

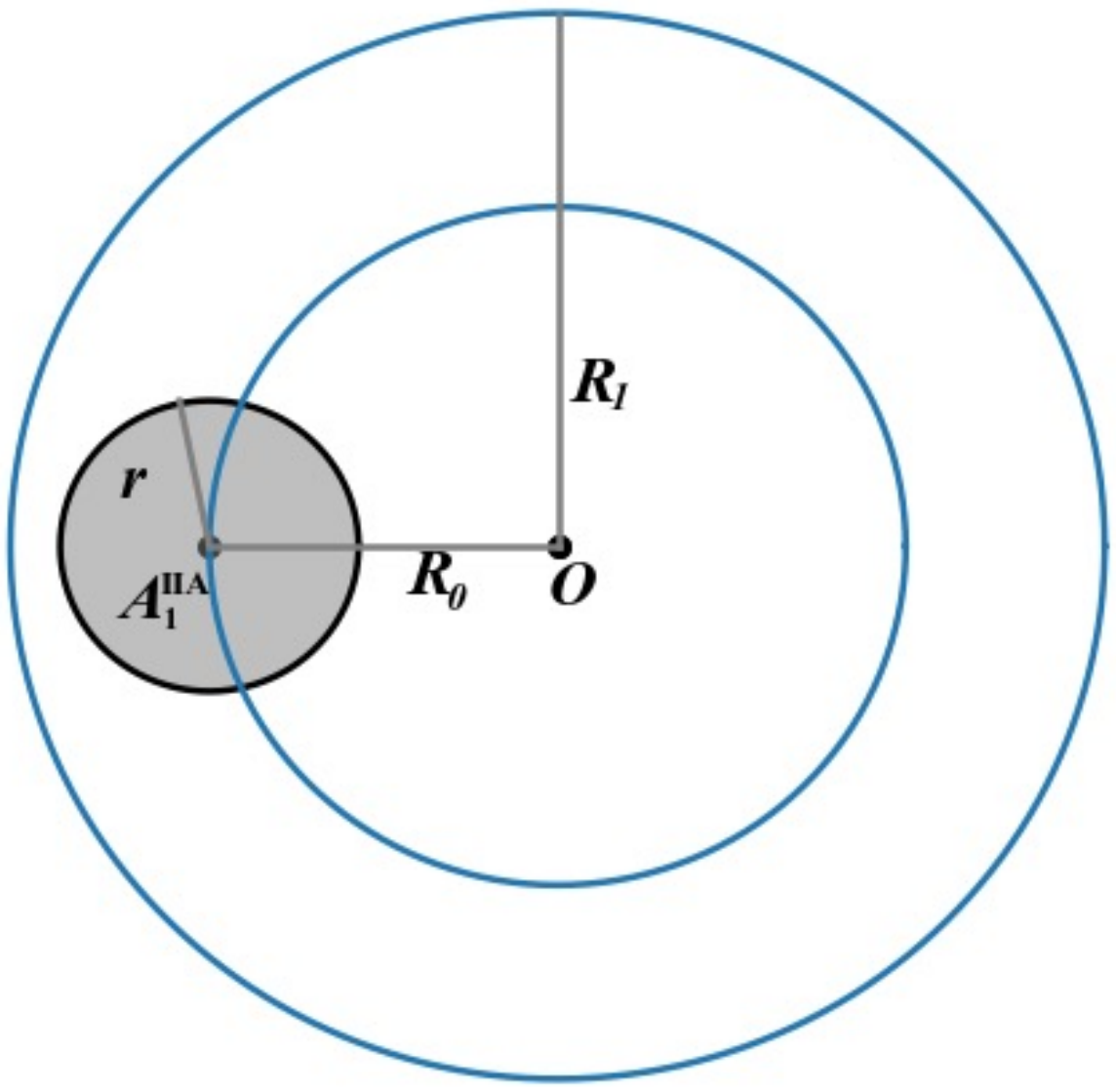

**Fig. S13: Variables involved in Case II(A).**

**Case II(B)**: $R_1 - R_0 < r \leq R_1 + R_0$. As shown in Fig. S14, we have

$$A_1^{\mathrm{IIB}}(r) = r^2\theta_1 + R_1^2\theta - 2\sqrt{p(p - R_0)(p - R_1)(p - r)},$$

where

$$\theta_1 = \frac{\arccos\,(r^2 + R_0 - R_1^2)}{2R_0 r},$$

$$\theta = \frac{\arccos\,(R_0^2 + R_1^2 - r^2)}{2R_0 R_1},$$

$$p = \frac{1}{2}(R_0 + R_1 + r).$$

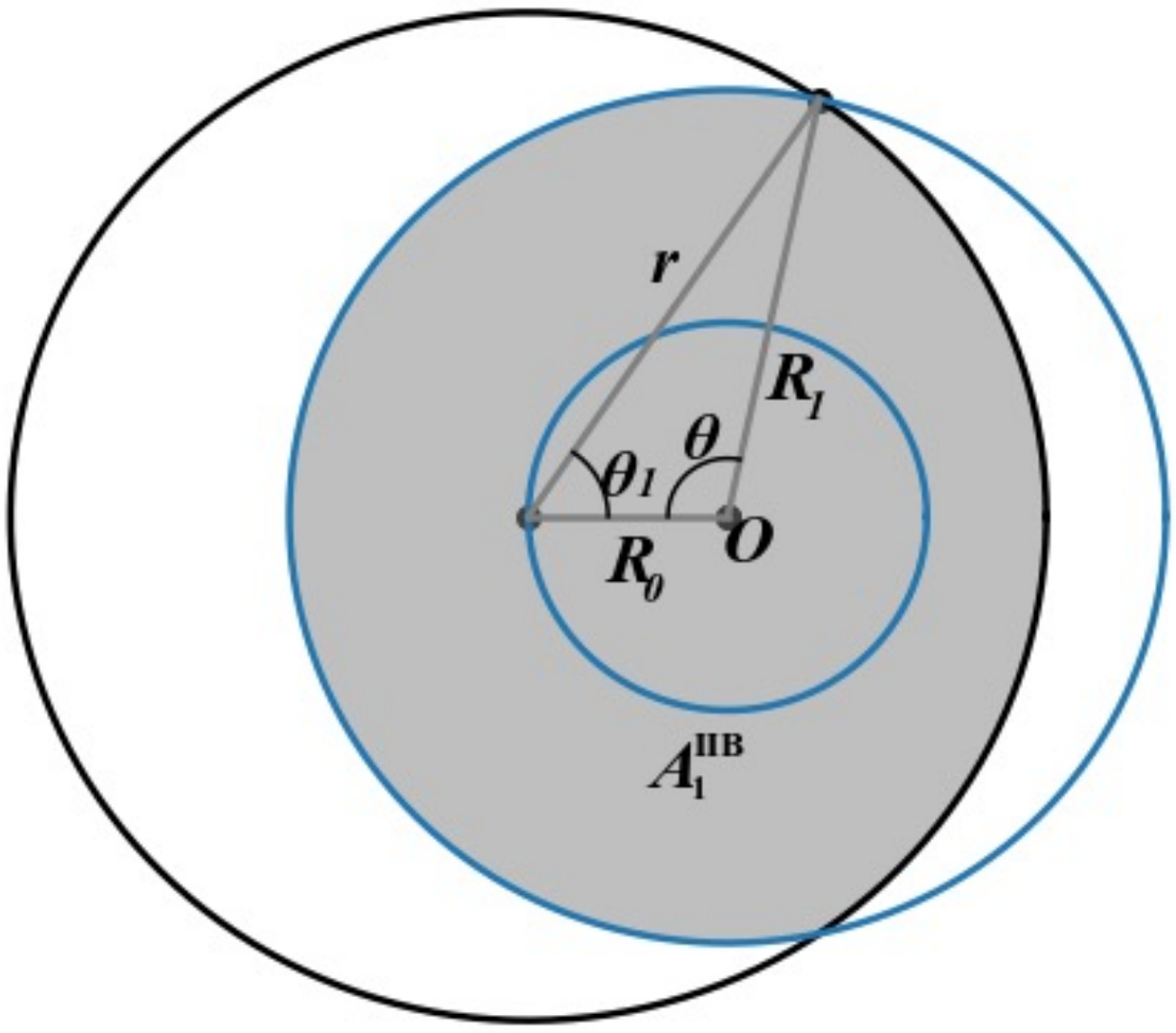

**Fig. S14: Variables involved in Case II(B).**

**Case II(C)**: $R_1 + R_0 \leq r$. As shown in Fig. S15, we have

$$A_1^{\mathrm{IIC}}(r) = \pi R_1^2.$$

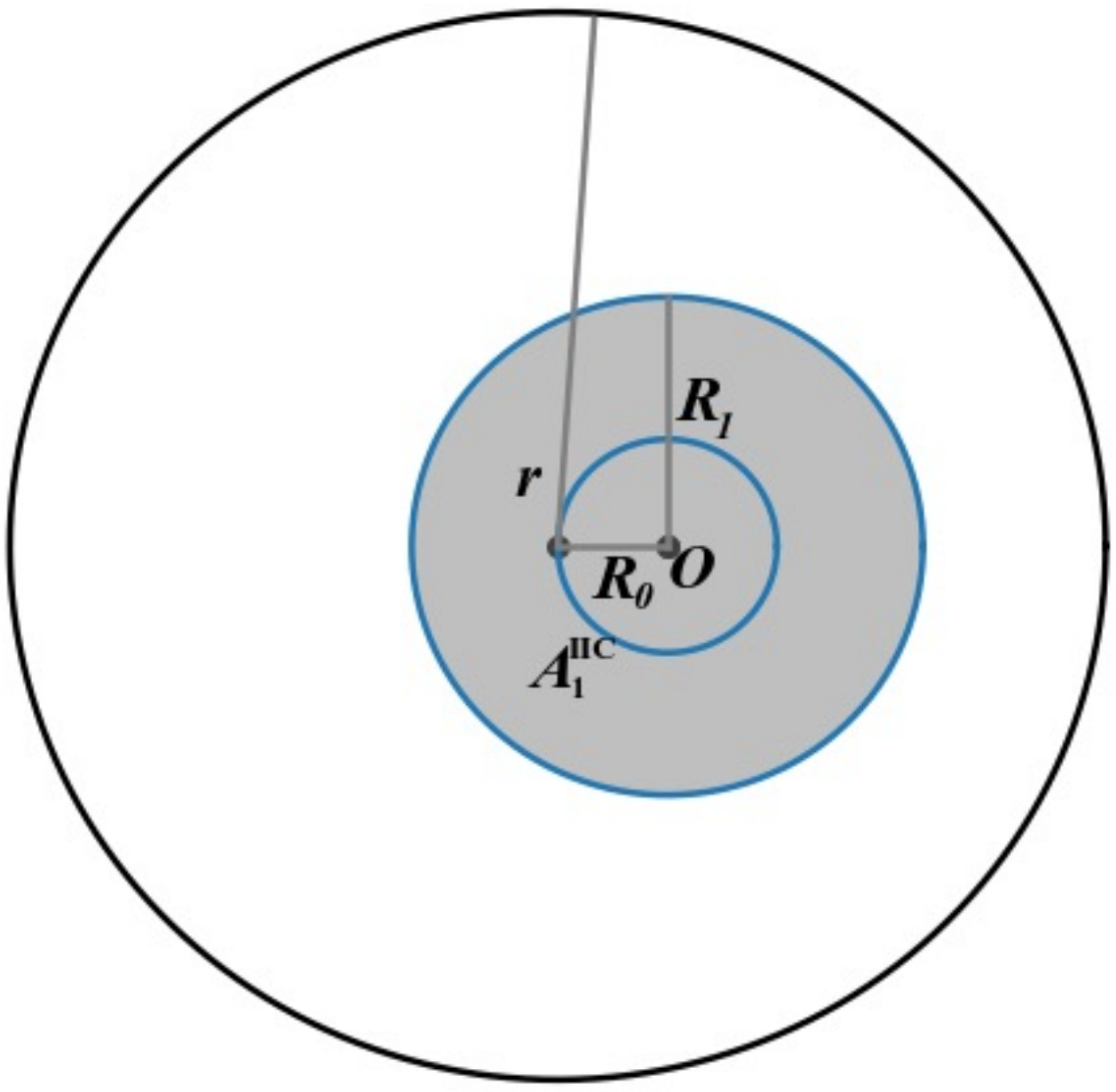

**Fig. S15: Variables involved in Case II(C).**

**Case II(D)**: $r < 2R_0$. As shown in Fig. S5, we have

$$A_2^{\mathrm{IID}}(r) = 2\theta R_0^2 + \left(\frac{\pi}{2} - \theta\right) r^2 - R_0 r \cos\theta,$$

where

$$\theta = \arcsin\left(\frac{r}{2R_0}\right).$$

**Case II(E)**: $r \geq 2R_0$. As shown in Fig. S6, we have

$$A_2^{\mathrm{IIE}}(r) = \pi R_0^2.$$

Next, we provide stability condition of the inner border. In particular, when individuals at the inner border receive equal excitatory and inhibitory forces, the inner border is stable. The stability condition is formulated as

$$\big(A_1(r_+) - A_2(r_+)\big)w_+ = \big(A_1(r_-) - A_2(r_-)\big)w_-,$$

where the left-hand side and the right-hand side are the excitatory and inhibitory force at the inner border, respectively.

From Case II(A) to Case II(E), there are fourteen possible combinations of force ranges centered at the inner border pertaining to equilibrium, as follows:

$$\begin{cases}
(A_1^{\mathrm{IIA}}(r_+) - A_2^{\mathrm{IID}}(r_+))w_+ = (A_1^{\mathrm{IIA}}(r_-) - A_2^{\mathrm{IID}}(r_-))w_-, & 2R_0 \geq r_-;\ R_1 - R_0 \geq r_- > r_+, \\
(A_1^{\mathrm{IIA}}(r_+) - A_2^{\mathrm{IID}}(r_+))w_+ = (A_1^{\mathrm{IIB}}(r_-) - A_2^{\mathrm{IID}}(r_-))w_-, & 2R_0 \geq r_-;\ R_1 + R_0 > r_- > R_1 - R_0 \geq r_+, \\
(A_1^{\mathrm{IIB}}(r_+) - A_2^{\mathrm{IID}}(r_+))w_+ = (A_1^{\mathrm{IIB}}(r_-) - A_2^{\mathrm{IID}}(r_-))w_-, & 2R_0 \geq r_-;\ r_1 + r_0 > r_- > r_+ > R_1 - R_0, \\
(A_1^{\mathrm{IIA}}(r_+) - A_2^{\mathrm{IID}}(r_+))w_+ = (A_1^{\mathrm{IIA}}(r_-) - A_2^{\mathrm{IIE}}(r_-))w_-, & r_- > 2R_0 \geq r_+;\ R_1 - R_0 > r_- > r_+, \\
(A_1^{\mathrm{IIA}}(r_+) - A_2^{\mathrm{IID}}(r_+))w_+ = (A_1^{\mathrm{IIB}}(r_-) - A_2^{\mathrm{IIE}}(r_-))w_-, & r_- > 2R_0 \geq r_+;\ R_1 + R_0 > r_- > R_1 - R_0 \geq r_+, \\
(A_1^{\mathrm{IIB}}(r_+) - A_2^{\mathrm{IID}}(r_+))w_+ = (A_1^{\mathrm{IIB}}(r_-) - A_2^{\mathrm{IIE}}(r_-))w_-, & r_- > 2R_0 \geq r_+;\ R_1 + R_0 > r_- > r_+ > R_1 - R_0, \\
(A_1^{\mathrm{IIA}}(r_+) - A_2^{\mathrm{IID}}(r_+))w_+ = (A_1^{\mathrm{IIC}}(r_-) - A_2^{\mathrm{IIE}}(r_-))w_-, & r_- > 2R_0 \geq r_+;\ r_- > R_1 + R_0 > R_1 - R_0 > r_+, \\
(A_1^{\mathrm{IIB}}(r_+) - A_2^{\mathrm{IID}}(r_+))w_+ = (A_1^{\mathrm{IIC}}(r_-) - A_2^{\mathrm{IIE}}(r_-))w_-, & r_- > 2R_0 \geq r_+;\ r_- > R_1 + R_0 > r_+ > R_1 - R_0, \\
(A_1^{\mathrm{IIA}}(r_+) - A_2^{\mathrm{IIE}}(r_+))w_+ = (A_1^{\mathrm{IIA}}(r_-) - A_2^{\mathrm{IIE}}(r_-))w_-, & r_- > r_+ > 2R_0;\ R_1 - R_0 > r_- > r_+, \\
(A_1^{\mathrm{IIA}}(r_+) - A_2^{\mathrm{IIE}}(r_+))w_+ = (A_1^{\mathrm{IIB}}(r_-) - A_2^{\mathrm{IIE}}(r_-))w_-, & r_- > r_+ > 2R_0;\ R_1 + R_0 > r_- > R_1 - R_0 \geq r_+, \\
(A_1^{\mathrm{IIB}}(r_+) - A_2^{\mathrm{IIE}}(r_+))w_+ = (A_1^{\mathrm{IIB}}(r_-) - A_2^{\mathrm{IIE}}(r_-))w_-, & r_- > r_+ > 2R_0;\ R_1 + R_0 > r_- > r_+ > R_1 - R_0, \\
(A_1^{\mathrm{IIA}}(r_+) - A_2^{\mathrm{IIE}}(r_+))w_+ = (A_1^{\mathrm{IIC}}(r_-) - A_2^{\mathrm{IIE}}(r_-))w_-, & r_- > r_+ > 2R_0;\ r_- > R_1 + R_0 > R_1 - R_0 > r_+, \\
(A_1^{\mathrm{IIB}}(r_+) - A_2^{\mathrm{IIE}}(r_+))w_+ = (A_1^{\mathrm{IIC}}(r_-) - A_2^{\mathrm{IIE}}(r_-))w_-, & r_- > r_+ > 2R_0;\ r_- > R_1 + R_0 > r_+ > R_1 - R_0, \\
(A_1^{\mathrm{IIC}}(r_+) - A_2^{\mathrm{IIE}}(r_+))w_+ = (A_1^{\mathrm{IIC}}(r_-) - A_2^{\mathrm{IIE}}(r_-))w_-, & r_- > r_+ > 2R_0;\ r_- > r_+ > R_1 + R_0.
\end{cases}$$

The transcendent equations have no analytical solutions, and we have to numerically solve the equations by enumerating all possible values of $R_0$ and $R_1$, for given $r_-/r_+$ and $w_+/w_-$. Different combinations of $R_0$ and $R_1$ are substituted into different equations to examine if the stability condition is satisfied. Finally, all stable combinations of $R_0$ and $R_1$ are sought out. We will show the graphical solution of the inner border in combination with that of the outer border later.

Next, we provide the stability condition of the outer border. Fig. S16 shows variables involved in our analysis, where $R_0$ and $R_1$ represent the radius of the inner and outer borders of the ring, respectively, and $r$ represents the range of individual interactions (either positive or negative). The gray area $A_4$ is the overlap between the interaction range (either positive or negative) centered at an arbitrary outer-border individual and the outer circle of the ring. The blue area $A_3$ is the overlap between the interaction range $r$ (either positive or negative) centered at an arbitrary outer-border individual and the inner circle of the ring. The difference $A_4 - A_3$ is the actual area where interactions occur. All five scenarios associated with the outer-ring stability are classified into Case III, as displayed in Table S1.

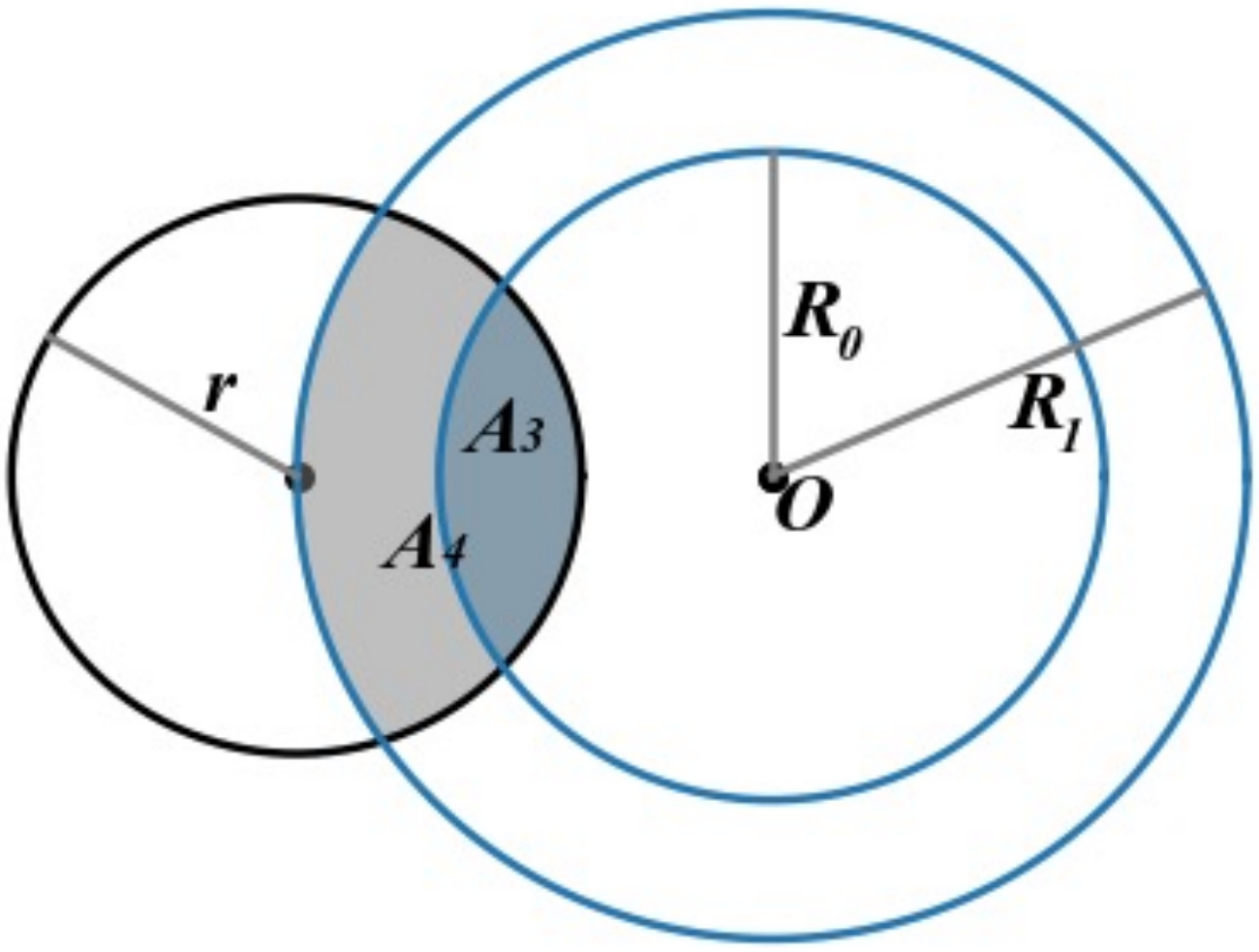

**Fig. S16. Variables involved in the analysis of outer-ring stability.**

**Case III(A):** $r \le R_1 - R_0$. As shown in Fig. S17, we have

$$A_3^{\mathrm{IIIA}}(r) = 0.$$

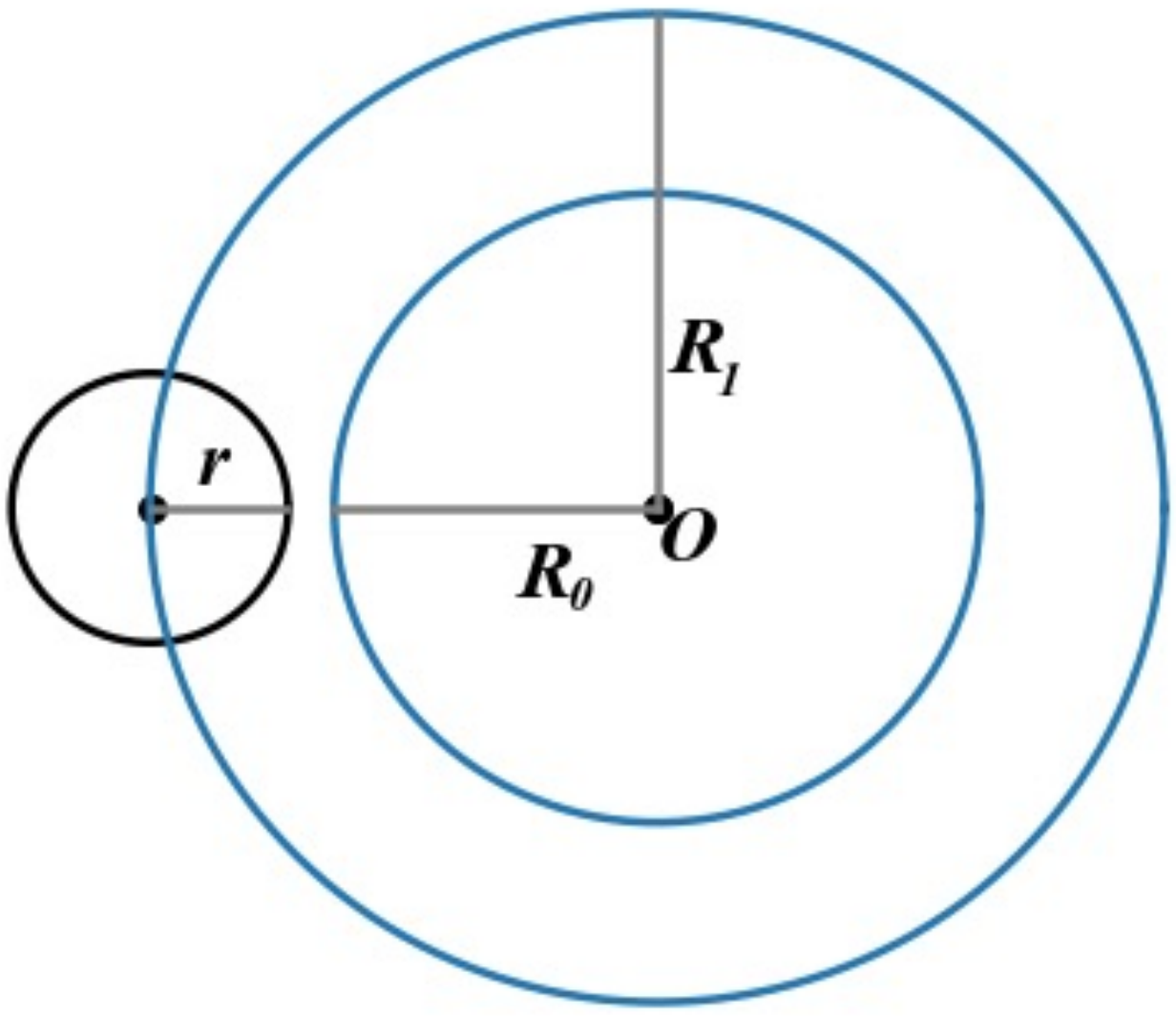

**Fig. S17: Variables involved in Case III(A).**

**Case III(B):** $R_1 - R_0 < r \le R_1 + R_0$. As shown in Fig. S18, we have

$$A_3^{\mathrm{IIIB}}(r) = R^2\theta_1 + R_0^2\theta - 2\sqrt{p(p - R_0)(p - R_1)(p - r)},$$

where

$$\theta_1 = \frac{\arccos\,(r^2 + R_1 - R_0^2)}{2R_1 r},$$

$$\theta = \frac{\arccos\,(R_0^2 + R_1^2 - r^2)}{2R_0 R_1},$$

$$p = \frac{1}{2}(R_0 + R_1 + r).$$

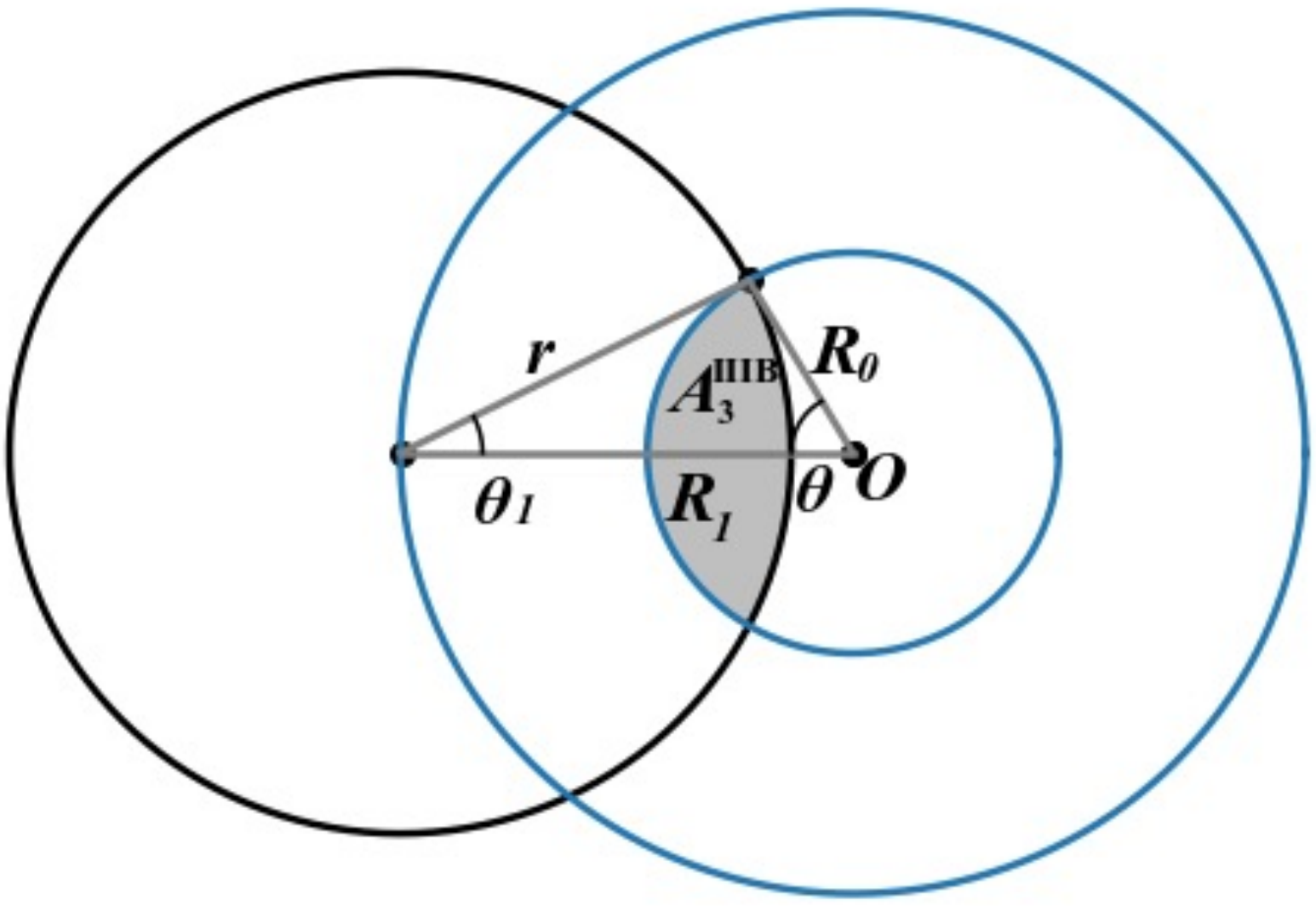

Fig. S18: Variables involved in Case III(B).

**Case III(C):** $R_1 + R_0 \leq r$. As shown in Fig. S19.

$$A_3^{\mathrm{IIIC}}(r) = \pi R_0^2.$$

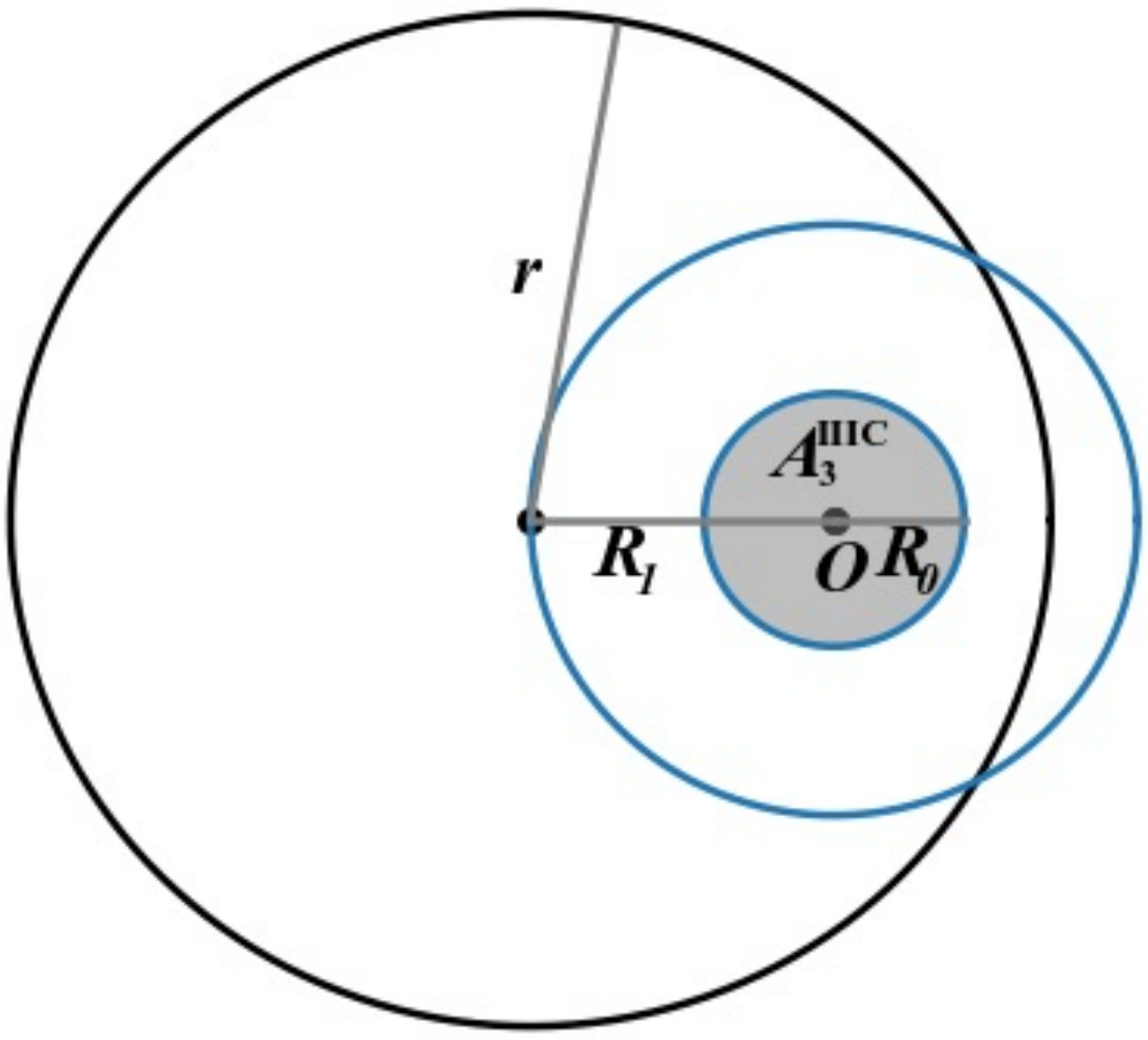

Fig. S19: Variables involved in Case III(C).

**Case III(D):** $r < 2R_1$. As shown in Fig. S20, we have

$$A_4^{\mathrm{IIID}}(r) = 2\theta R_1^2 + \left(\frac{\pi}{2} - \theta\right) r^2 - R_1 r \cos\theta,$$

where

$$\theta = \arcsin\left(\frac{r}{2R_1}\right).$$

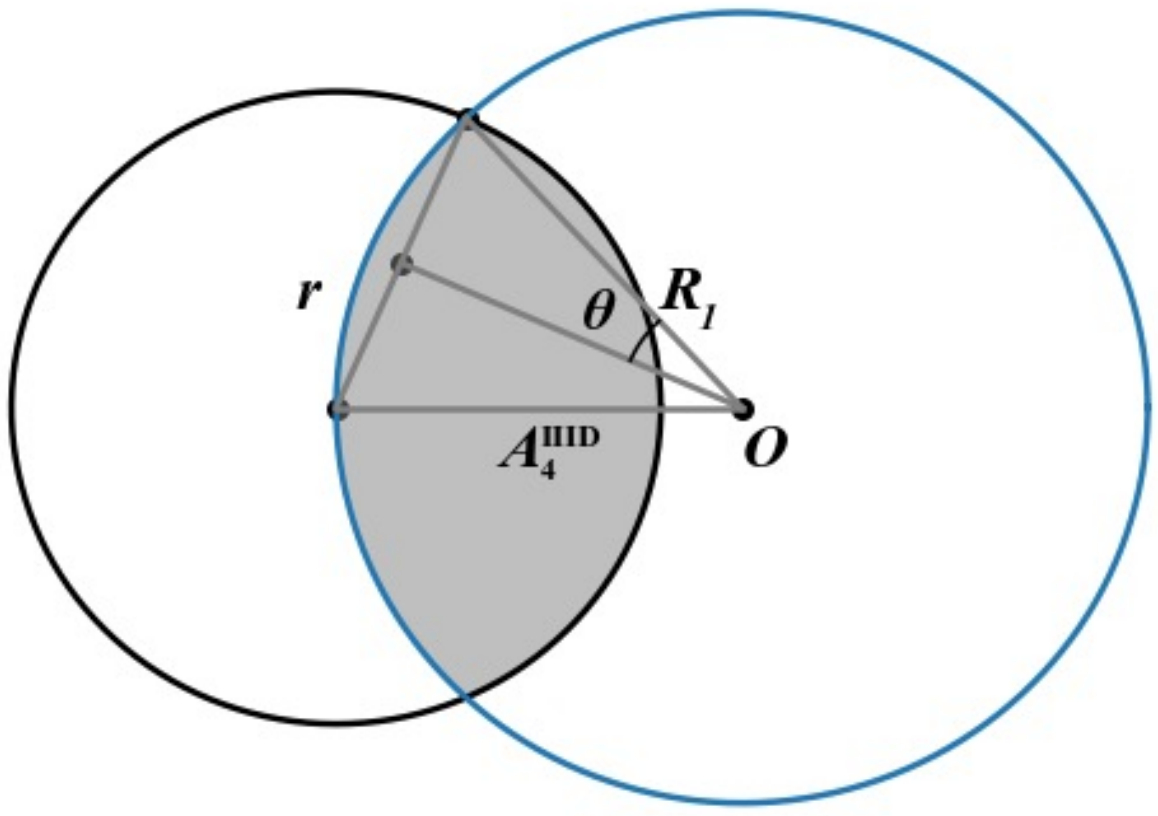

**Fig. S20: Variables involved in Case III(D).**

**Case III(E):** $r \geq 2R_1$. As shown in Fig. S21, we have

$$A_4^{\mathrm{IIIE}}(r) = \pi R_1^2.$$

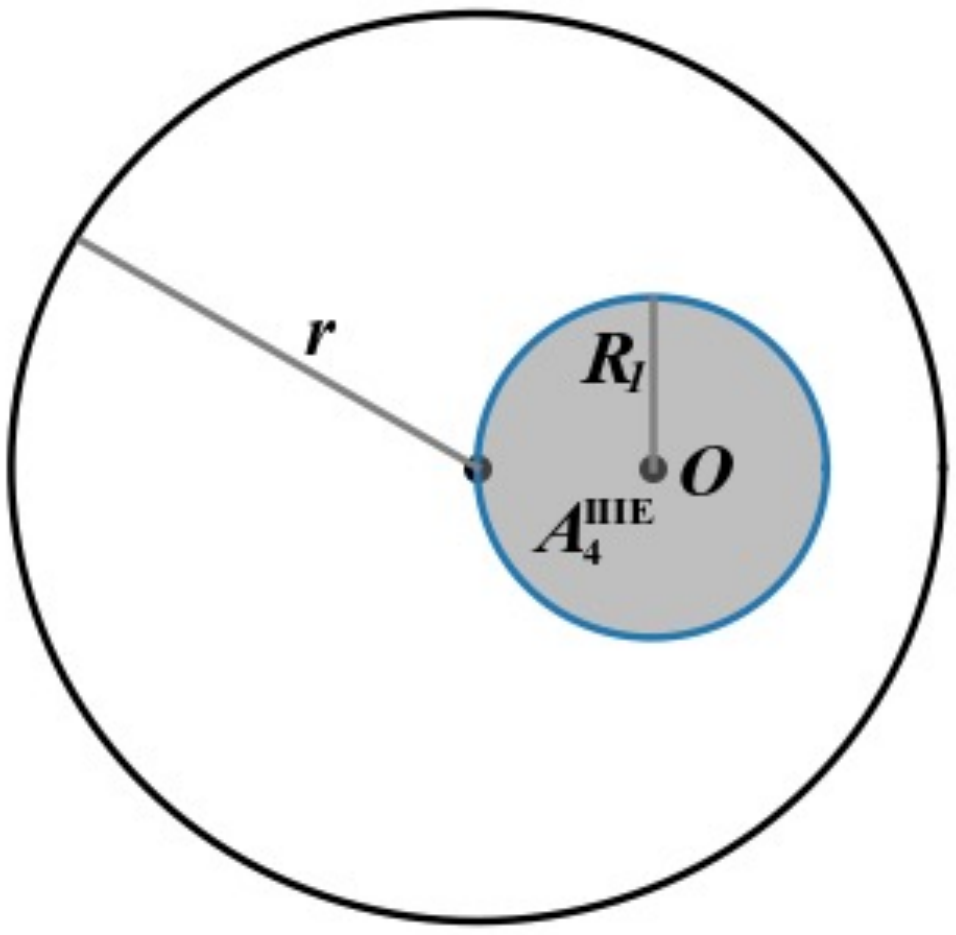

**Fig. S21: Variables involved in Case III(E).**

Next, we provide the stability condition of the outer border. Specifically, if individuals at the outer border receive equal excitatory and inhibitory forces, the outer border is stable. The stability condition is formulated as

$$\big(A_4(r_+) - A_3(r_+)\big)w_+ = \big(A_4(r_-) - A_3(r_-)\big)w_-,$$

where the left-hand side and the right-hand side are the excitatory and inhibitory force at the outer border, respectively.

From Case III(A) to Case III(E), there are ten possible combinations of force ranges centered at the outer border related with equilibrium, as follows:

$$\begin{cases}
(A_3^{\mathrm{IIIA}}(r_+) - A_4^{\mathrm{IIID}}(r_+))w_+ = (A_3^{\mathrm{IIIA}}(r_-) - A_4^{\mathrm{IIID}}(r_-))w_-, & 2R_0 \geq r_-;\ \ R_1 + R_0 \geq r_- > r_+\,, \\
(A_3^{\mathrm{IIIA}}(r_+) - A_4^{\mathrm{IIID}}(r_+))w_+ = (A_3^{\mathrm{IIIB}}(r_-) - A_4^{\mathrm{IIID}}(r_-))w_-, & 2R_0 \geq r_-;\ \ R_1 + R_0 > r_- > R_1 - R_0 \geq r_+, \\
(A_3^{\mathrm{IIIB}}(r_+) - A_4^{\mathrm{IIID}}(r_+))w_+ = (A_3^{\mathrm{IIIB}}(r_-) - A_4^{\mathrm{IIID}}(r_-))w_-, & 2R_0 \geq r_-;\ \ R_1 + R_0 > r_- > r_+ > R_1 - R_0, \\
(A_3^{\mathrm{IIIA}}(r_+) - A_4^{\mathrm{IIID}}(r_+))w_+ = (A_3^{\mathrm{IIIC}}(r_-) - A_4^{\mathrm{IIID}}(r_-))w_-, & 2R_0 \geq r_-;\ \ r_- > R_1 + R_0 > R_1 - R_0 > r_+, \\
(A_3^{\mathrm{IIIB}}(r_+) - A_4^{\mathrm{IIID}}(r_+))w_+ = (A_3^{\mathrm{IIIC}}(r_-) - A_4^{\mathrm{IIID}}(r_-))w_-, & 2R_0 \geq r_-;\ \ r_- > R_1 + R_0 > r_+ > R_1 - R_0, \\
(A_3^{\mathrm{IIIC}}(r_+) - A_4^{\mathrm{IIID}}(r_+))w_+ = (A_3^{\mathrm{IIIC}}(r_-) - A_4^{\mathrm{IIID}}(r_-))w_-, & 2R_0 \geq r_-;\ \ r_- > r_+ > R_1 + R_0, \\
(A_3^{\mathrm{IIIA}}(r_+) - A_4^{\mathrm{IIID}}(r_+))w_+ = (A_3^{\mathrm{IIIC}}(r_-) - A_4^{\mathrm{IIIE}}(r_-))w_-, & r_- > 2R_0 \geq r_+;\ \ r_- > R_1 + R_0 > R_1 - R_0 > r_+, \\
(A_3^{\mathrm{IIIB}}(r_+) - A_4^{\mathrm{IIID}}(r_+))w_+ = (A_3^{\mathrm{IIIC}}(r_-) - A_4^{\mathrm{IIIE}}(r_-))w_-, & r_- > 2R_0 \geq r_+;\ \ r_- > R_1 + R_0 > r_+ > R_1 - R_0, \\
(A_3^{\mathrm{IIIC}}(r_+) - A_4^{\mathrm{IIID}}(r_+))w_+ = (A_3^{\mathrm{IIIC}}(r_-) - A_4^{\mathrm{IIIE}}(r_-))w_-, & r_- > 2R_0 \geq r_+;\ \ r_- > r_+ > R_1 + R_0, \\
(A_3^{\mathrm{IIIC}}(r_+) - A_4^{\mathrm{IIIE}}(r_+))w_+ = (A_3^{\mathrm{IIIC}}(r_-) - A_4^{\mathrm{IIIE}}(r_-))w_-, & r_- > r_+ > 2R_0;\ \ r_- > r_+ > R_1 + R_0.
\end{cases}$$

The transcendent equations have no analytical solutions, and we have to numerically solve the equations by enumerating all possible values of $R_0$ and $R_1$, for any given $r_-/r_+$ and $w_+/w_-$. Different combinations of $R_0$ and $R_1$ are substituted into relevant equations to examine if the stability condition is satisfied. Finally, all stable combinations of $R_0$ and $R_1$ are sought out.

Fig. S22 shows an example of the graphical solution of the stable inner- and outer-ring conditions for $w_+/w_- = 2.1$ and $r_-/r_+ = 1.5$. Likewise, we obtain graphical solutions of stability conditions in the critical region $(w_+/w_-)_{\text{ring}} < w_+/w_- < (w_+/w_-)_\infty$.

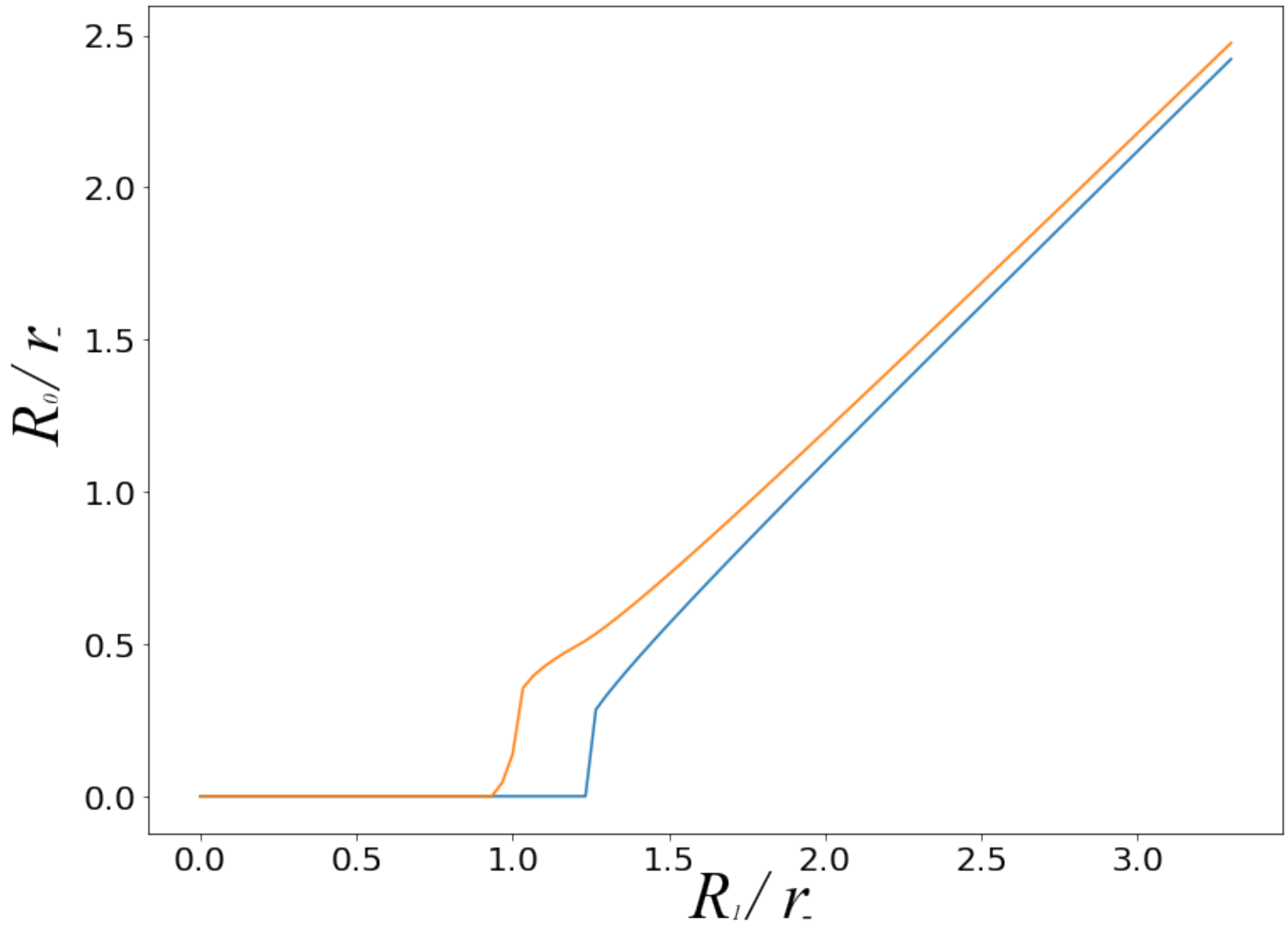

**Fig. S22: Numerical solution of the radius of rings under a specific strength ratio.** The yellow curve represents the values of $R_0/r_-$ and $R_1/r_-$ that satisfy the stability condition of the inner border for $w_+/w_- = 2.1$, and the blue curve represents the values of $R_0/r_-$ and $R_1/r_-$ that satisfy stability condition of the outer border.

Finally, the stability of a ring is determined by both inner- and outer-ring stabilities simultaneously. In particular, if both the inner and outer rings are stable simultaneously, the ring is stable; otherwise, that either inner- or outer-ring is unstable indicates the ring is unstable. In other words, the ring is stable if and only if the graphical solutions of inner and out ring have interactions. We compute differences between the two graphical solutions, given $r_{-}/r_{+} = 1.5$ and $w_{+}/w_{-} = 2.1$. As shown in Fig. S23, no intersection arises, demonstrating that the ring is unstable for the given parameter values.

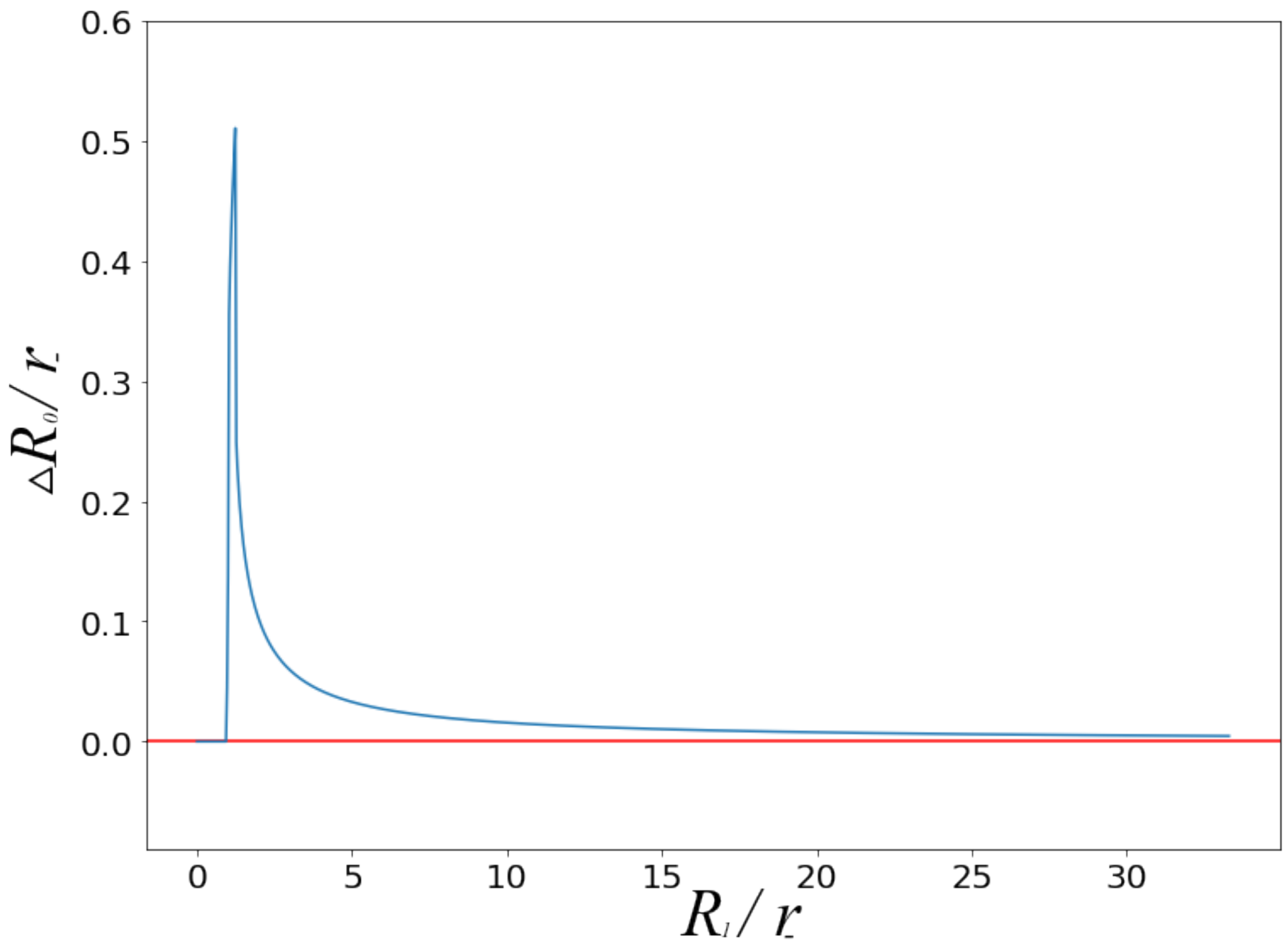

**Fig. S23: The differences between the numerical solutions of the inner- and outer-ring stability given $w_{+}/w_{-} = 2.1$.** The blue curve is the differences between the two graphical solutions in Fig. S22.

Subsequently, we enumerate all possible values of $w_+/w_-$ and $r_-/r_+$ to examine relevant ring stability. As shown in Fig. S24, given $r_-/r_+ = 1.5$, the absence of an intersection curve indicates rings are unstable. In the same vein, for any combinations of $w_+/w_-$ and $r_-/r_+$, our analyses demonstrate that rings are always unstable, regardless of interaction parameters.

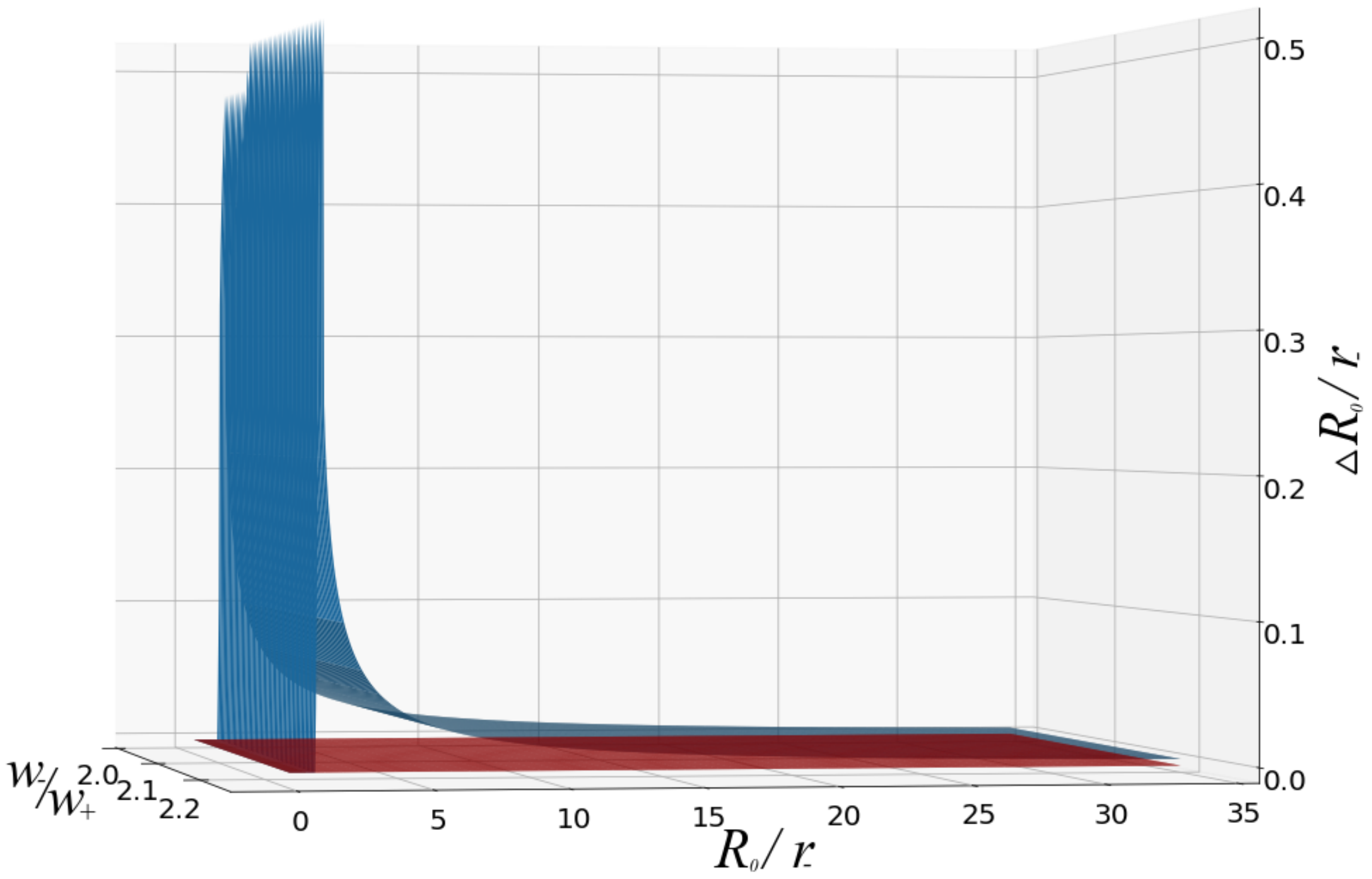

**Fig. S24: The differences between the numerical solutions of the inner- and outer-ring stability for different values of $w_+/w_-$.** The blue surface represents the differences between the two graphical solutions of inner- and outer-ring stability for a range of $w_+/w_-$.

**Table S1: Area under different conditions.**

| Case | Subcase | Condition | Area | Figure |
|---|---|---|---|---|
| I: At spot border | A | $r < 2R_0$ | $A^{\mathrm{IA}}(r) = 2\theta R_0^2 + \left(\frac{\pi}{2} - \theta\right) r^2 - R_0 r \cos\theta,$ $\theta = \arcsin\left(\frac{r}{2R_0}\right).$ | Fig. S5 |
| | B | $r \geq 2R_0$ | $A^{\mathrm{IB}}(r) = \pi R_0^2.$ | Fig. S6 |
| II: At inner border | A | $r \leq R_1 - R_0$ | $A_1^{\mathrm{IIA}}(r) = \pi r^2.$ | Fig. S13 |
| | B | $R_1 - R_0 < r \leq R_1 + R_0$ | $A_1^{\mathrm{IIB}}(r) = r^2\theta_1 + R_1^2\theta - 2\sqrt{p(p - R_0)(p - R_1)(p - r)},$ $\theta_1 = \frac{\arccos\,(r^2 + R_0 - R_1^2)}{2R_0 r},$ $\theta = \frac{\arccos\,(R_0^2 + R_1^2 - r^2)}{2R_0 R_1},$ $p = \frac{1}{2}(R_0 + R_1 + r).$ | Fig. S14 |
| | C | $R_1 + R_0 \leq r$ | $A_1^{\mathrm{IIC}}(r) = \pi R_1^2.$ | Fig. S15 |
| | D | $r < 2R_0$ | Identical to Case I-A | Identical to Case I(A) |
| | E | $r \geq 2R_0$ | Identical to Case I-B | Identical to Case I(B) |
| III: At outer border | A | $r \leq R_1 - R_0$ | $A_3^{\mathrm{IIIA}}(r) = 0.$ | Fig. S17 |
| | B | $R_1 - R_0 < r \leq R_1 + R_0$ | $A_3^{\mathrm{IIIB}}(R) = R^2\theta_1 + R_0^2\theta - 2\sqrt{p(p-R_0)(p-R_1)(p-r)}$, $\theta_1 = \frac{\arccos(r^2 + R_1 - R_0^2)}{2R_1 r}$, $\theta = \frac{\arccos(R_0^2 + R_1^2 - r^2)}{2R_0 R_1}$, $p = \frac{1}{2}(R_0 + R_1 + r)$. | Fig. S18 |
| | C | $R_1 + R_0 \leq r$ | $A_3^{\mathrm{IIIC}}(r) = \pi R_0^2$. | Fig. S19 |
| | D | $r < 2R_0$ | $A_4^{\mathrm{IIID}}(r) = 2\theta R_1^2 + \left(\frac{\pi}{2} - \theta\right) r^2 - R_1 r \cos\theta$, $\theta = \arcsin\left(\frac{r}{2R_1}\right)$. | Fig. S20 |
| | E | $r \geq 2R_0$ | $A_4^{\mathrm{IIIE}}(r) = \pi R_1^2$. | Fig. S21 |

# Theory of finite-size space

Let us derive the size of unstable rings in a finite-size circle space with radius $L$. Because rings are unstable, the outer border of a ring reaches the boundary of the finite circle space, i.e., $R_1 = L$. We insert $R_1$ into the stablilty condition of inner ring, yielding the radius of inner ring $R_0$. A graphical solution of $R_0$ is shown in Fig. S25.

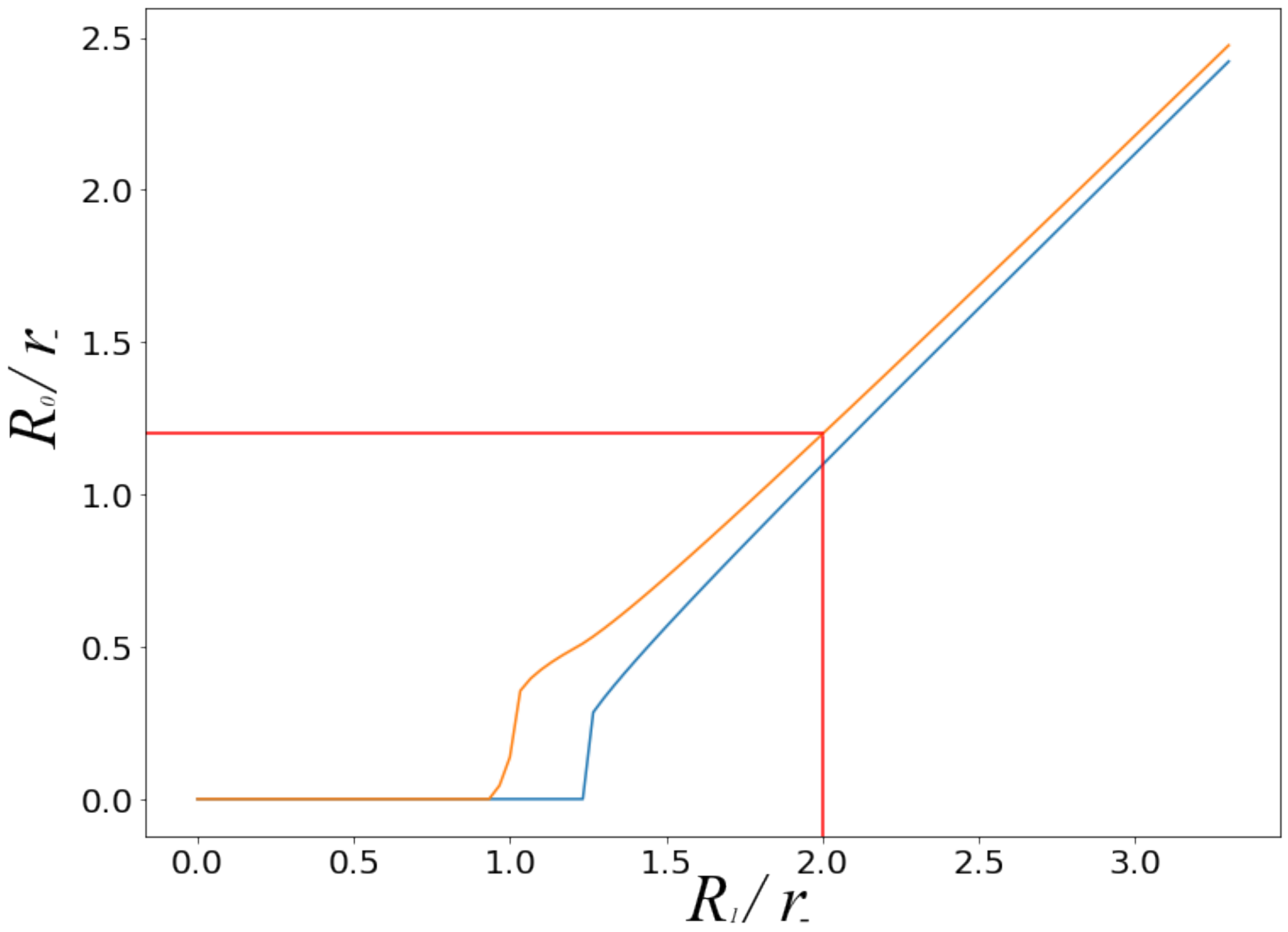

**Fig. S25: The graphical solution of the inner-ring radius in a finite-scale space.** The yellow and blue curve represents the stability condition of the inner border and the outer border, respectively. The vertical red line is outer-ring radius $R_1$ as determined by the space scale, and its intersection with the yellow curve gives the numerical solution of $R_0$, the horizontal red line.

A comparison between the theoretical results via numerical solutions and simulation results for different values of $w_+/w_-$ is shown in Fig. 10(b) in the main text. In Fig. S26, we make a comparison with respect to different values of $r_-/r_+$. The numerical solutions of theory are in good agreement with the simulation results, as shown in Fig. S26.

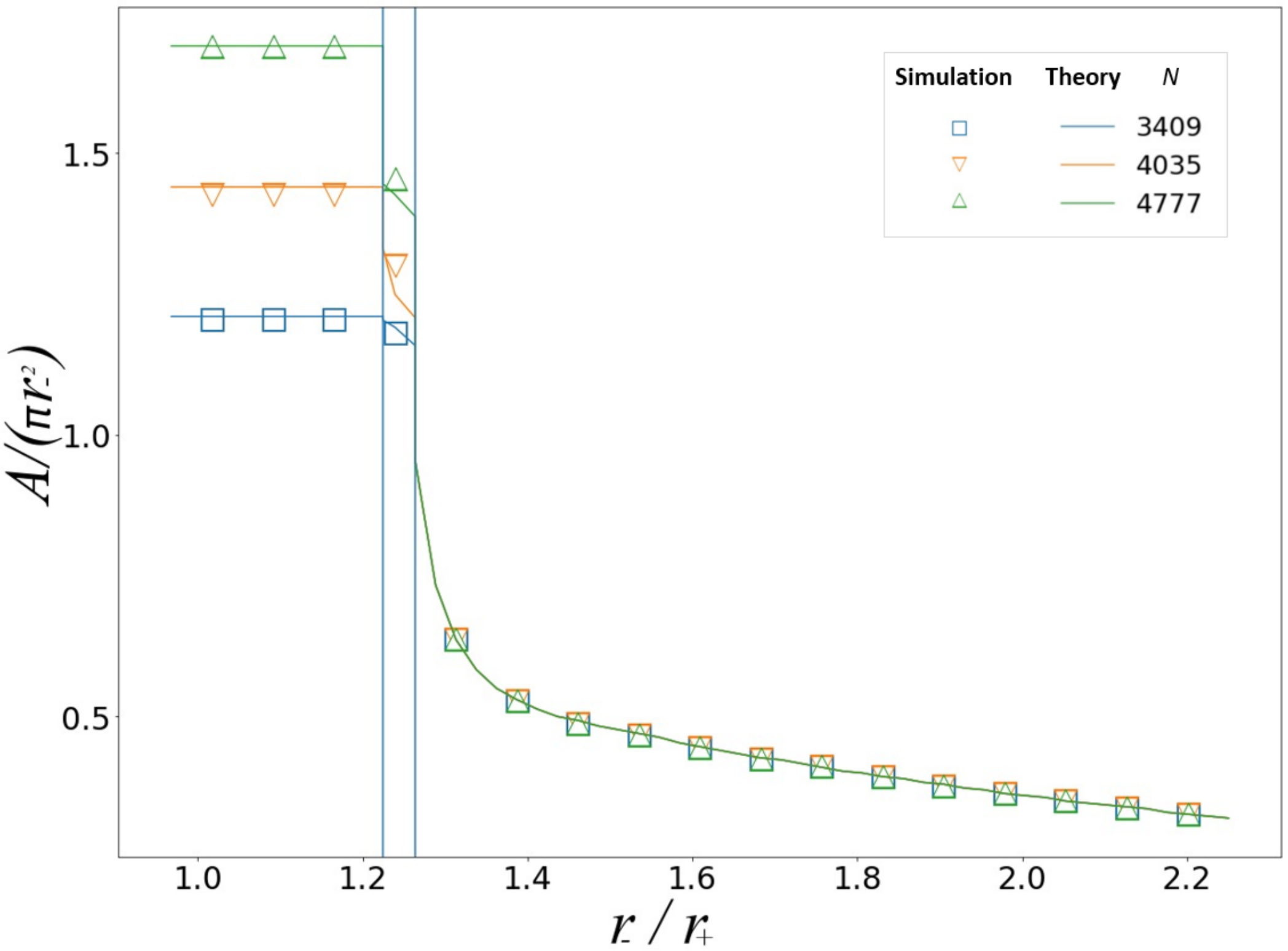

**Fig. S26: phase transitions of Turing patterns.** The area $A$ of a single spot normalized by the area $\pi r_-^2$ of inhibitory range as a function of the radius ratio $r_-/r_+$ initiated from a single excited individual. We obtain each simulation result by 1000 independent realizations. The vertical axis represents the relative value of spot area $A/(\pi r_-^2)$.

# Analyzing chaos in dynamic patterns

## Onset of chaos in one-dimension physical space

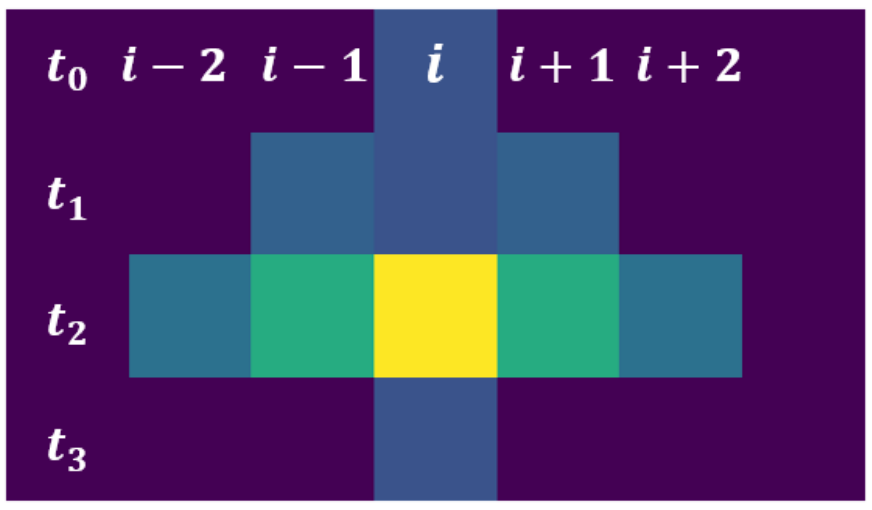

| | | | | |
|---|---|---|---|---|
| $0$ | $0$ | $x$ | $0$ | $0$ |
| $0$ | $w_+x$ | $x$ | $w_+x$ | $0$ |
| $w_+^2x$ | $2w_+x$ | $x+w_+^2x$ | $2w_+x$ | $w_+^2x$ |
| $0$ | $0$ | $x$ | $0$ | $0$ |

**Fig. S27: Four steps of pattern evolution from a single excited individual in one-dimension physical space.** Five individuals near individual $i$ are denoted as $i-2$ to $i+2$, and four steps from $t_0$ to $t_3$ are involved in our analysis. Different colors denote states of individuals and the initial state of individual $i$ is assumed to be $s_i(t_0)=x$. States of individuals at different steps are calculated and displayed in the matrix on the right-hand side.

We derive the onset of chaos from a silent phase, as illustrated in Fig. 11(a) in the main text. Actually, the onset $w_+^o$ depends on the strength $w_-$ of negative force. Note that insofar as $w_+$ exceeds the onset point, there arise chaotic patterns from an entirely silent space. Thus, the onset of chaos equals the onset of pattern spreading, and we derive the latter as an alternative to simplify our analyses. Suppose that the evolution initiates from a single excited individual $i$ with its state $s_i(t_0)=x$. In order to trigger pattern spreading, $i$' s state ought to increase or at least keep the same after some steps. Based on simulation results, we find that the states at the third step plays a key role in determining the fate of a pattern. In particular, as shown in Fig. S27, if $s_i(t_3)$ is less than $s_i(t_0)$, the initial pattern will eventually disappear some steps later; otherwise, a spreading pattern emerges. We write down iterative equations from step zero to step three around the location of $i$, and compare $s_i(t_3)$ with $s_i(t_0)$ to formulate the onset of chaos and pattern spreading. A few individuals in the vicinity of $i$ are involved (see Fig. S27), and the iteration equations are as follows:

$$
\begin{cases}
s_{i-2}(t_0) = 0, \\
s_{i-1}(t_0) = 0, \\
s_i(t_0) = x, \\
s_{i+1}(t_0) = 0, \\
s_{i+2}(t_0) = 0, \\
s_{i-1}(t_1) = s_{i-1}(t_0) + w_+ s_{i-2}(t_0) + w_+ s_i(t_0), \\
s_i(t_1) = s_i(t_0) + w_+ s_{i-1}(t_0) + w_+ s_{i+1}(t_0), \\
s_{i+1}(t_1) = s_{i+1}(t_0) + w_+ s_i(t_0) + w_+ s_{i+2}(t_0), \\
s_{i-2}(t_1) = 0, \\
s_{i+2}(t_1) = 0, \\
s_{i-1}(t_2) = s_{i-1}(t_1) + w_+ s_{i-2}(t_1) + w_+ s_i(t_1), \\
s_i(t_2) = s_i(t_1) + w_+ s_{i-1}(t_1) + w_+ s_{i+1}(t_1), \\
s_{i+1}(t_2) = s_{i+1}(t_1) + w_+ s_i(t_1) + w_+ s_{i+2}(t_1), \\
s_i(t_3) = s_i(t_2) - w_- s_{i-1}(t_2) - s_{i+1}(t_2).
\end{cases}
$$

From the equations, we obtain some critical intermediate variables, as follows.

$$
\begin{gathered}
s_{i-1}(t_1) = w_+ x, \\
s_i(t_1) = x, \\
s_{i+1}(t_1) = w_+ x, \\
s_{i-1}(t_2) = 2w_+ x, \\
s_i(t_2) = x + w_+^2 x, \\
s_{i+1}(t_2) = 2w_+ x, \\
s_i(t_3) = x + w_+^2 x - 2w_+ w_- x.
\end{gathered}
$$

The onset of pattern spreading (chaos) requires that $s_i(t_3)$ equals $s_i(t_0)$, i.e.,

$$
s_i(t_3) = x + w_+^2 x - 2w_+ w_- x = x = s_i(t_0),
$$

which finally yields the onset of chaos

$$
w_+^o = 2w_-.
$$

The analytical result of the onset of chaos is in good agreement with simulation results, as shown in Fig. 11(b) in the main text.

# Phase transition between chaos and fractals in one-dimension physical space

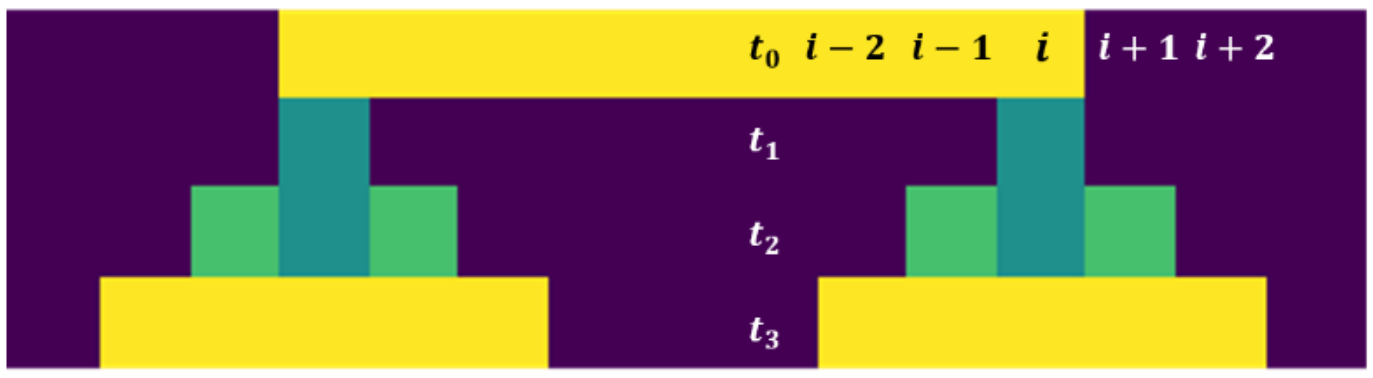

| | | | | |
|---|---|---|---|---|
| 1 | 1 | 1 | 0 | 0 |
| 0 | 0 | $1-w_-$ | 0 | 0 |
| 0 | $(1-w_-)w_+$ | $1-w_-$ | $(1-w_-)w_+$ | 0 |
| 1 | 1 | 1 | 1 | 1 |

**Fig. S28: Four steps of pattern evolution initiated from a single excited individual in one-dimension physical space.** Five individuals in the vicinity of $i$ are denoted as $i-2$ to $i+2$, and four steps from $t_0$ to $t_3$ are involved in our analysis. Different colors denote states of individuals and the initial state of individual $i$ is $s_i(t_0)=1$. States of individuals at different steps are displayed in the matrix on the right-hand side.

According to Fig. 11(a) in the main text, when $w_+$ exceeds a second phase transition point, the chaotic phase turns to a fractal phase composed of four different states of individuals. Here we calculate the second phase transition between chaos and fractals. Analogous to the onset of chaos, the early evolution of a small number of individuals determines the fate of evolution. In particular, whether states of individuals repeat after some steps is the key to predicting the phase of patterns. If states return after some steps, the system is in a fractal phase with self-similarity; otherwise, the absence of repetition indicates the system enters a chaotic phase. We find that the process for examining evolutionary outcomes involves five individuals near individual $i$, as shown in Fig. S28, and four steps from step zero to step three are sufficient to determine the evolution fate. The relevant iterative equations are as follows.

$$\begin{cases} s_{i-2}(t_0)=1, \\ s_{i-1}(t_0)=1, \\ s_i(t_0)=1, \\ s_{i+1}(t_0)=0, \\ s_{i+2}(t_0)=0, \\ s_{i-1}(t_1)=s_{i-1}(t_0)-w_-s_{i-2}(t_0)-w_-s_i(t_0), \\ s_i(t_1)=s_i(t_0)-w_-s_{i-1}(t_0)-w_-s_{i+1}(t_0), \\ s_{i+1}(t_1)=s_{i+1}(t_0)-w_-s_i(t_0)-w_-s_{i+2}(t_0), \\ s_{i+2}(t_1)=0, \\ s_i(t_2)=s_i(t_1)+w_+s_{i-1}(t_1)+w_+s_{i+1}(t_1), \\ s_{i+1}(t_2)=s_{i+1}(t_1)+w_+s_i(t_1)+w_+s_{i+2}(t_1), \\ s_{i+2}(t_2)=s_{i+2}(t_1)+w_+s_{i+1}(t_1)+w_+s_{i+3}(t_1), \\ s_{i+3}(t_2)=0, \\ s_{i+2}(t_3)=s_{i+2}(t_2)+w_+s_{i+1}(t_2)+w_+s_{i+3}(t_2). \end{cases}$$

We obtain some important intermediate variables, as follows.

$$
\begin{gathered}
s_{i-1}(t_1) = s_{i+1}(t_1) = s_{i+2}(t_1) = 0, \\
s_i(t_1) = 1 - w_-, \\
s_i(t_2) = 1 - w_-, \\
s_{i+1}(t_2) = (1 - w_-)w_+, \\
s_{i+2}(t_2) = 0, \\
s_{i+2}(t_3) = (1 - w_-)w_+^2 = 1.
\end{gathered}
$$

The condition of maintaining fractal properties is that $s_{i+2}(t_3)$ repeats the initial state $s_i(t_0)$, i.e.,

$$
s_{i+2}(t_3) = (1 - w_-)w_+^2 = s_i(t_0) = 1,
$$

which yields the phase transition point between chaos and fractals as

$$
w_+^c = \sqrt{\frac{1}{1 - w_-}}.
$$

The analytical result of the phase transition between chaos and fractals is in good agreement with simulation results, as shown in Fig. 11(b) in the main text.

# Onset of chaos in two-dimension physical space

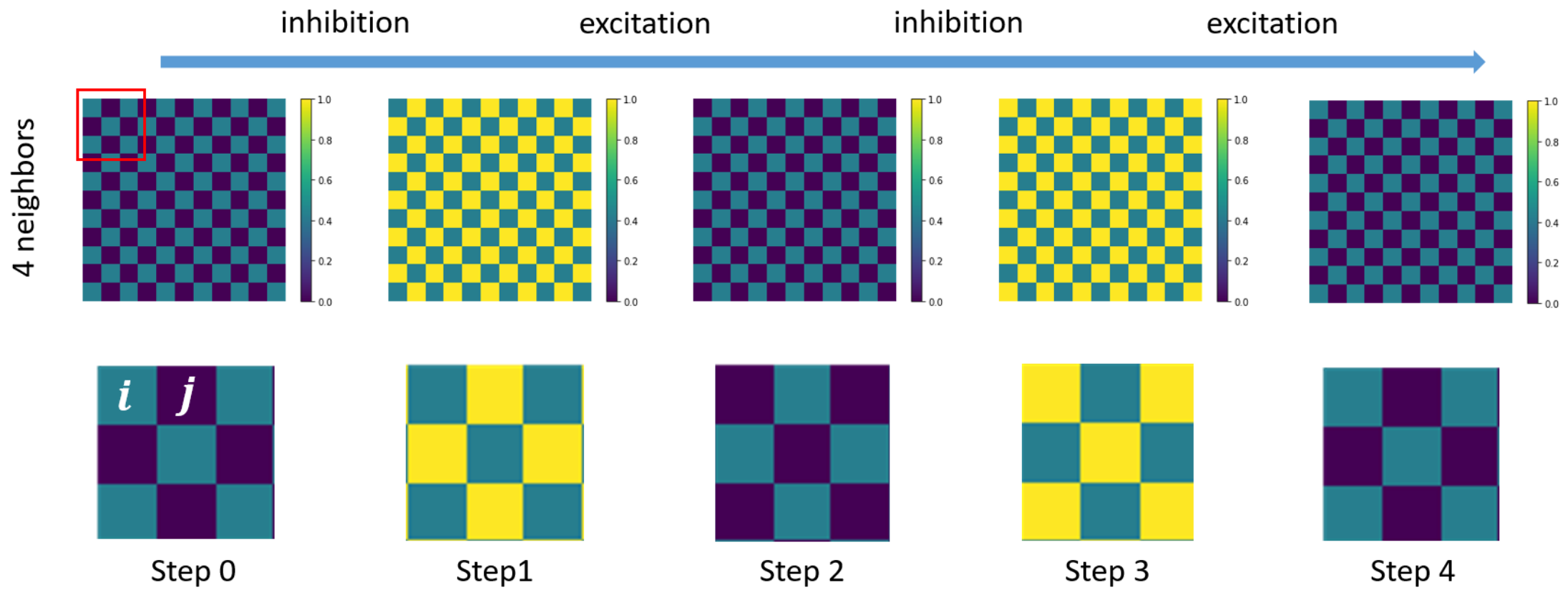

Step 0:

| | | |
|---|---|---|
| $x$ | $0$ | $x$ |
| $0$ | $x$ | $0$ |
| $x$ | $0$ | $x$ |

Step 1:

| | | |
|---|---|---|
| $x$ | $4w_+x$ | $x$ |
| $4w_+x$ | $x$ | $4w_+x$ |
| $x$ | $4w_+x$ | $x$ |

Step 2:

| | | |
|---|---|---|
| $0$ | $x$ | $0$ |
| $x$ | $0$ | $x$ |
| $0$ | $x$ | $0$ |

Step 3:

| | | |
|---|---|---|
| $4w_+x$ | $x$ | $4w_+x$ |
| $x$ | $4w_+x$ | $x$ |
| $4w_+x$ | $x$ | $4w_+x$ |

Step 4:

| | | |
|---|---|---|
| $x$ | $0$ | $x$ |
| $0$ | $x$ | $0$ |
| $x$ | $0$ | $x$ |

**Fig. S29: Five steps of pattern evolution initiated from a single excited individual in two-dimension physical space.** Two locations denoted as $i$ and $j$, and five steps denoted as $t_0$ to $t_5$, are involved in our analysis. Different colors denote states of individuals. States of the individuals in the focused area at different steps are displayed in the matrices at the bottom row.

As shown in Fig. 11(c) and (e) in the main text, there exist onsets of chaos in two-dimension physical space from a silent space. Similar to the approach applied to one-dimension physical space, we provide analytical results in two-dimension space. The onset of chaos is as well identical to the onset of pattern spreading. There are two scenarios, four- and eight-neighbor spaces, corresponding to two different evolutionary processes. For the four-neighbor scenario, repetitions of individual states occur every five steps if patterns successfully spread. And from step zero to step two, we are able to anticipate the fate of a pattern. As shown in Fig. S29, nine individuals near $i$ are involved, and we assume the initial state of $i$ to be $s_i(t_0) = x$. The relevant iterative equations are as follows:

$$\begin{cases} s_i(t_0) = x, \\ s_j(t_0) = 0, \\ s_j(t_1) = s_j(t_0) + 4w_+ s_i(t_0), \\ s_i(t_1) = s_i(t_0) + 4w_+ s_j(t_0), \\ s_j(t_2) = s_j(t_1) - 4w_- s_i(t_1). \end{cases}$$

We thus have

$$s_i(t_1) = x,$$
$$s_j(t_1) = 4w_+x.$$

The key requirement of a successful spreading is

$$s_j(t_2) = 4w_+x - 4w_-x = x = s_i(t_0),$$

which gives the onset of chaos for four-neighbor spaces as

$$w_+^o = w_- + \frac{1}{4}.$$

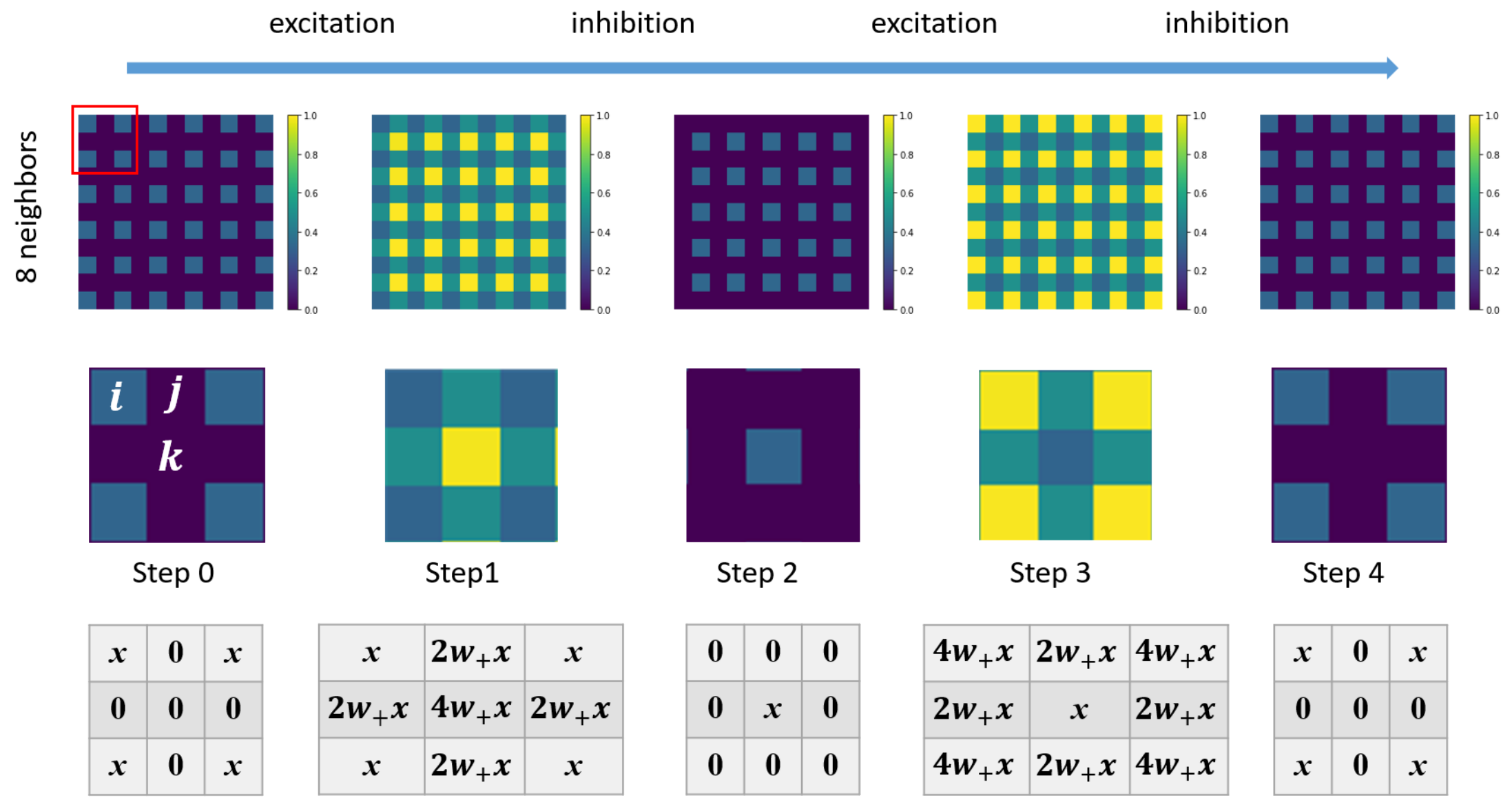

**Fig. S30: Five steps of pattern evolution initiated from a single excited individual in two-dimension physical space.** Three locations denoted as $i$, $j$ and $k$, and five steps denoted as from $t_0$ to $t_5$, are involved in evolution. Different colors denote states of individuals. States of individuals in the focused area at different steps are displayed in the matrices at the bottom row.

As shown in Fig. S30, for eight neighbors, the iterative equations are

$$\begin{cases} s_i(t_0) = x, \\ s_j(t_0) = 0, \\ s_k(t_0) = 0, \\ s_j(t_1) = s_j(t_0) + 2w_+ s_i(t_0) + 2w_+ s_k(t_0) + 4w_+ s_j(t_0), \\ s_i(t_1) = s_i(t_0) + 4w_+ s_j(t_0) + 4w_+ s_k(t_0), \\ s_k(t_1) = s_k(t_0) + 4w_+ s_i(t_0) + 4w_+ s_j(t_0), \\ s_k(t_2) = s_k(t_1) - 4w_- s_i(t_1) - 4w_- s_j(t_1). \end{cases}$$

From the equations, we obtain some crucial intermediate variables, as

$$s_i(t_1) = x,$$
$$s_j(t_1) = 2w_+ x,$$
$$s_k(t_1) = 4w_+ x.$$

The condition of the onset of chaos and spreading is

$$s_k(t_2) = 4w_+ x - 8w_+ w_- x - 4w_- x = x = s_i(t_0),$$

which yields the onset

$$w_+^o = \frac{1 + 4w_-}{4 - 8w_-}.$$

The analytical results of the onset of chaos in both the four- and eight-neighbor physical space are in good agreement with simulation results, as shown in Fig. 11(d) and (f) in the main text.

# Analyzing primary features of traveling waves

## Plane-wave velocity

Plane-wave velocity is measured by the distance that wavefront travels in a unit time. Predicting the next-step wavefront is key to calculating the distance and velocity. The location of next-step wavefront is determined by the joint force from the current wavefront. Beyond the next-step wavefront, the current joint force decreases abruptly, recovering a silent space. Insofar as we identify the location with the sharp transition of joint forces, we find the next-step wavefront. The task becomes how to analyze the transition of joint force based on interaction range and states of individuals.

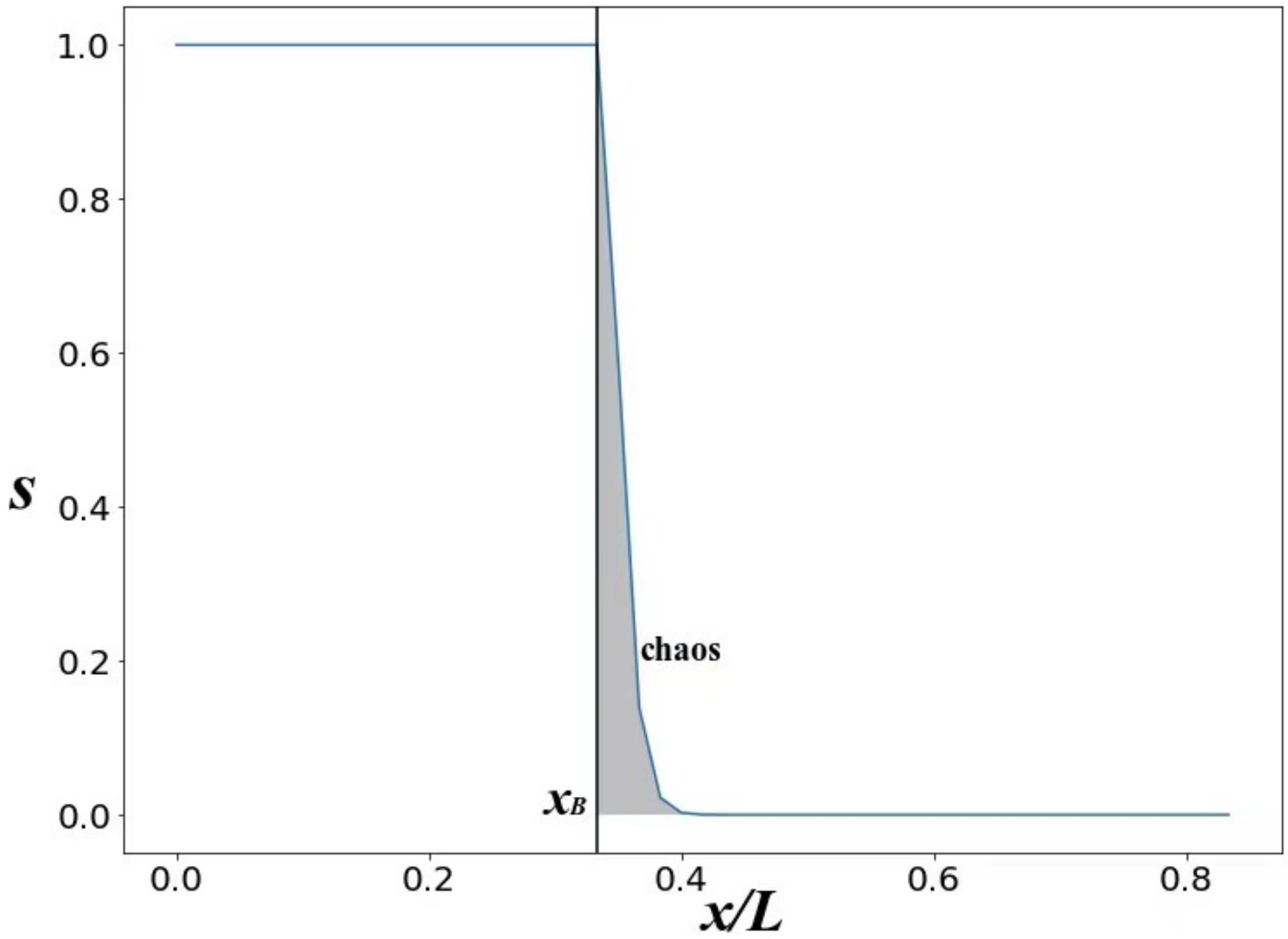

**Fig. S31: Average spatial distribution of the state.** Interaction ranges $r_+ = r_- = 5$. $w_+ = w_- = 0.2$. The space scale $L = 60$.

Fig. S31 shows the states of individuals close to the current wavefront. Without loss of generality, we focus on type-B individuals and see their wavefront. Indeed, there is an abrupt transition of states at $x_B$ from the upper bound to the lower bound at the wavefront. The transition stems from the ceiling effect of bounded state. Moreover, we find that the gray range on the right-hand side of $x_B$ is chaotic (as exemplified in Fig. 11(d) in the main text). Fortunately, the effect of the chaotic region on the next-step wavefront is negligible compared to the vast majority of other individuals whose states reach the upper bound. Therefore, we omit the chaotic gray region to calculate the forces from individuals near the idealized wavefront, in order to identify the next-step wavefront.

To be concrete, we denote individual state at location $x$ at step $t$ as $s(t,x)$. The individual states near the simplified wavefront can be described as a step function:

$$s(t,x) = \begin{cases} 1, & x \leq x_B, \\ 0, & x > x_B. \end{cases}$$

This approximation allows us to analyze the transition of forces at the next-step wavefront. Let us recall the definition of joint force, i.e.,

$$F_i(t) = \sum_{|\boldsymbol{r}_j - \boldsymbol{r}_i| < r_+} s_j(t)\, w_+ \;\; - \sum_{|\boldsymbol{r}_k - \boldsymbol{r}_i| < r_-} s_k(t)\, w_-.$$

According to the definition, we need to identify all individuals who contribute positive and negative forces to the next-step wavefront. For simplicity, we assume that the width of plane waves is significantly large, such that the inhibition from another type of individuals to the next-step wavefront is negligible. In general, all individuals who exert excitatory forces to location $x_B + d$ belong to the gray area in Fig. S32, where the black circle is the range of excitatory force and only the individuals within the gray area exert forces to location $x_B + d$. The states of the individuals in this area reach the upper bound. Thus, the joint positive force, defined as the sum of the product of individual states and strength, equals the product of the number of individuals and strength $w_+$ in the gray area. And the number of individuals, according to the discrete-space arrangement, is just the size of the gray area. Taken together, the problem of calculating joint force at a location thus transfers to solving area size. The gray region depends on the distance of the current wavefront $x_B$ from the location $x_B + d$. Area $A$ of the gray region can be formulated with respect to $d$ as (see Fig. S32)

$$A(d) = r_+\theta - d\sqrt{r_+^2 - d^2},$$

where

$$\theta = \arccos\left(\frac{d}{r_+}\right).$$

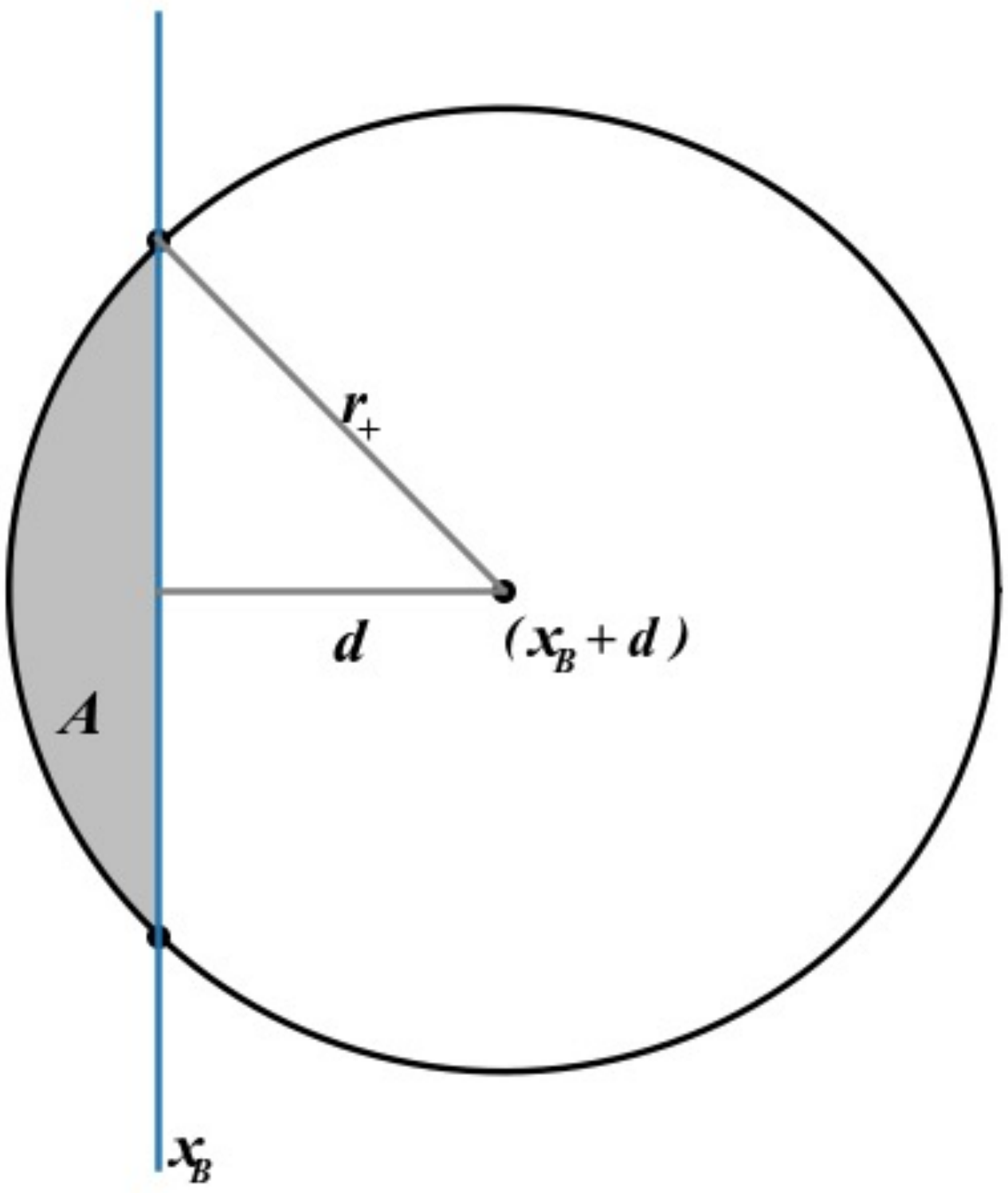

**Fig. S32: Illustration of interaction area and relevant variables for a plane wave.**

Because the joint force $F$ equals the product of the number of individuals and strength $w_+$ in the gray area, we have

$$F(t, x_B + d) = A(d_c)w_+.$$

Next, we calculate the state value of the individuals at position $x_B + d$ at the next time $t + 1$, based on the definition of state, i.e.,

$$s(t + 1, x_B + d) = s(t, x_B + d) + f[F(t, x_B + d)],$$

where $s(t, x_B + d)$ is $0$ at time $t$ (see Fig. S31). Thus, we have

$$s(t+1, x_B+d) = f(A(d)w_+),$$

where $f(A(d)w_+)$ is independent of $t$. Due to the sharp transition of states at wavefront, we can exploit the formula of $s(t+1, x_B+d)$ to identify the location of the next-step wavefront. Specifically, at the next-step wavefront, we have

$$\begin{cases} s(t+1, x_B+d_c) = 1, \\ s(t+1, x) < 1, \quad \text{for} \ \ x_B+d_c < x, \end{cases}$$

where $x_B+d_c$ is the location of the next-step wavefront, and $d_c$ is the moving distance of the wavefront from step $t$ to $t+1$. These equations yield

$$f(A(d_c)w_+) = 1.$$

Moreover, note that the upper bound of function $f$ has no ceiling effect at any wavefront, we thus have

$$f(A(d_c)w_+) = A(d_c)w_+ = 1,$$

which yields the moving distance $d_c$, and the velocity of plane wave

$$v = d_c.$$

We can numerically solve the velocity from the transcendent equation

$$A(d_c)w_+ = \left(r_+ \cdot \arccos\left(\frac{d_c}{r_+}\right) - d\sqrt{r_+^2 - d_c^2}\right)w_+ = 1.$$

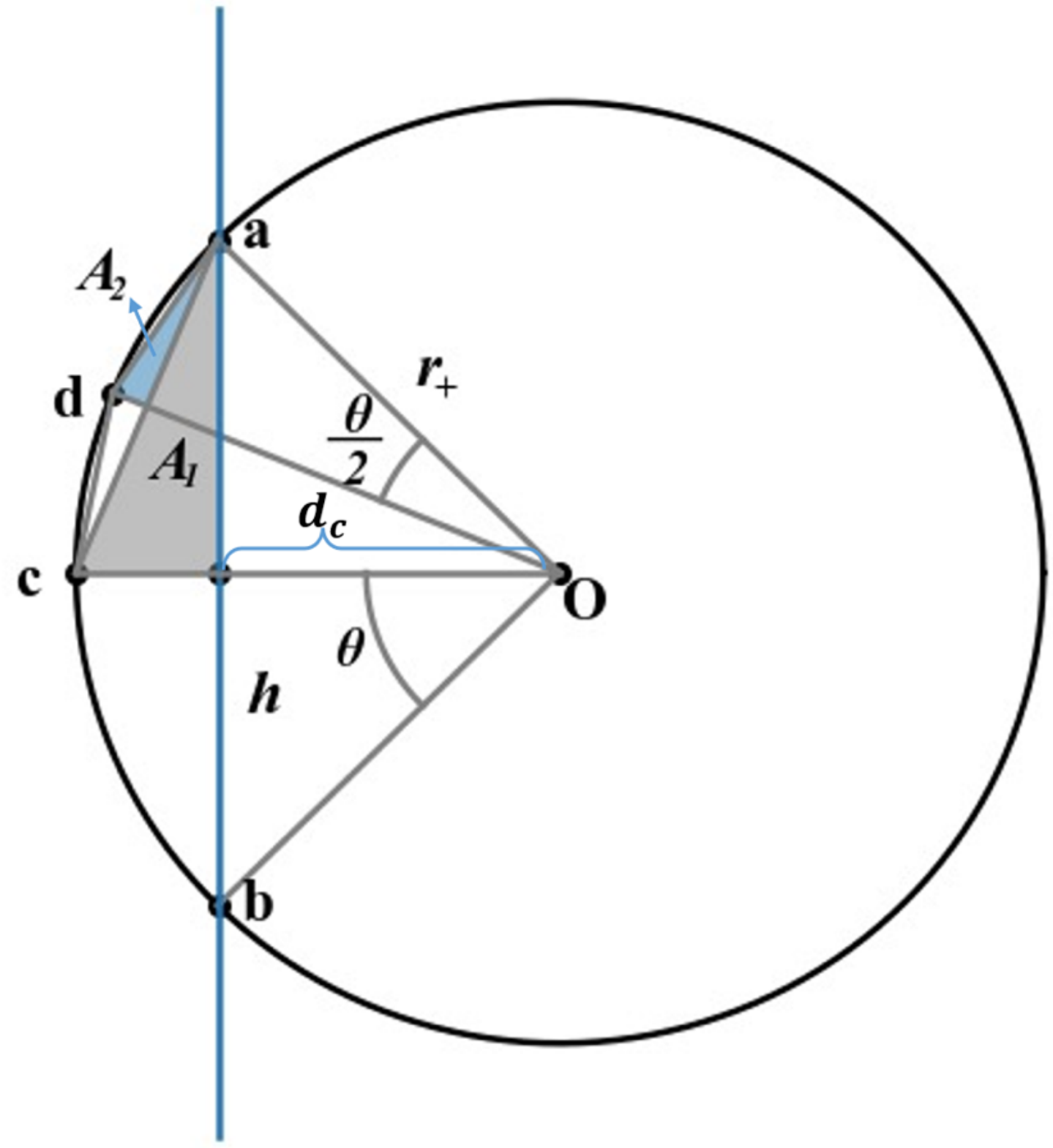

**Fig. S33: Variables involved in the approximation of interaction area for analytically solving plane-wave velocity.**

In order to better understand properties of the velocity, we apply some reasonable approximations to the transcendent equation to obtain analytical expression of the interaction area $A$. In particular, as show in Fig. S33, we have

$$\begin{aligned} A &= 2A_1 + 2A_{\widehat{ac}} \\ &= 2A_1 + 2^2A_2 + 2^2A_{\widehat{ad}} \\ &= 2A_1 + 2^2A_2 + \cdots + 2^nA_n, \end{aligned}$$

where $A_{\widehat{ac}}$ is the area under arch $\widehat{ac}$, and $A_{\widehat{ad}}$ is the area under arch $\widehat{ad}$.

For sufficiently large $r_+$, $\theta$ is sufficiently small, such that we have

$$A_1 = \frac{1}{2}r_+ \sin\theta \cdot r_+(1-\cos\theta) \approx \frac{1}{4}r_+^2\theta^3,$$

$$A_2 = \frac{1}{2}r_+ \sin\frac{\theta}{2} \cdot r_+\left(1-\cos\frac{\theta}{2}\right) \approx \frac{1}{4}r_+^2\left(\frac{\theta}{2}\right)^3 = \left(\frac{1}{2}\right)^3 A_1,$$

$$\vdots$$

$$A_n = \frac{1}{2}r_+ \sin\frac{\theta}{2^{n-1}} \cdot r_+\left(1-\cos\frac{\theta}{2^{n-1}}\right) \approx \frac{1}{4}r_+^2\left(\frac{\theta}{2^{n-1}}\right)^3 = \left(\frac{1}{2^{n-1}}\right)^3 A_1.$$

Thus, the interaction area $A$ becomes

$$\begin{aligned} A &= 2A_1 + 2^2 A_2 + \cdots + 2^n A_n \\ &= 2A_1 + 2A_1 \cdot \left(\frac{1}{4}\right)^1 + \cdots + 2A_1 \cdot \left(\frac{1}{4}\right)^n \\ &= \frac{8}{3} A_1 \\ &= \frac{4}{3} h(r_+ - d_c) \\ &= \frac{4}{3} \sqrt{{r_+}^2 - d_c^2} \cdot (r_+ - d_c). \end{aligned}$$

For sufficiently large $d_c$, we can write $d_c = r_+ - \Delta$, where $\Delta \ll d_c$ and $\Delta \ll r_+$. Therefore, we have

$$\begin{aligned} A &= \frac{4}{3} \Delta \sqrt{\Delta(2r_+ + \Delta)} \\ &\approx \frac{4}{3} \Delta \sqrt{\Delta(2r_+)}. \end{aligned}$$

Inserting the formula of $A$ to the transcendent equation

$$A w_+ = 1,$$

we have

$$\Delta = \sqrt[3]{\frac{1}{\left(\frac{4}{3} w_+\right)^2 \cdot 2r_+}},$$

and thus the velocity of plane waves becomes

$$\begin{aligned} v &= d_c \\ &= r_+ - \Delta \\ &= r_+ - \sqrt[3]{\frac{9}{32 r_+ \cdot w_+^2}}. \end{aligned}$$

Figure 12(a) in the main text shows a good agreement between this analytical result and simulation results of the plane-wave velocity.

# Target-wave velocity

Let us derive the velocity of target waves. The target-wave velocity is measured by the distance that wavefront travels in a unit time. Similar to plane waves, our task is how to derive the transition of joint force based on interaction range and states of individuals.

Without loss of generality, we focus on type-B individuals and see their wavefront. Indeed, there is an abrupt transition of states at $l$ from the upper bound to the lower bound at the wavefront. Similar to plane waves, we denote individual state at location $x$ at step $t$ as $s(t,x)$, and the individual states near the simplified wavefront can be expressed as a step function:

$$s(t,x) = \begin{cases} 1, & x \le l, \\ 0, & x > l. \end{cases}$$

This approximation allows us to analyze the transition of forces at the next-step wavefront. Let us recall the definition of joint force, i.e.,

$$F_i(t) = \sum_{|\boldsymbol{r}_j - \boldsymbol{r}_i| < r_+} s_j(t)\, w_+ \; - \sum_{|\boldsymbol{r}_k - \boldsymbol{r}_i| < r_-} s_k(t)\, w_-.$$

According to the definition, we need to identify all individuals who exert positive and negative forces to the next-step wavefront. For simplicity, we assume that the width of plane waves is significantly large, such that the inhibition from another type of individuals to the next-step wavefront is negligible. In general, all individuals who exert excitatory forces to location $l+d$ belong to the gray area in Fig. S34, where the black circle is the range of excitatory force and only the individuals within the gray area exert forces to location $l+d$. The states of individuals in this area reach the upper bound. Thus, the joint positive force, defined as the sum of the product of individual states and strength, equals the product of the number of individuals and strength $w_+$ in the gray area. And the number of individuals, according to the discrete-space arrangement, is just the size of the gray area. Taken together, the problem of deriving joint force at a location transfers to how to solve area size. To be specific, we focus on the location at $l+d$, and the area $A$ of the gray region can be formulated with respect to $l$ and $d$ as (see Fig. S34)

$$A(l,d) = l^2\theta_0 + r_+^2\theta_1 - 2\sqrt{p(p-l-d)(p-l)(p-r_+)},$$

where

$$\theta_0 = \arccos\left(\frac{l^2 + (l+d)^2 - r_+^2}{2l(l+d)}\right),$$
$$\theta_1 = \arccos\left(\frac{r_+^2 + (l+d)^2 - l^2}{2d(l+d)}\right),$$
$$p = \frac{1}{2}(2l + d + r_+).$$

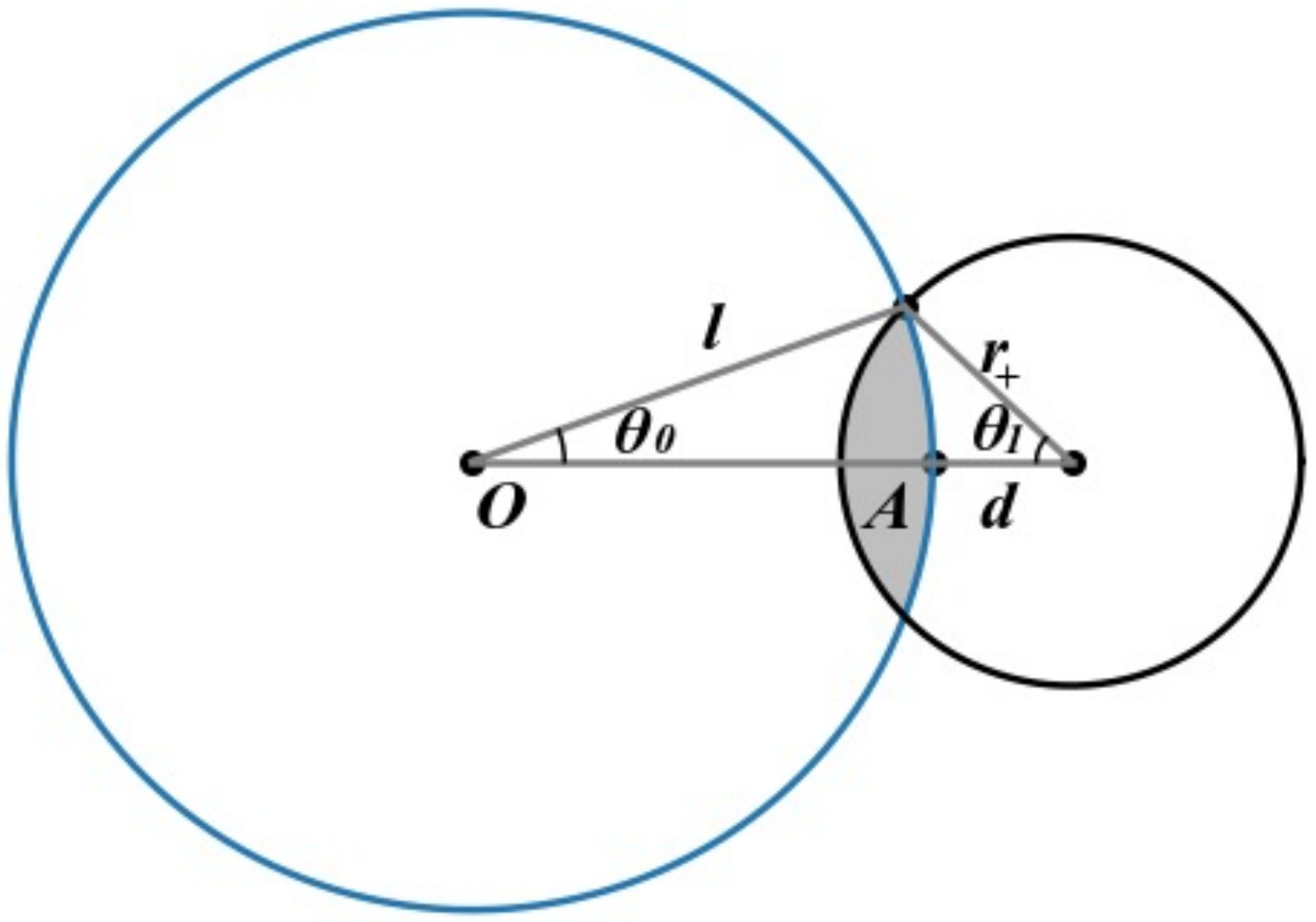

**Fig. S34: Illustration of interaction area and relevant variables for a target wave.**

Because the joint force $F$ equals the product of the number of individuals and strength $w_+$ in the gray area, we have

$$F(t, l+d) = A(l,d)w_+.$$

Next, we calculate the state value of the individuals at position $l+d$ at the next time $t+1$, based on the definition of state, i.e.,

$$s(t+1, l+d) = s(t, l+d) \; + \; f[F(t, l+d)],$$

where $s(t, l+d)$ is $0$ at time $t$. Thus, we have

$$s(t+1, l+d) = f(A(l, d)w_+),$$

where $f(A(l, d)w_+)$ is independent of $t$. Due to the sharp transition of states at the wavefront, we can exploit the formula of $s(t+1, l+d_c)$ to identify the location of the next-step wavefront. Specifically, at the next-step wavefront, we have

$$\begin{cases} s(t+1, l+d_c) = 1, \\ s(t+1, x) < 1, \quad \text{for} \;\; x > l + d_c, \end{cases}$$

where $l + d_c$ is the location of the next-step wavefront, and $d_c$ is the moving distance of the wavefront from step $t$ to $t+1$. These equations yield

$$f(A(l, d_c)w_+) = 1.$$

Moreover, note that the upper bound of function $f$ has no ceiling effect at any wavefront. Thus, we have

$$f(A(l, d_c)w_+) = A(l, d_c)w_+ = 1,$$

which yields the moving distance $d_c$, and the velocity of plane wave

$$v = d_c.$$

The velocity can be numerically solved from the transcendent equation

$$A(l, d_c)w_+ = l^2\theta_0 + r_+^2\theta_1 - 2\sqrt{p(p-l-d_c)(p-l)(p-r_+)},$$

where

$$\theta_0 = \arccos\left(\frac{l^2 + (l + d_c)^2 - r_+^2}{2l(l + d_c)}\right),$$
$$\theta_1 = \arccos\left(\frac{r_+^2 + (l + d_c)^2 - l^2}{2d_c(l + d_c)}\right),$$
$$p = \frac{1}{2}(2l + d_c + r_+).$$

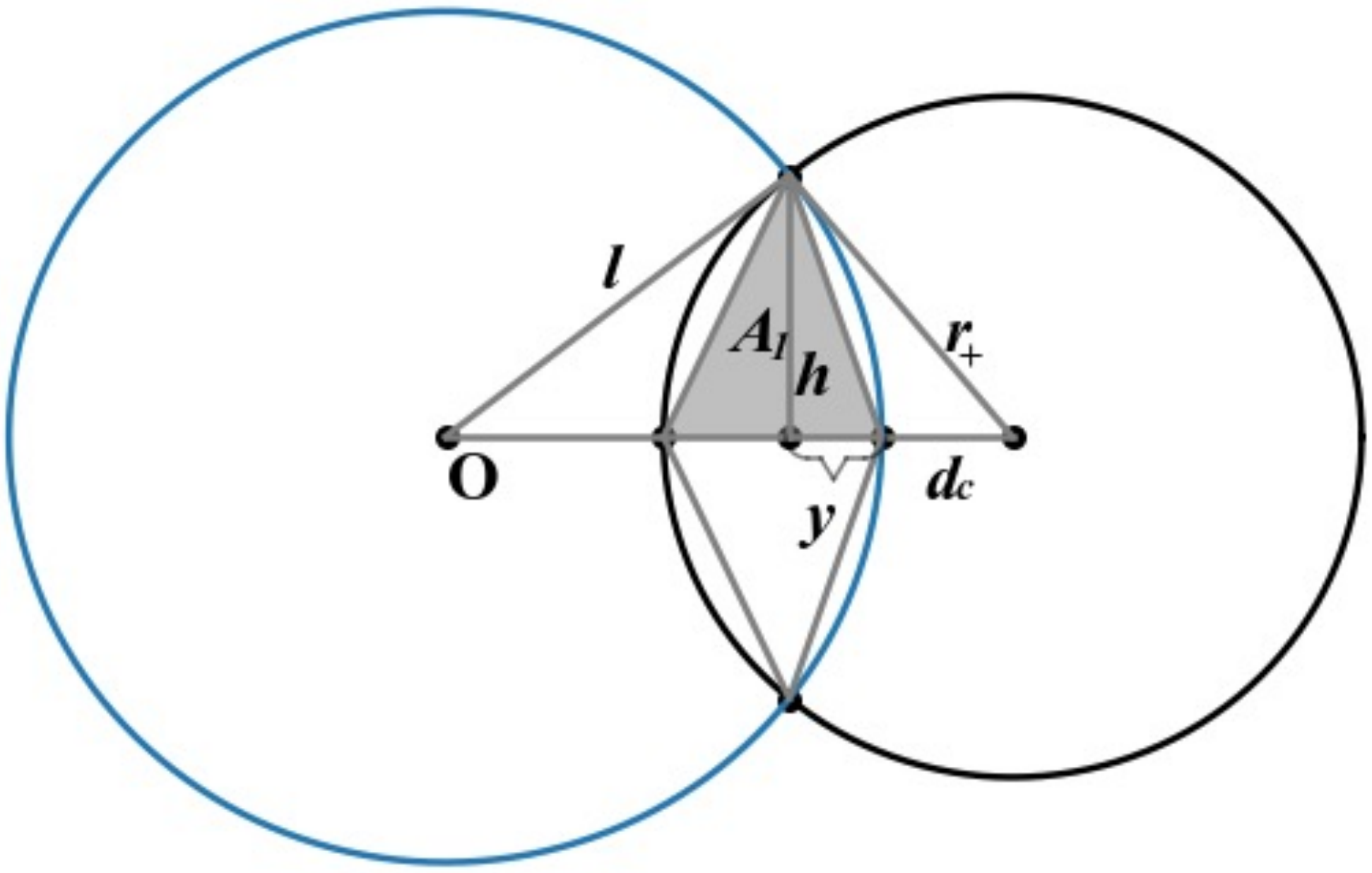

**Fig. S35: Variables involved in the approximation of interaction area for analytically solving target-wave velocity.**

To better understand properties of the velocity, we apply some reasonable approximations to the transcendent equation to obtain analytical results of the velocity. As show in Fig. S35, for sufficiently large $r_+$, similar to plane waves, we have

$$A = \frac{8}{3}A_1 = \frac{4}{3}h(r_+ - d_c).$$

In order to calculate $h$, we need to solve $y$ through a relationship of triangles

$$r_+^2 - (d_c + y)^2 = l^2 - (l - y)^2,$$

which leads to

$$y = \frac{r_+^2 - d_c^2}{2(d_c + l)}.$$

And then

$$h^2 = r_+^2 - (d_c + y)^2 = r_+^2 - \left(\frac{d_c^2 + r_+^2 + 2ld_c}{2d_c + 2l}\right)^2.$$

For sufficiently large $d_c$, we can write $d_c = r_+ - \Delta$, where $\Delta \ll d_c$ and $\Delta \ll r_+$, which yields

$$\begin{aligned} h^2 &= r_+^2 - \left[r_+^2 - \frac{2l\Delta - \Delta^2}{2(l + r_+ - \Delta)}\right]^2. \\ &\approx r_+^2 - \left[r_+^2 - \frac{l\Delta}{l + r_+}\right]^2 \\ &= r_+^2 - r_+^2\left(1 - \frac{l}{l + r_+}\frac{\Delta}{r_+}\right)^2. \\ &\approx r_+^2 - r_+^2\left(1 - 2\frac{l}{l + r_+}\frac{\Delta}{r_+}\right) \\ &= 2\frac{l\Delta}{r_+(l + r_+)}. \end{aligned}$$

Inserting

$$A = \frac{4}{3}h(r_+ - d_c)$$

into the transcendent equation

$$A \cdot w_{+} = 1,$$

we have

$$\sqrt{2\frac{l\Delta}{r_{+}(l+r_{+})}} \cdot \Delta = \frac{1}{\frac{4}{3}w_{+}},$$

which yields

$$\Delta = \sqrt[3]{\frac{l+r_{+}}{2r_{+}l\left(\frac{4}{3}w_{+}\right)^{2}}},$$

and the target-wave velocity is

$$\begin{aligned} v &= d_c \\ &= r_{+} - \Delta \\ &= r_{+} - \sqrt[3]{\frac{9(l+r_{+})}{32r_{+} \cdot l \cdot w_{+}^{2}}}. \end{aligned}$$

The asymptotic line of $v$ for sufficiently large $l$ is

$$v \approx r_{+} - \sqrt[3]{\frac{9}{32r_{+} \cdot w_{+}^{2}}}.$$

Figure 12(b) in the main text shows a good agreement between this analytical prediction and simulation results of the plane-wave velocity.

## Radial velocity of spiral waves

Let us derive the radial velocity of spiral waves. As shown in Fig. S36, when far from the center, an isometric spiral resembles a circle. Thus, at locations far from the center, the radial velocity of spiral waves is approximately equal to the velocity of target waves:

$$v = r_+ - \sqrt[3]{\frac{9(l + r_+)}{32r_+ \cdot l \cdot w_+^2}}.$$

And for sufficiently long distances $l$ from the center

$$v = r_+ - \sqrt[3]{\frac{9}{32r_+ \cdot w_+^2}}.$$

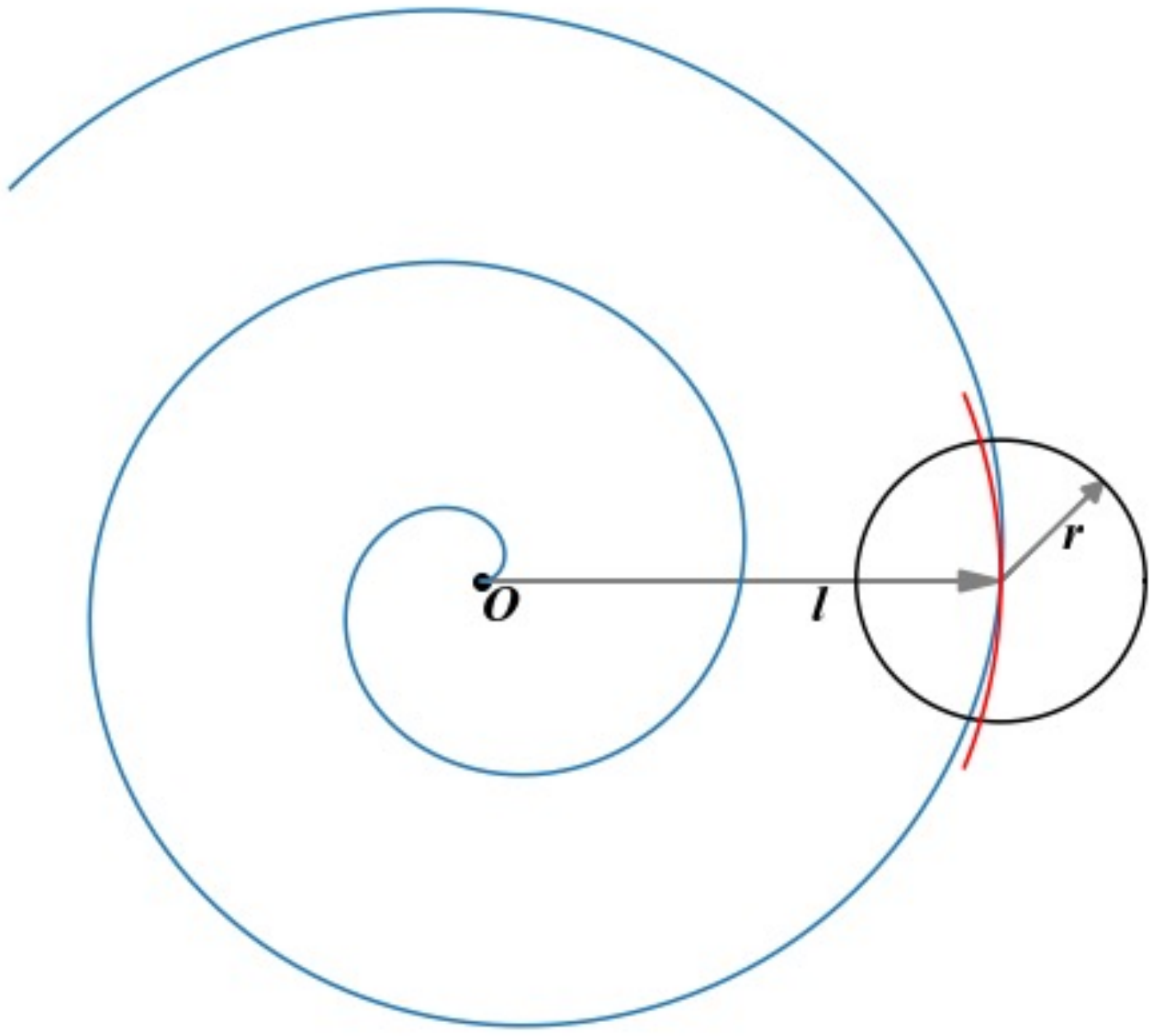

**Fig. S36: Illustration of the approximation of spiral to a circle curve at a distance far from the spiral center.**

## Rotational velocity of spiral waves

The rotational velocity is defined as the rotational angle of wavefront around the spiral center in one time step. In particular, the next-step wavefront around the center is at the equilibrium locations between the current excitatory and inhibitory forces. In this regard, the key is how to find the next-step equilibrium based on the current-step spatial configuration. Simulation results shed light on this issue. To be concrete, we found that around the center, three types of individuals form a regular Y shape boundary, as shown in Fig. S37(a). Moreover, associated with the Y shape, Fig. S37(b) shows the actual occupied areas of different individuals with overlap areas among them. The regular spatial configuration allows us to calculate the next-step equilibrium, i.e., the angle bisector between the boundaries of two adjacent individual areas, as shown in Fig. S37(b). The angle bisectors corresponding to force equilibrium become the next-step wavefront, as shown in Fig. S37(c). According to the relevant geometrical features, we can simply obtain the unit-time rotational angle

$$\Delta\theta = \frac{\pi}{6}.$$

The comparison between this theoretical result of rotational velocity and simulation results is shown in Fig. 12(d) in the main text.

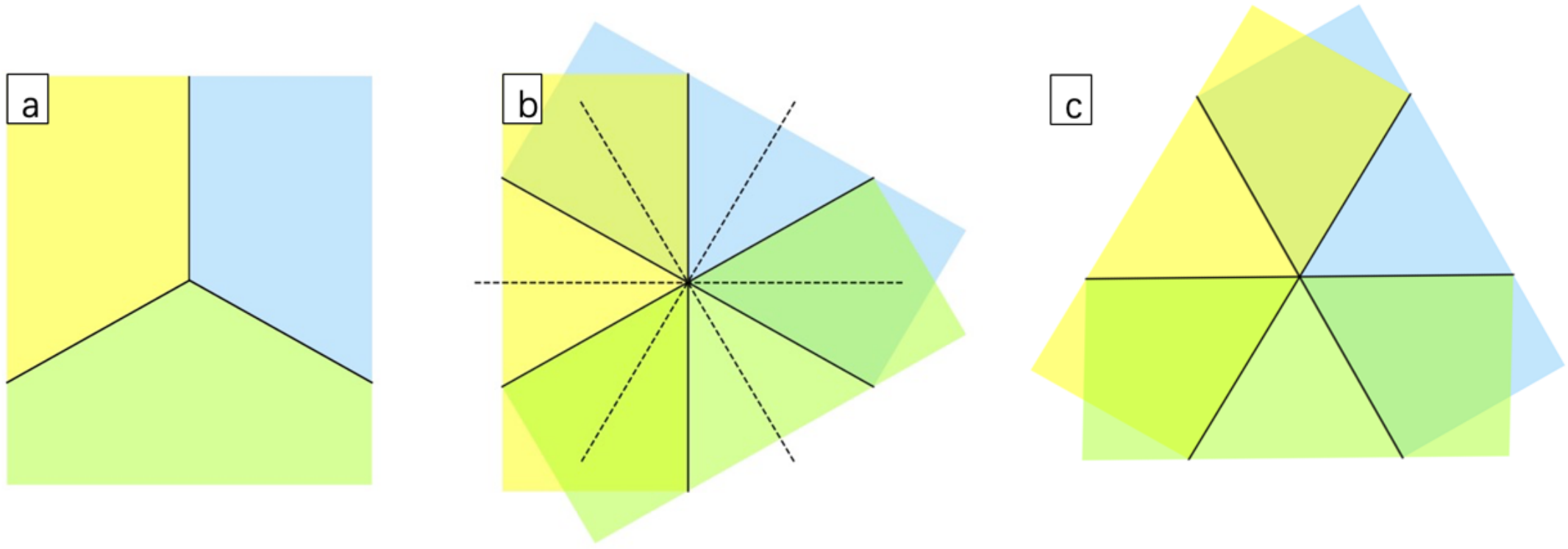

**Fig. S37: Illustrations of rotation around a spiral center.** (**a**) The spatial relationship among three types of individuals at the spiral center. (**b**) The actual occupied areas of three types of individuals around the spiral center and the equilibrium locations (dashed lines) among them. (**c**) The next-step occupied areas of three types of individuals.

## Spacing of isometric spiral waves

The spacing of isometric spiral waves can be inferred by the fact that when the spiral center makes a full rotation, the outer wavefront travels a unit interval between adjacent spiral coils. And each of the unit intervals contains three bands of individuals with the width of each band equaled to $1/3$ of the interval. Thus, the unit interval can be obtained from the product of radial velocity and the time of a full rotation. The width of each individual band is $1/3$ of the interval, i.e.,

$$\begin{aligned}\Delta l &= \frac{1}{3}\cdot\frac{2\pi}{\Delta\theta}\cdot v(l)\\ &= 4\left(r_+ - \sqrt[3]{\frac{9(l+r_+)}{32r_+\cdot l\cdot w_+^2}}\right).\end{aligned}$$

We compare theoretical result with simulation results in Fig. 12(e) in the main text. Except quite close to the spiral center, the analytical prediction is in good agreement with simulation results of the spacing of isometric spiral waves.

## Spacing of target waves

The spacing of target waves depends on the mutation rate at the target center and the radial velocity of target waves. Note that at the target center, after a time interval, a new type of individual emerges and begins to expand outward. And after the same time intervals, the center is occupied by another type of individuals, and a new band of individuals arises. Thus, the spacing is the traveling distance within the unit time interval $\Delta t$ of mutation, i.e., the product of the traveling velocity $v$ and $\Delta t$. Specifically, at a location with distance $l$ from the target center, we have

$$\begin{aligned}\Delta l(l) &= \Delta t \cdot v(l) \\ &= \Delta t\left(r_+ - \sqrt[3]{\frac{9(l + r_+)}{32 r_+ \cdot l \cdot w_+^2}}\right).\end{aligned}$$

The average spacing of target waves is equal to the average traveling velocity multiplied by the mutation rate. We approximate the average velocity using the velocity at a distance sufficiently far from the center, i.e., $l \to \infty$:

$$\Delta l = \Delta t\left(r_+ - \sqrt[3]{\frac{9}{32 r_+ \cdot w_+^2}}\right).$$

The analytical estimations of target-wave spacing are in good agreement with simulation results, as shown in Fig. 12(c) in the main text.

# The general-term formula of a modified Fibonacci sequence for expanding spirals

When we employed the Fibonacci sequences to fit the seashell spirals and the spirals in Fig. 8(e) in the main text, we found that the spirals generated by the standard Fibonacci sequence expand too fast in radial directions. We believe the absence of death in the Fibonacci sequence accounts for this disagreement. In this regard, we introduce a death process into the standard Fibonacci sequence to obtain a modified sequence. The spacing-expanding spiral generated by the modified sequence is in good agreement with the seashell spirals and the spirals in Fig. 8(e) in the main text.

Let us recall the generation process of the standard Fibonacci sequence and elaborate on the significance of introducing death into it. First, we briefly introduce the process of generating the standard Fibonacci sequence. Suppose a species has a one-step maturation period. After maturation, a pair of individuals reproduce one pair of offspring per step. The series starts from a single pair of individuals at the first step, denoted by one. At the second step, the first pair of individuals mature without reproduction. Thus, the number of pairs is still one. At the third step, the matured pair at the second step reproduce one pair of offspring, and the total number of pairs becomes two. At the fourth step, the pair born at the third step mature without reproduction, and the already matured pair reproduce one pair of offspring. The total number of pairs is three. Subsequently, following this rule, we obtain the Fibonacci sequence $1, 1, 2, 3, 5, 8, \ldots$

We introduce lifespan and death of individuals into the standard Fibonacci sequence. Let the lifespan of individuals be three steps. Any individuals born at step $t$ will die at step $t+3$ after reproduction. At the first step, the series starts from a single pair of individuals, denoted by one. At the second step, the pair mature without reproduction, and the total number of pairs is one. At the third step, the pair matured at the second step reproduce one pair of offspring, and the total number of pairs is two. At the fourth step, the pair born at the first step die after reproduction, and the pair born at the third step mature. The total number of pairs is two. At the fifth step, the pair matured at the fourth step reproduce one pair of offspring, and the pair born at the fourth step mature. The total number of pairs becomes three. According to the rule, we obtain the modified Fibonacci sequence $1, 1, 2, 2, 3, 4, 5, 7, \ldots$

Let us derive the recursive formula $F_t$ of the modified Fibonacci sequence. We firstly provide the relationship between the total number of pairs $F_t$ and the number of newborn pairs in previous steps. We denote the number of newborn pairs at step $t$ as $G_t$. For instance, at step four, $F_4 = G_2 + G_3 + G_4 = 0 + 1 + 1 = 2$, regardless of the newborn pairs at the first step. This is because the pairs born at the first step have already died, and have no effect on $F_4$. In a similar vein, we have $F_5 = G_3 + G_4 + G_5 = 3$. Thus, the general formula of the relationship between $F_t$ and $G_t$ is

$$F_t = G_t + G_{t-1} + G_{t-2},$$

Secondly, we derive the general recursive formula of $G_t$. Note that only those pairs matured before step $t$ contribute to $G_t$, and the pairs ought to be born before step $t-1$. Thus, $G_t$ might be related with $G_{t-2}$, $G_{t-3}$, $G_{t-4}$, and so on. Moreover, because of the three-step lifespan, the pairs of individuals born before $t-3$ have died at step $t$, and have no effect on $G_t$. Taken together, we simply have

$$G_t = G_{t-2} + G_{t-3}.$$

Based on the general formula of the relationship between $F_t$ and $G_t$, and the general recursive formula of $G_t$, we are able to derive the general recursive formula of $F_t$. Specifically,

$$\begin{aligned} F_{t+1} &= G_{t+1} + G_t + G_{t-1} \\ &= G_{t-1} + G_{t-2} + G_t + G_{t-1} \\ &= F_t + F_{t-1} - G_{t-2} - G_{t-3} \\ &= F_t + F_{t-1} - G_{t-3} - G_{t-4} - G_{t-5} \\ &= F_t + F_{t-1} - F_{t-3}. \end{aligned}$$

Thus, the general recursive formula of the modified Fibonacci sequence is

$$F_{n+1} = F_n + F_{n-1} - F_{n-3}.$$

Subsequently, we derive the general-term formula of the modified sequence by employing the same approach for deriving the standard Fibonacci sequence. To be concrete, we write

$$F_{n+4} - F_{n+3} - F_{n+2} + F_n = 0.$$

This constant-coefficient fourth-order linear recurrence equation has its characteristic equation

$$x^4 - x^3 - x^2 + 1 = 0.$$

The characteristic equation offers four characteristic roots that are the particular solutions of the general term of the modified Fibonacci sequence. The four particular solutions are as follows:

$$
\begin{cases}
\theta = 1, \\
\alpha = \sqrt[3]{\frac{1}{2}+\sqrt{\frac{23}{108}}}+\sqrt[3]{\frac{1}{2}-\sqrt{\frac{23}{108}}}, \\
\beta = \frac{-1+\sqrt{3}i}{2}\cdot\sqrt[3]{\frac{1}{2}+\sqrt{\frac{23}{108}}}-\frac{1+\sqrt{3}i}{2}\cdot\sqrt[3]{\frac{1}{2}-\sqrt{\frac{23}{108}}}, \\
\gamma = -\frac{1+\sqrt{3}i}{2}\cdot\sqrt[3]{\frac{1}{2}+\sqrt{\frac{23}{108}}}+\frac{-1+\sqrt{3}i}{2}\cdot\sqrt[3]{\frac{1}{2}-\sqrt{\frac{23}{108}}}.
\end{cases}
$$

Thus, the general-term formula of the modified Fibonacci sequence is a linear combination of the four particular solutions. We can thus assume the general term to be

$$F_n = C_1\theta^n + C_2\alpha^n + C_3\beta^n + C_4\gamma^n.$$

Due to $\theta^n = 1,$ and

$$F_1 = C_1 + C_2\alpha + C_3\beta + C_4\gamma = 1,$$

the general-term formula can be simplified to

$$F_n = C_2(\alpha^n - \alpha) + C_3(\beta^n - \beta) + C_4(\gamma^n - \gamma) + 1.$$

By substituting $F_2 = 1, F_3 = 2, F_4 = 2$ into the general-term formula, we obtain three equations for solving the coefficients:

$$
\begin{cases}
C_2(\alpha^2 - \alpha) + C_3(\beta^2 - \beta) + C_4(\gamma^2 - \gamma) = 0, \\
C_2(\alpha^3 - \alpha) + C_3(\beta^3 - \beta) + C_4(\gamma^3 - \gamma) = 1, \\
C_2(\alpha^4 - \alpha) + C_3(\beta^4 - \beta) + C_4(\gamma^4 - \gamma) = 1,
\end{cases}
$$

which yields coefficients

$$
\begin{cases}
C_2 = \frac{1}{(\alpha-\beta)(\alpha-\gamma)(\alpha-1)}, \\
C_3 = \frac{1}{(\beta-\alpha)(\beta-\gamma)(\beta-1)}, \\
C_4 = \frac{1}{(\gamma-\alpha)(\gamma-\beta)(\gamma-1)}.
\end{cases}
$$

By substituting the coefficients into the general-term formula, we finally have

$$F_n = \frac{(\alpha^n - \alpha)}{(\alpha - \beta)(\alpha - \gamma)(\alpha - 1)} + \frac{(\beta^n - \beta)}{(\beta - \alpha)(\beta - \gamma)(\beta - 1)} + \frac{(\gamma^n - \gamma)}{(\gamma - \alpha)(\gamma - \beta)(\gamma - 1)} + 1,$$

where

$$\begin{cases} \alpha = \sqrt[3]{\frac{1}{2} + \sqrt{\frac{23}{108}}} + \sqrt[3]{\frac{1}{2} - \sqrt{\frac{23}{108}}}, \\ \beta = \frac{-1 + \sqrt{3}\mathrm{i}}{2} \cdot \sqrt[3]{\frac{1}{2} + \sqrt{\frac{23}{108}}} - \frac{1 + \sqrt{3}\mathrm{i}}{2} \cdot \sqrt[3]{\frac{1}{2} - \sqrt{\frac{23}{108}}}, \\ \gamma = -\frac{1 + \sqrt{3}\mathrm{i}}{2} \cdot \sqrt[3]{\frac{1}{2} + \sqrt{\frac{23}{108}}} + \frac{-1 + \sqrt{3}\mathrm{i}}{2} \cdot \sqrt[3]{\frac{1}{2} - \sqrt{\frac{23}{108}}}. \end{cases}$$

We elaborate on the process of fitting shell spirals based on the modified Fibonacci sequence. The process of generating a spiral from the modified Fibonacci sequence differs significantly from the golden spiral based on the standard Fibonacci sequence. In particular, for the golden spiral, the spiral center changes between two consecutive sequence numbers. By contrast, we fix the spiral center and allow a certain rotational angle between two adjacent sequence numbers. Each sequence number corresponds to a distance from the spiral center to the spiral curve. There is a zoom factor that multiplies the sequence numbers (distances) to better fit the shell spirals. The distances between two adjacent sequence numbers (distances) within a certain rotational angle are simulated (fitted) by the cubic spline interpolation method, as shown in Fig. 12(f) in the main text.

There are two typical rotational angles between two adjacent sequence numbers, i.e., $\pi$ and $\pi/2$, which correspond to the outer- and inner-shell spirals, respectively. As shown in Fig. 8 in the main text, both the inner and outer spirals of three kinds of shells are in good agreement with our theoretical spirals based on the modified Fibonacci sequence. In addition, the theory also agrees well with our simulated spirals according to the asymmetric self-organization, as shown in Fig. 12(f) in the main text. On the contrary, the golden spiral based on the standard Fibonacci sequence expands too fast to mimic either inner or outer spirals. Moreover, we have also examined spirals with a fixed center based on the standard Fibonacci sequence. There is still a significant disagreement between the center-fixed golden spiral and the real shell spirals.

# Additional patterns

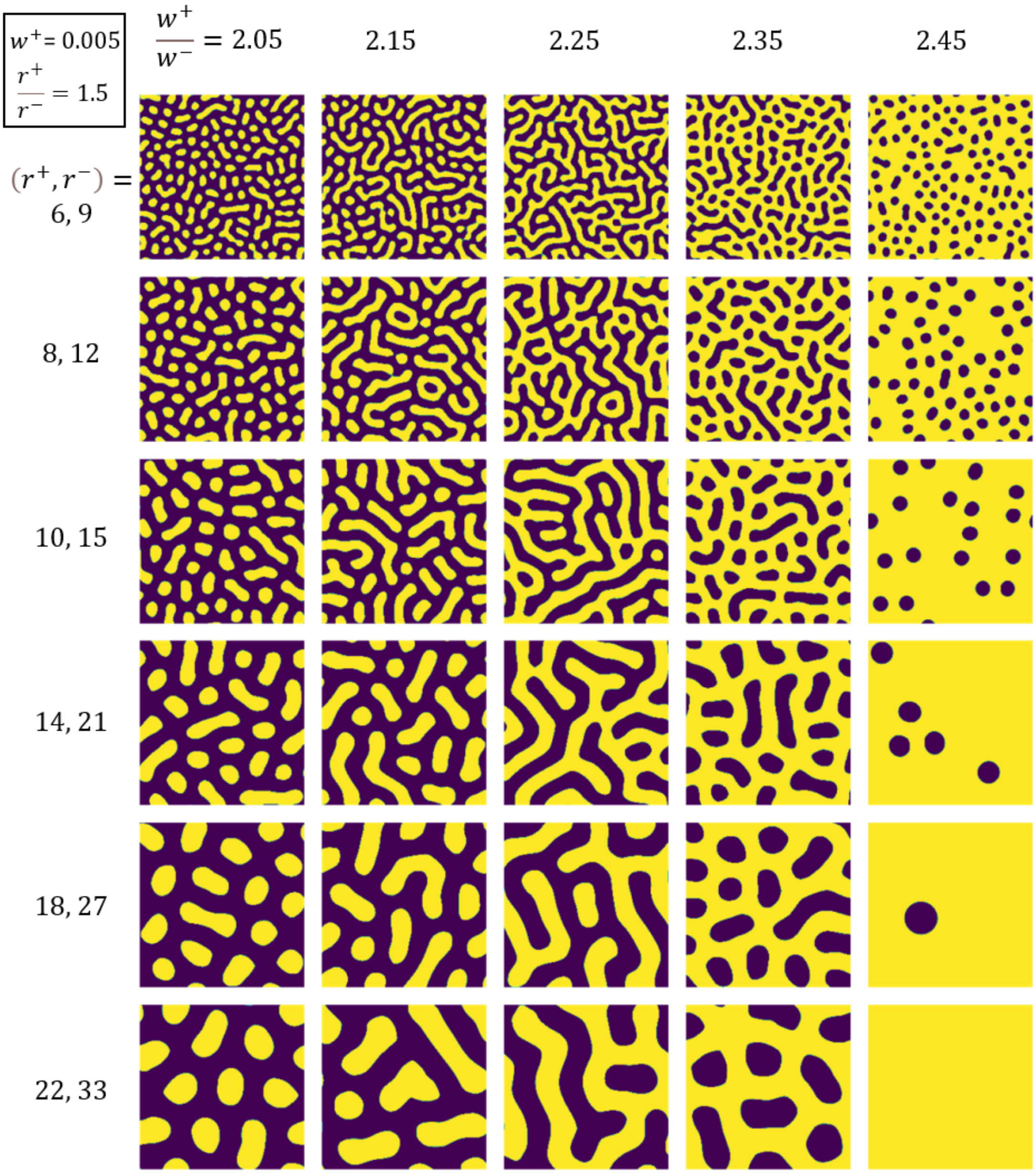

**Fig. S38: Random Turing patterns.**

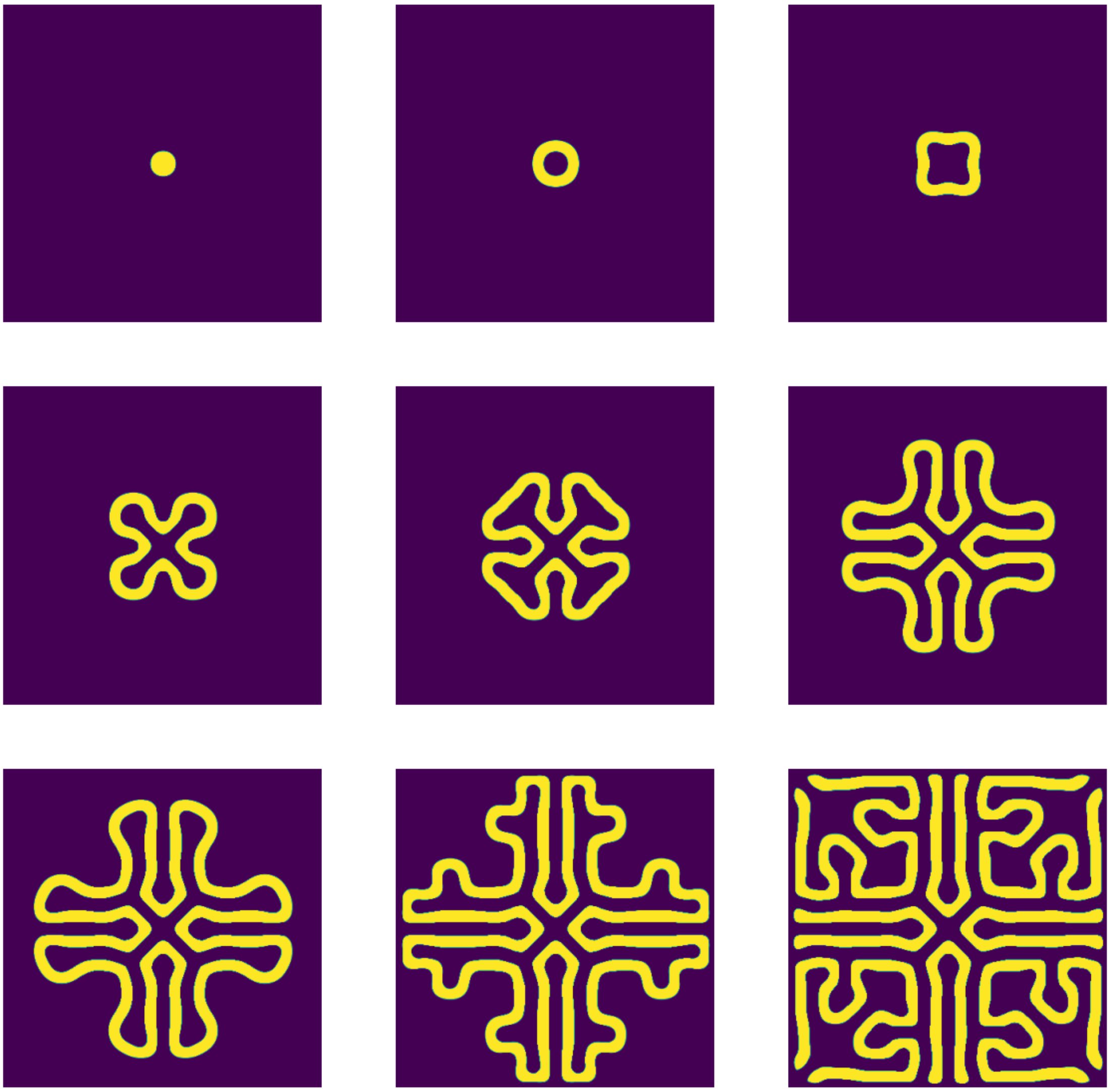

**Fig. S39: Pattern evolution near the first phase transition point.** $w_+ = 0.01$, $w_- = 0.0048$, $r_+ = 12$ and $r_- = 18$, $N = 401 \times 401$. The snapshots are from step 400 to 3600.

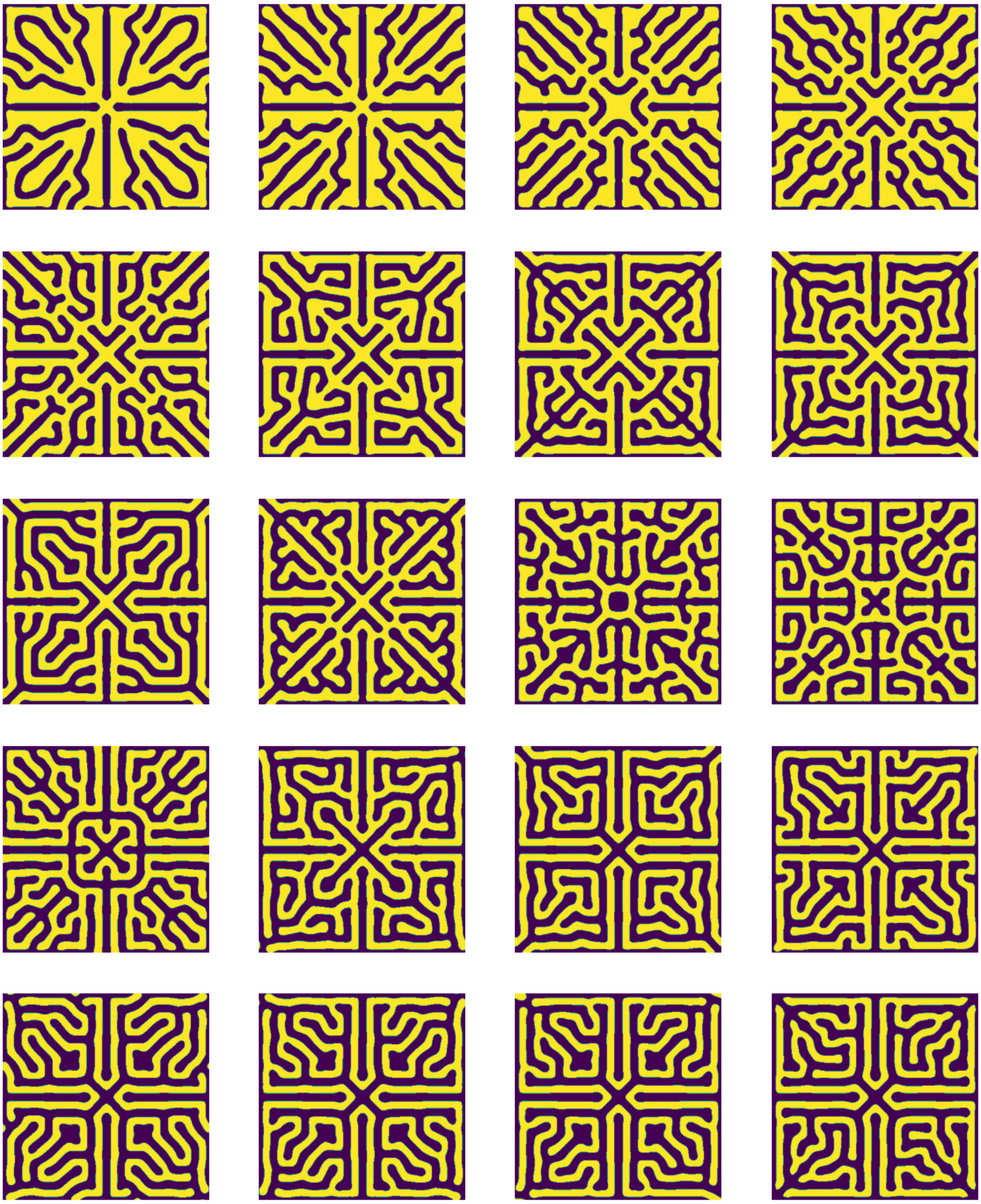

**Fig. S40: Steady Turing patterns between the first and second phase transition point.** Strength ratio $w_+/w_-$ is from $2.51$ to $2.16$, $r_+ = 12$, $r_- = 18$, $N = 401 \times 401$. The patterns are recorded after 5000 steps.

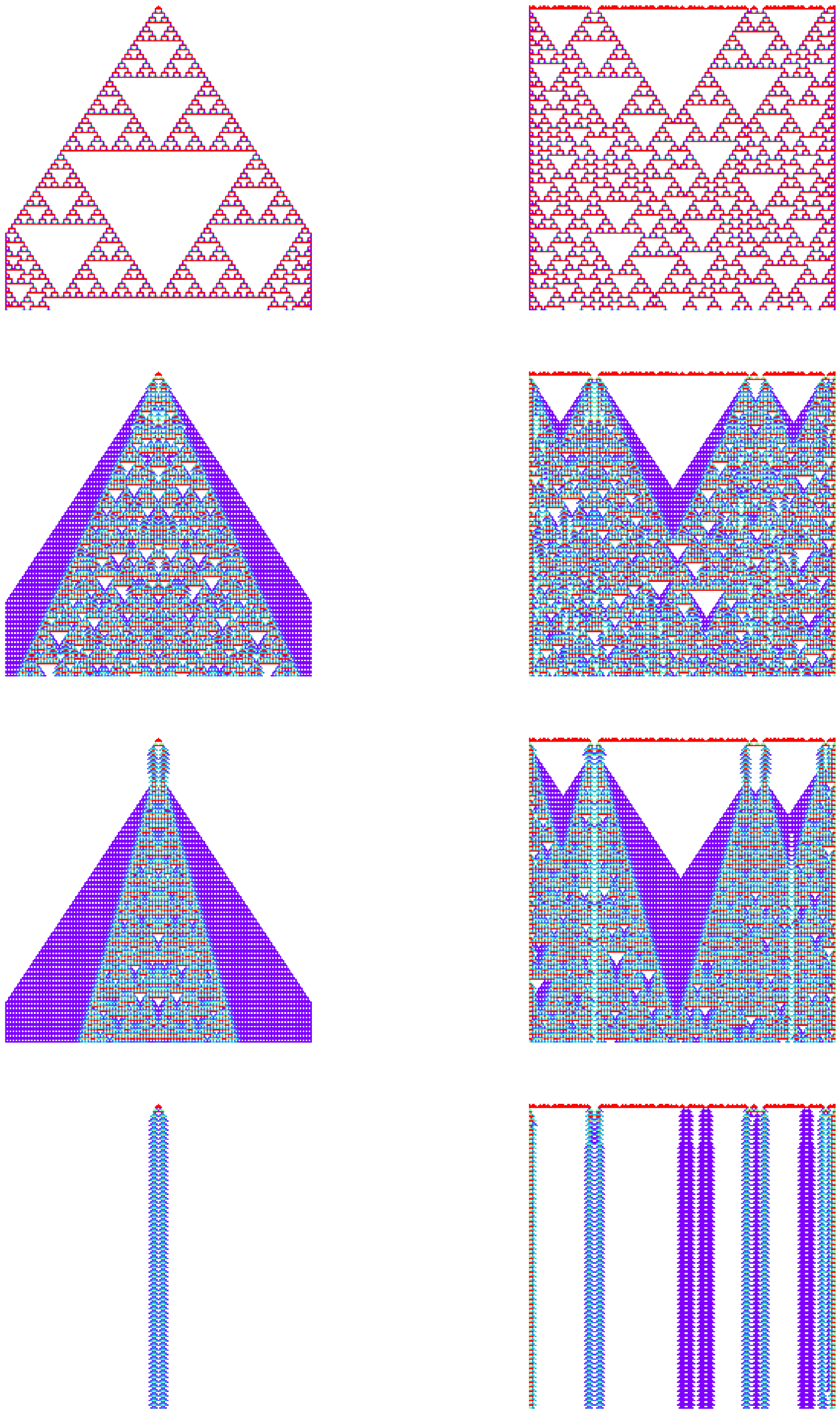

**Fig. S41: Fractal and chaotic patterns in one-dimension physical space.** The panels on the left-hand side panels initiate from a single excited individual at the center; the right-hand side panels initiate from the same random initial condition. From the top to the bottom, $w_+$ is from $\sqrt{2}$ to $\sqrt{2}-0.42$, $w_- = 0.5$, $r_+ = 12$, $r_- = 18$ and $N = 201 \times 201$. The color denotes state values of individuals.

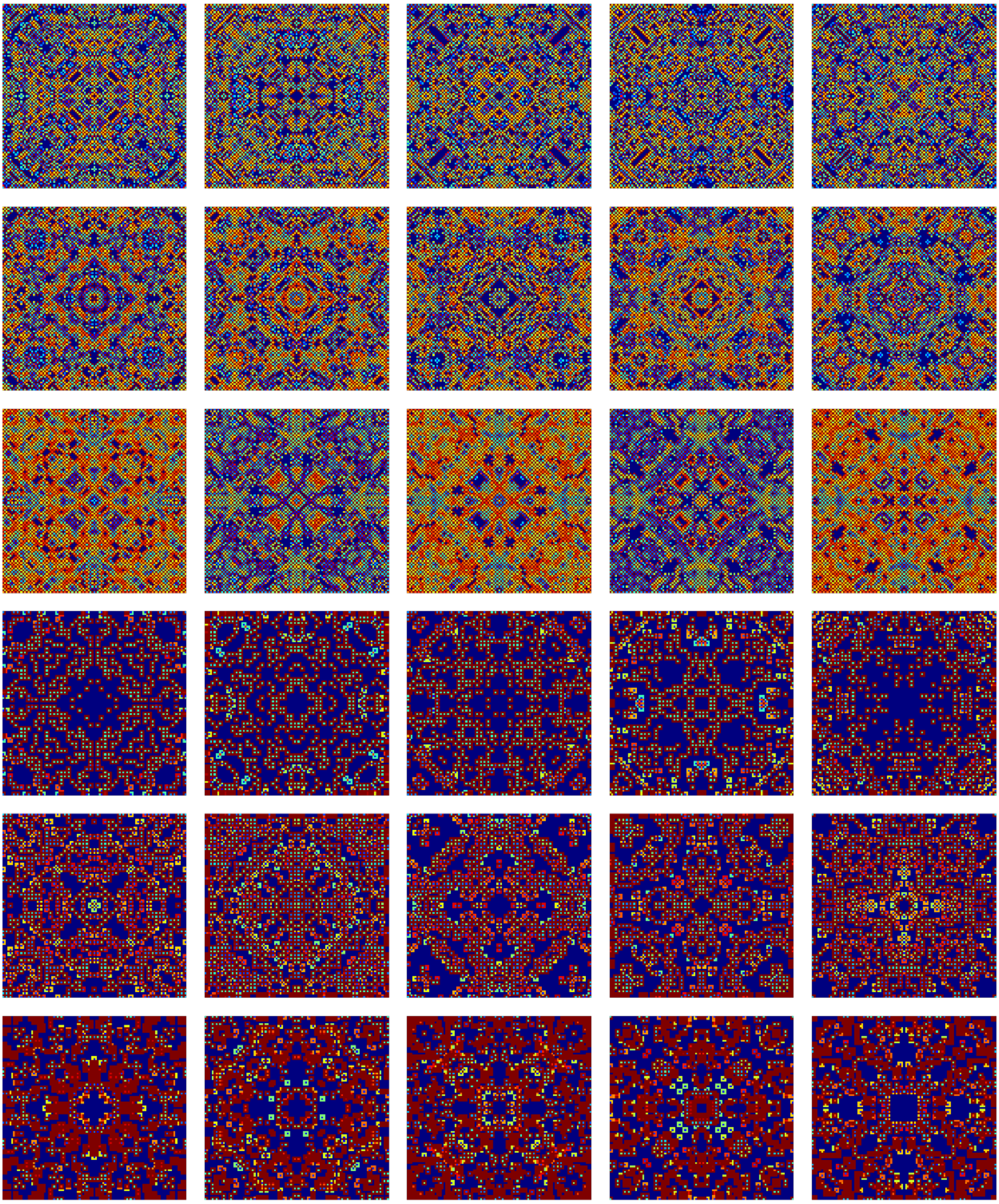

**Fig. S42: Chaotic kaleidoscopes in two-dimension physical space from breaking time symmetry.** The first to the third rows are associated with four neighbors and the fourth to the fifth rows are associated with eight neighbors. In each row, parameter values are the same and the panels are five randomly selected snapshots. From the top to the bottom rows, $w_+$ and $w_-$ are $(2, -1/2)$, $(2.5, -1/2.5)$, $(3, -1/3)$, $(4, -1/4)$, $(4.5, -1/4.5)$, $(5, -1/5)$, respectively. In all the panels, $r_+ = r_- = 1$, and $N = 101 \times 101$. The color denotes state values of individuals.

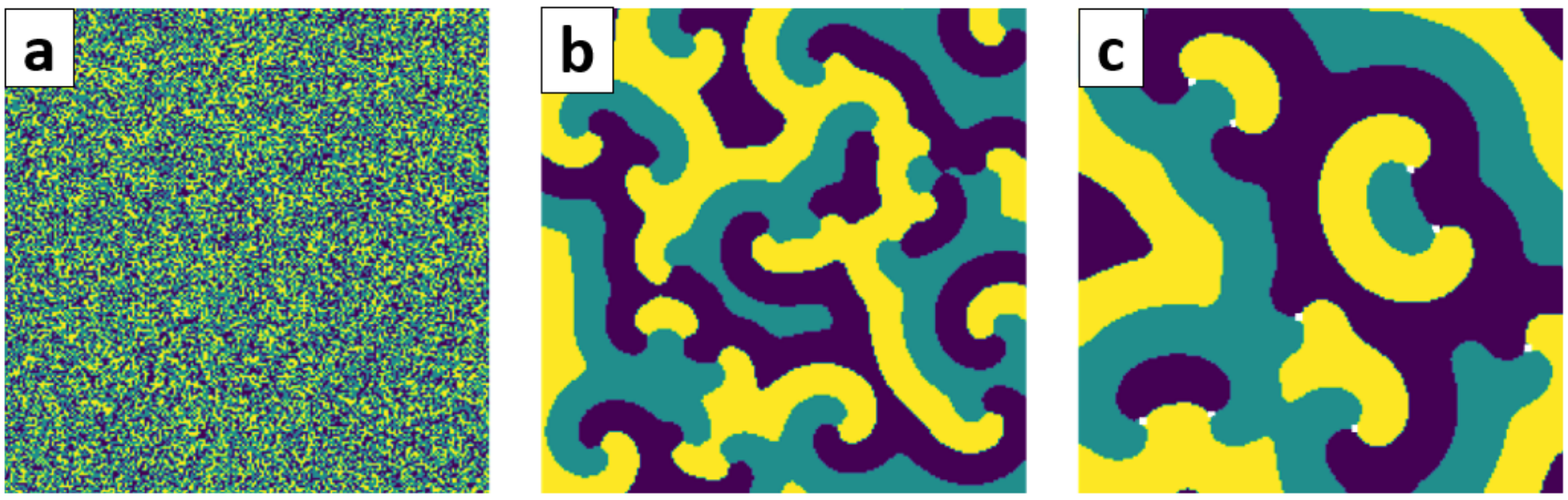

**Fig. S43: Random spirals from bidirectional-interaction asymmetry.** (**a**) Random initial condition, (**b**) smaller spirals generated from parameter value $r_+ = r_- = 3$, and (**c**) larger spirals from parameter value $r_+ = r_- = 6$. In all the panels, $w_+ = w_- = 0.1$ and $N = 201 \times 201$.

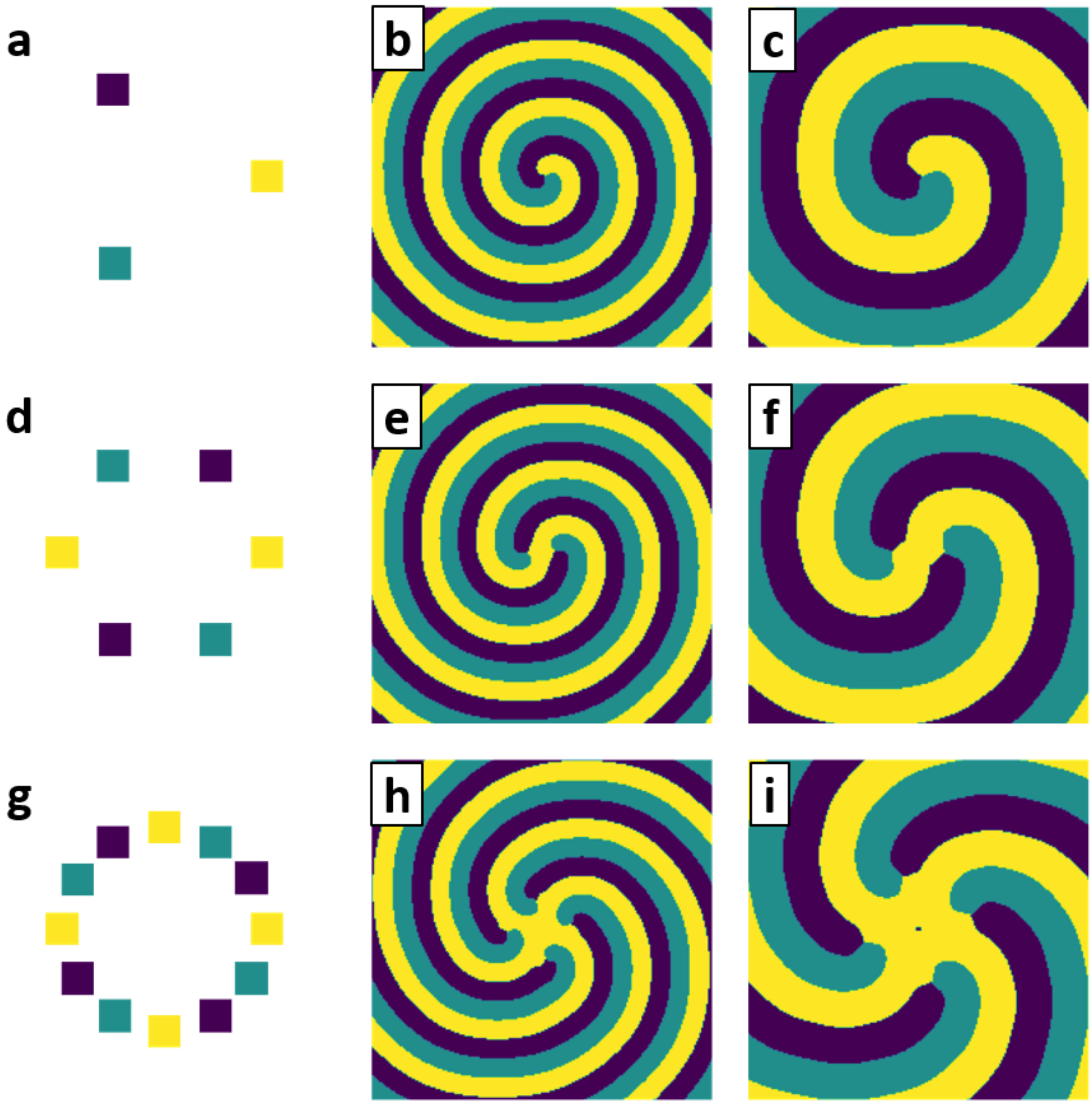

**Fig. S44: Multi-armed spiral waves from the combination of bidirectional-interaction asymmetry and initial-state asymmetry.** (**a**) to (**c**), (**d**) to (**f**) and (**g**) to (**h**) shows a single-armed spirals, two-armed spirals and four-armed spirals, respectively. Their initial configurations are shown in (**a**), (**d**) and (**g**), respectively. In (**b**) (**e**) (**h**), $r_+ = r_- = 3$, and in (**c**) (**f**) (**i**) $r_+ = r_- = 6$. In all the panels, $w_+ = w_- = 0.1$ and $N = 201 \times 201$.

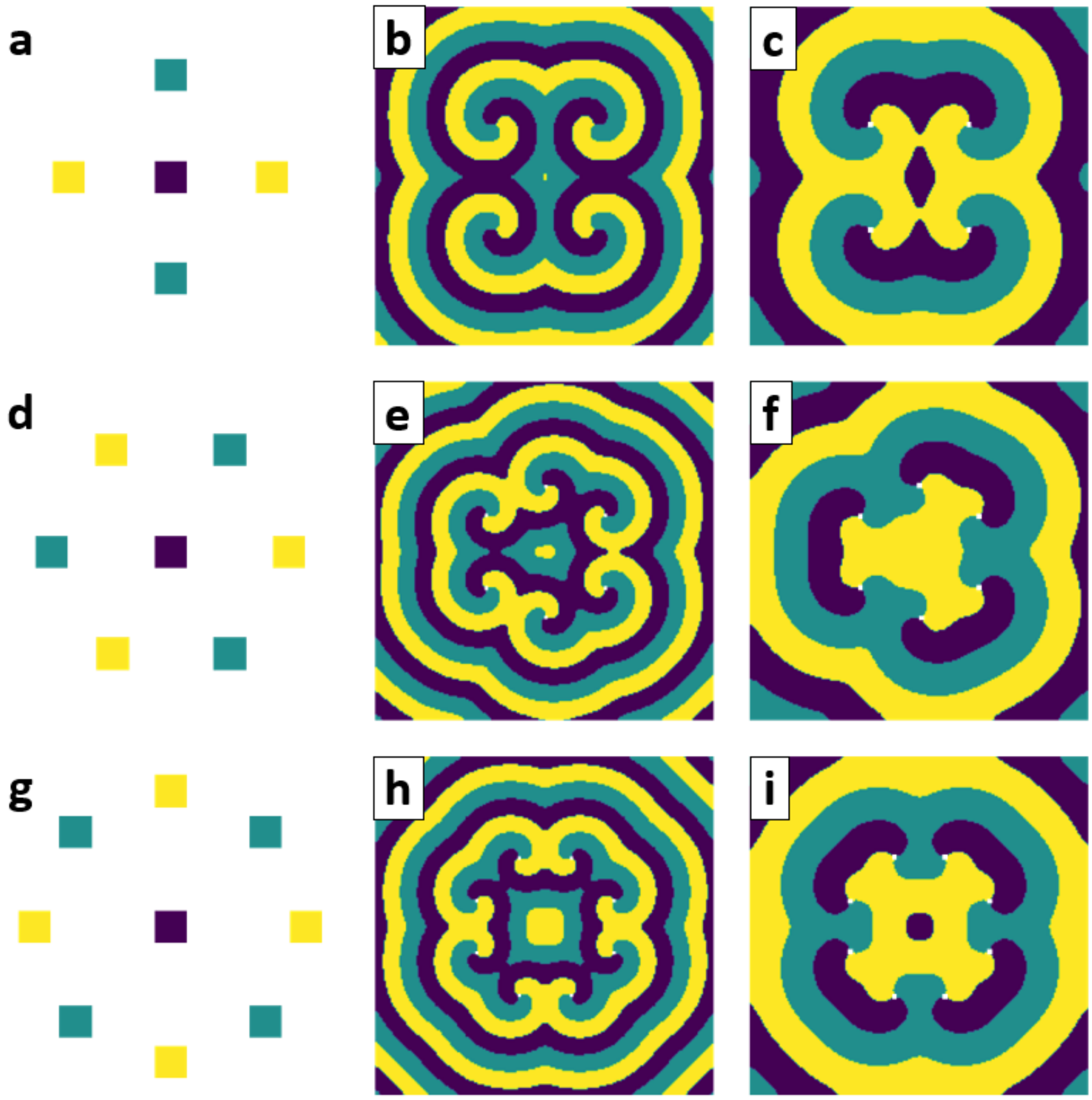

**Fig. S45: Spiral pairs from the combination of bidirectional-interaction asymmetry and initial-state asymmetry.** (**a**) to (**c**), (**d**) to (**f**) and (**g**) to (**h**) shows two-pair, three-pair and four-pair spirals, respectively. Their initial configurations are shown in (**a**), (**d**) and (**g**), respectively. In (**b**), (**e**) and (**h**), $r_+ = r_- = 3$, and in (**c**), (**f**) (**i**) $r_+ = r_- = 6$. In all the panels, $w_+ = w_- = 0.1$ and $N = 201 \times 201$.

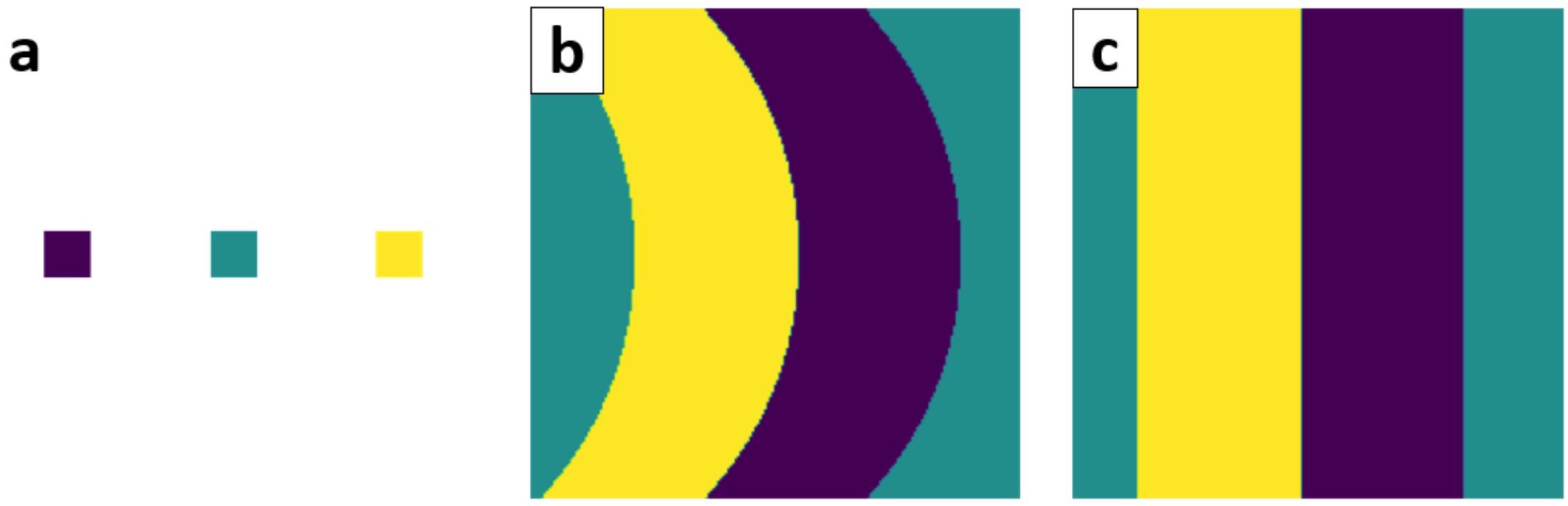

**Fig. S46: Plane waves from the combination of bidirectional-interaction asymmetry and initial-state asymmetry.** (**a**) Initial configuration, (**b**) a temporal pattern and (**c**) a steady pattern of plane waves after a sufficient number of steps. Parameter values are $r_+ = r_- = 6$, $w_+ = w_- = 0.1$ and $N = 201 \times 201$.

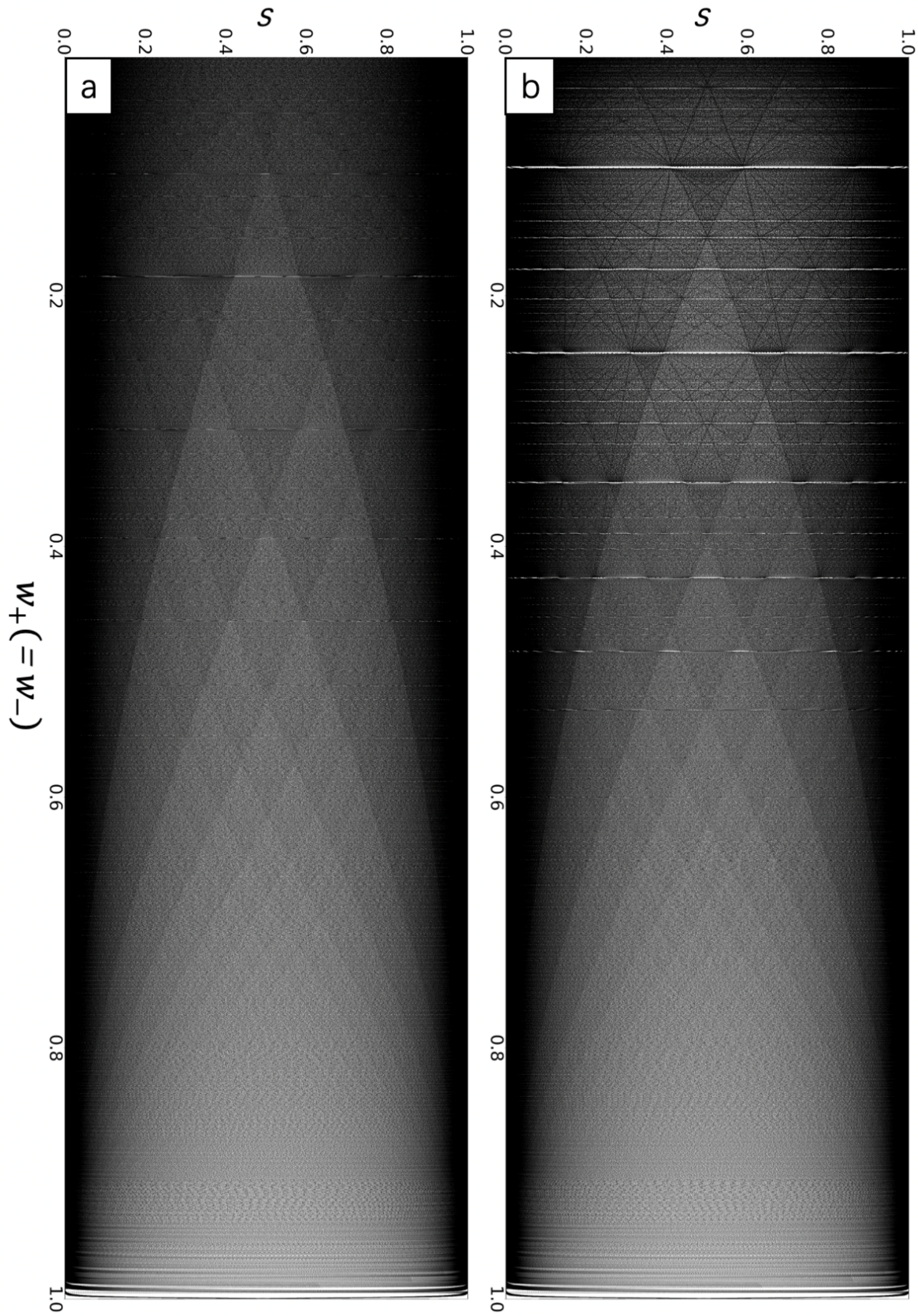

**Fig. S47: Chaos in state-parameter diagram of plane waves.** (**a**) (**b**) All states $s$ as a function of $w_+$ for (**a**) a plane wave with $r_+ = r_- = 1$, (**b**) a plane wave with $r_+ = r_- = 2$. $s$ is recorded during 500 steps after the first 1000 steps.

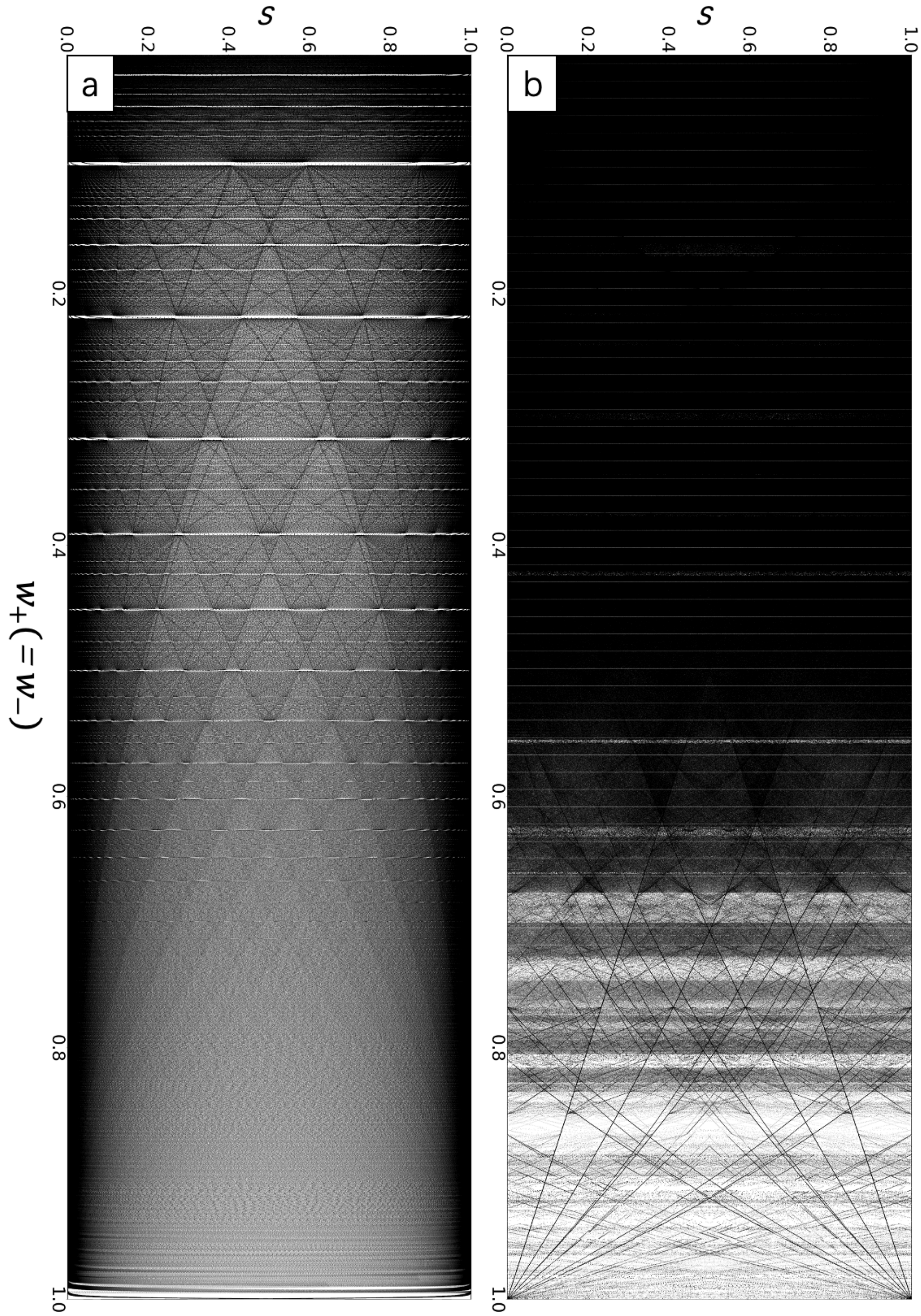

**Fig. S48: Chaos in state-parameter diagram of traveling waves.** (**a**) (**b**) All states $s$ as a function of $w_+$ for (**a**) a plane wave with $r_+ = r_- = 3$, (**b**) a spiral wave with $r_+ = r_- = 2$. $s$ is recorded during 500 steps after the first 1000 steps.